\documentclass[twocolumn]{aastex631}
\usepackage{amsmath,amsfonts,amssymb,graphicx,chngcntr,multirow, float,booktabs, afterpage}
\usepackage[]{hyperref}

\hypersetup{colorlinks=true}

\received{}
\revised{}
\accepted{}

\shorttitle{TDE calorimetry}

\begin{document}

\title{Tidal disruption event Calorimetry: Observational constraints on the physics of TDE optical flares}

\author{A. Mummery}
\affiliation{School of Natural Sciences, Institute for Advanced Study, 1 Einstein Drive, Princeton, NJ 08540, USA }
\author[0000-0002-4670-7509]{B.~D.~Metzger}
\affiliation{Department of Physics and Columbia Astrophysics Laboratory, Columbia University, New York, NY 10027, USA }
\affiliation{Center for Computational Astrophysics, Flatiron Institute, 162 5th Ave, New York, NY 10010, USA}
\author{S. van Velzen}
\affiliation{Leiden Observatory, Leiden University, Postbus 9513, 2300 RA Leiden, NL }
\author[0000-0002-5063-0751]{M. Guolo}
\affiliation{ Bloomberg Center for Physics and Astronomy, Johns Hopkins University, 3400 N. Charles St., Baltimore, MD 21218, USA }

\begin{abstract}
    Tidal disruption events are routinely discovered as bright optical/UV flares, the properties of which 
    are now well categorized on the population level. The underlying physical processes that produce the evolution of their X-ray emission and their long-lasting UV/optical plateau are well understood; however, the origin of their early-time optical/UV emission remains the subject of much debate and uncertainty. In this paper we propose and perform ``Calorimetric'' tests of published theories of these optical flares, contrasting theoretical predictions for the scaling of the radiated energy and peak luminosity of these flares with black hole mass (something which is predicted by each theory), with the observed (positive) black hole mass scaling. No one theory provides a satisfactory description of observations at all black hole mass scales. Theories relating to the reprocessing of an Eddington-limited compact accretion disk, or emission (energy) released in the formation of a Keplerian disk near the circularisation radius, perform best, but require extending. Models whereby the optical/UV flare are directly produced by shocks between debris streams (e.g., {\tt TDEmass}), or the efficient reprocessing of the fallback rate (e.g., {\tt MOSFIT}, or any other model in which $L \propto \dot{M}_{\mathrm{fb}}$), are ruled out at high $(>5\sigma)$ significance by the data. 
\end{abstract}

\section{Introduction}
Tidal disruption events (TDEs) are multi wavelength transients powered by the destruction of a star which has been scattered so close to a massive black hole that it enters its own tidal radius. These events offer great potential as probes of quiescent black hole demographics, are natural discovery channels for intermediate mass black holes, and are the only supermassive black hole systems where accretion flows evolve on $\sim$ year timescales. Understanding their multi wavelength emission properties, and relating these properties to the underlying physics of the events, remains an open and consequential problem in astrophysics. 

Time-domain surveys are rapidly expanding the known samples of TDE flares, such that statistically robust conclusions about the key observables and correlations with other properties of the TDE or environment are now becoming possible. A ``canonical'' TDE is  typically discovered as a luminous optical flare \citep[e.g.,][]{vanVelzen21,Yao23}, which may or may not coincide with rapid X-ray brightening \citep[e.g.,][]{Guolo24}. Optical luminosities of known TDEs peak at $L \sim 10^{42}-10^{45}\,\mathrm{erg\,s^{-1}}$ and decline on month-to-year  timescales \citep{Hammerstein23,Yao23}. 

While the intrinsic rate of X-ray and optically discovered TDEs are comparable \citep{Yao23,Sazonov21},  observational capabilities (i.e., survey efficiency not  intrinsic properties) mean that optical discoveries dominate the TDE population numerically, a fact that is only going to become more consequential as we enter the Rubin/LSST era \citep{Bricman20}. 

While data is abundant, consensus on the underlying physics producing these optical/UV flares is lacking. Indeed, the emission mechanisms at work in these optical/UV TDE flares remain a point of strong debate within the community. While these flares are well understood to be inconsistent with thermal emission from a thin accretion disk, little other consensus remains. Possibilities include: stream-stream collisions \citep[e.g.,][]{Piran+15, Dai+15,Ryu+20a, Krolik24}, reprocessing of black hole accretion by accretion disk winds \citep[e.g.,][]{Strubbe&Quataert09,MetzgerStone16, Piro&Lu20}, the reprocessing of the fallback rate by an extended photosphere \citep{Mockler19}, or the gravitational contraction of an extended pressure-supported quasi-spherical ``envelope'' to form a rotationally-supported disk at the circularization radius \citep{Metzger22}.

While the emission at early times in an optically-bright TDE is a point of much debate, at late times ($\Delta t \gtrsim 1\, {\rm yr}$) the optical/UV emission transitions to a plateau (weakly time dependent) phase \citep[][]{vanVelzen19, MumBalb20a, Mummery_et_al_2024}, which  arises naturally from a spreading disk formed from the stellar debris (this disk also produces the soft X-ray observed at all times, e.g., \citealt{Guolo25c}).   Importantly, this framework enables an independent estimate of the black hole mass at the center, which is known to agree with the $M_\bullet-\sigma$ value \citep{Mummery_et_al_2024} and the $M_\bullet-M_{\rm bulge}$ value \citep{Ramsden25}. With an efficient way of measuring black hole masses in TDEs (with optical flares), it is now possible to pose precise questions of the measured correlation between the properties of the early flare and the black hole masses inferred to be at the center of the events. 

Few theoretical models for the early time emission in TDE flares have been developed sufficiently deeply to make robust predictions for the observed light curve and spectral shape/evolution of the flare. However, the {\it integrated} energy released during this early phase, and the peak of the luminosity reached during the flare are predicted in each model framework. Thus, these observations (and their correlations with black hole mass) provide a clean  data-driven probe of the physics of the early flare.  It is the purpose of this paper to apply these ``calorimetric'' tests and reduce the size of the current parameter space of models (by ruling some out) which seek to describe these flares. 



The layout of this paper is as follows. In section \ref{sec:daat} we introduce the data set used in this work, and derive the observational constraints which  must be satisfied by models of the optical/UV flare in TDEs. In section \ref{sec:models} we collate all models from the literature, and derive (or reproduce) their predictions for the peak luminosity and radiated energy of the early flare. Model predictions are compared to the data in section \ref{sec:comp}, and we discuss the results in section \ref{sec:disc}. We conclude in section \ref{sec:conc}. In Appendix \ref{app:Z} we demonstrate that our results are robust against any Malmquist biases, and in Appendix \ref{app:M} we show that our results are robust to the choice of black hole mass inference technique. Appendix \ref{app:acc} shows how to relate the fallback rate to the accretion rate onto the black hole. 

\textcolor{white}{...}

\textcolor{white}{...}

\section{TDE Calorimetry}\label{sec:daat}
\subsection{Definition and calculation}
For a sample of optically-selected TDE flares, we calculate the total radiated energy during the early UV/optical phase (prior to the plateau) as follows:
\begin{equation}\label{Edef}
E_{\rm rad} \equiv \int_{-\infty}^{+\infty}L_{\rm BB}(t)\, {\rm d}t,
\end{equation}
where we calculate $L_{\rm BB}$ for each TDE by fitting the multi-band light curves of the system with a phenomenological model comprising of a Gaussian rise and exponential decay.  
Specifically, these phenomenological models have the functional form  
\begin{equation}
    L_{\rm rise}(\nu, t) = L_{\rm pk} \times f_{\rm rise}(t) \times \frac{\nu B(\nu, T)}{\nu_0 B(\nu_0, T)} , 
\end{equation}
and 
\begin{equation}
    L_{\rm decay}(\nu, t) = L_{\rm pk} \times f_{\rm decay}(t) \times \frac{\nu B(\nu, T)}{\nu_0 B(\nu_0, T)} , 
\end{equation}
where $B(\nu, T)$ is the Planck function, and $\nu_0 = 6 \times 10^{14}$ Hz is a reference frequency (approximately equal to the ZTF $g$-band). The amplitude $L_{\rm pk}$ and temperature $T$ are common to both the rise and decay models. The functions $f_{\rm rise}(t)$ and $f_{\rm decay}(t)$ are defined so as to have a maximum amplitude of unity, so that $L_{\rm pk}$ remains the physical peak amplitude. The rise model is that of a Gaussian rise
\begin{equation}\label{gauss_rise}
    f_{\rm rise}(t) = \exp\left( - {\left(t - t_{\rm peak}\right)^2 \over 2\sigma_{\rm rise}^2}\right) ,
\end{equation}
with fitting parameters $t_{\rm peak}$ and $\sigma_{\rm rise}$. The decay is modeled with an exponential 
\begin{equation}\label{pl_decay}
    f_{\rm decay}(t) = \exp\left(-{(t-t_{\rm peak})\over\tau_{\rm decay}}\right),
\end{equation}
with fitting parameters $t_{\rm peak}$, $\tau_{\rm decay}$. The parameter $t_{\rm peak}$ is by default assumed to be common between the rise and decay models, and if no rise model is specified the parameter is fixed to $t_{\rm peak} = 0$.  

We fit, following the statistical procedure spelt out in \cite{Mummery_et_al_2024}, the above profile to all TDEs in the {\tt manyTDE}\footnote{https://github.com/sjoertvv/manyTDE} database. We also include a constant ``plateau'' luminosity at late ($t\geq t_{\rm peak}$) times, with 
\begin{equation}
    L_{\rm plat}(t) = L_P \times \frac{\nu B(\nu, T_P)}{\nu_0 B(\nu_0, T_P)} ,
\end{equation}
which ensures that the luminosity/energy in the early flare is not contaminated by the emission from the accretion flow (which can be well approximated by a constant over the timescales of interest $\Delta t\sim$ years). The plateau luminosity $L_P$ is typically $\sim {\cal O}(1\%)$ of the peak luminosity $L_{\rm pk}$ \citep{Mummery_et_al_2024} so including this component at all times post-peak induces minimal contamination in $L_{\rm pk}$. 

We use the parameter $L_{\rm pk}$ and the mean temperature $T$ to obtain the integrated blackbody luminosity at peak ($L_{\rm BB}$).  
For a handful of our sources the temperature is very poorly constrained, in these cases the correction from spectral luminosity to blackbody luminosity is poorly constrained, and large uncertainties appear in the value of $L_{\rm BB}$ (despite them having well constrained  intra-band optical luminosities).

Performing the calorimetry integral (eq. \ref{Edef}) analytically, we find 
\begin{eqnarray}
    E_{\rm rad} = L_{\rm pk} \times k(T) \times \left[\tau_{\rm decay} + \sqrt{\pi\over 2} \sigma_{\rm rise}\right] ,
\end{eqnarray}
where $k(T)$ is the temperature-dependent correction factor for each TDE, which for our blackbody assumption is  
\begin{equation}
    k(T) \equiv {\int_0^\infty B_\nu(\nu, T) \, {\rm d}\nu \over \nu_0 B_\nu(\nu_0, T)} .
\end{equation}
For sources with no observed rise, we take $\sigma_{\rm rise}=0$. The observed energy in these sources is necessarily a lower-limit. Indeed, it is highly unlikely that the actual spectral energy distributions of TDE optical/UV flares are perfectly described by a blackbody function, and it is likely that the inferred energy/bolometric luminosity of the sources constrained here are also formal lower-limits. 

It is relatively (observationally) simple to see why this is – one infers a temperature $T$ of this early emission which is typically slowly evolving and in the ballpark of $\log_{10} T \sim 4.5$ (Kelvin). The corresponding spectral (blackbody) peak is thus  $\nu_{\rm peak} = b k T/ h \approx 2\times 10^{15}$ Hz (where $b = 3 + W(-3/e^3) \approx 2.82$ is the usual Wien-displacement constant, and $W$ is the Lambert W function). This frequency is above the typical observing bands used to infer the temperature. Therefore one is inferring a temperature from curvature in the spectrum, and almost never from an unambiguous peak frequency. This lack of spectral coverage combined with dust absorption in the host galaxy can lead to a systematic underestimation of the blackbody temperature \citep{Gezari+09}. Indeed \citep{Guolo25c} shows that there is dust in many TDE host galaxies 
which can induce curvature in optical/UV spectra, and spurious temperature constraints. As we are treating the integrated energies/peak blackbody luminosities as lower limits (in the sense that models which underpredict perform worse than those which overpredict) this dust contamination does not meaningfully impact our results.  

\subsection{Statistical correlations with black hole mass}
For every TDE in our sample, as well as the peak blackbody luminosity $L_{\rm BB, pk} \equiv L_{\rm pk} \times k(T)$ and radiated energy $E_{\rm rad}$, we also have multiple independent estimates of the black hole mass in the event. Three of these black hole mass estimates come from established galactic scaling relationships between the black hole mass and galaxy mass, bulge mass or velocity dispersion \citep[we use the scaling relationships presented in][for the velocity dispersion and galaxy mass and the \citealt{Kormendy13} bulge-mass scaling relationship, see Appendix \ref{app:M}]{Greene20}. The fourth comes from the event itself, whereby we use the plateau luminosity $L_P$ to infer the black hole mass using spreading-disk models \citep[see][for full details]{Mummery_et_al_2024}. These mass estimates are all consistent (Appendix \ref{app:M}). 

These black hole mass estimates allow us to pose two key calorimetric questions, namely how do $L_{\rm BB, pk}$ and $E_{\rm rad}$ scale with the mass of the black hole in the center of the event? It is a comparison of these empirical scaling relationships with theoretical estimates which forms the fundamental basis of our colorimetric tests.

\begin{figure*}
    \centering
    \includegraphics[width=0.45\linewidth]{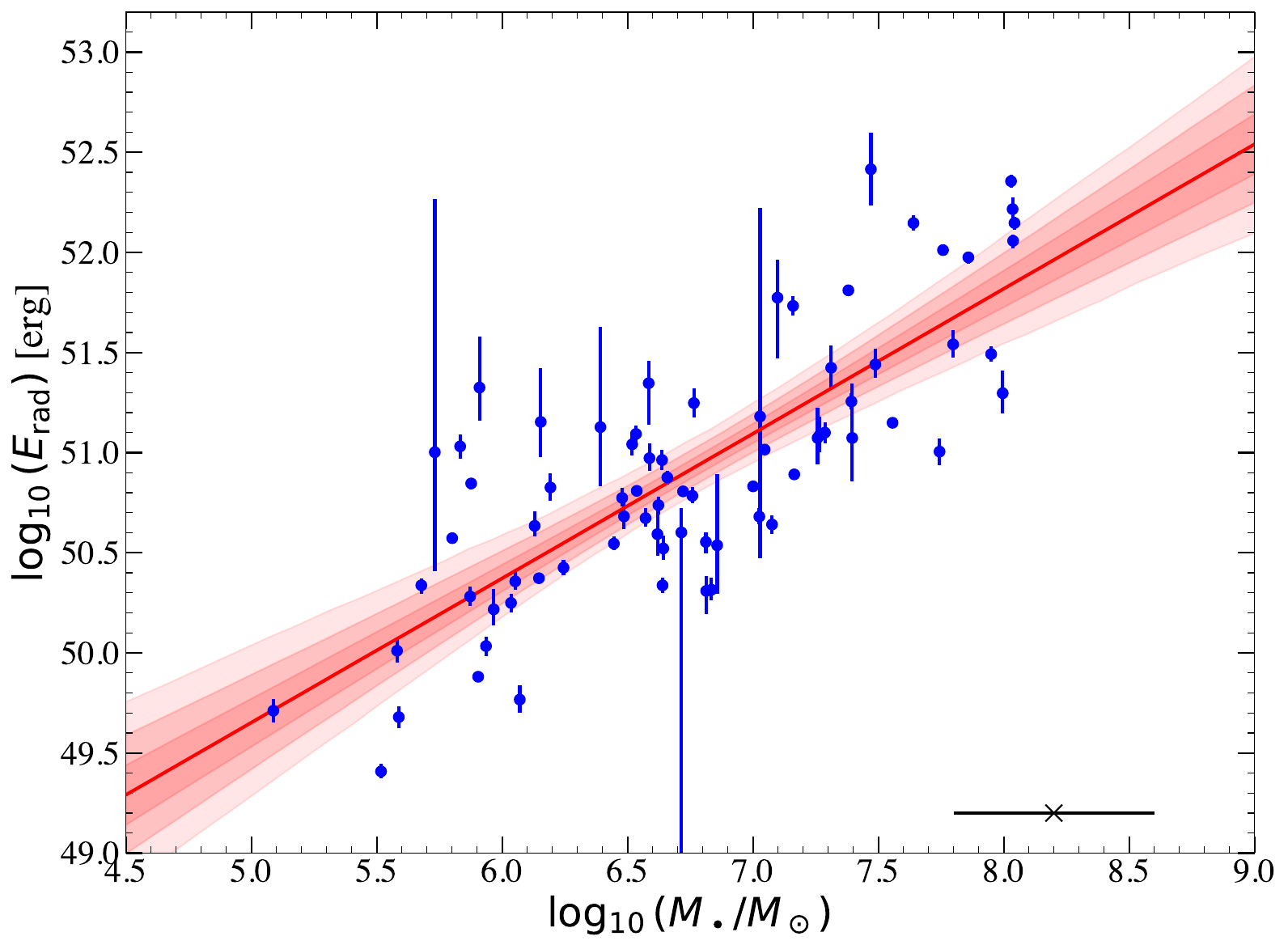}
    \includegraphics[width=0.45\linewidth]{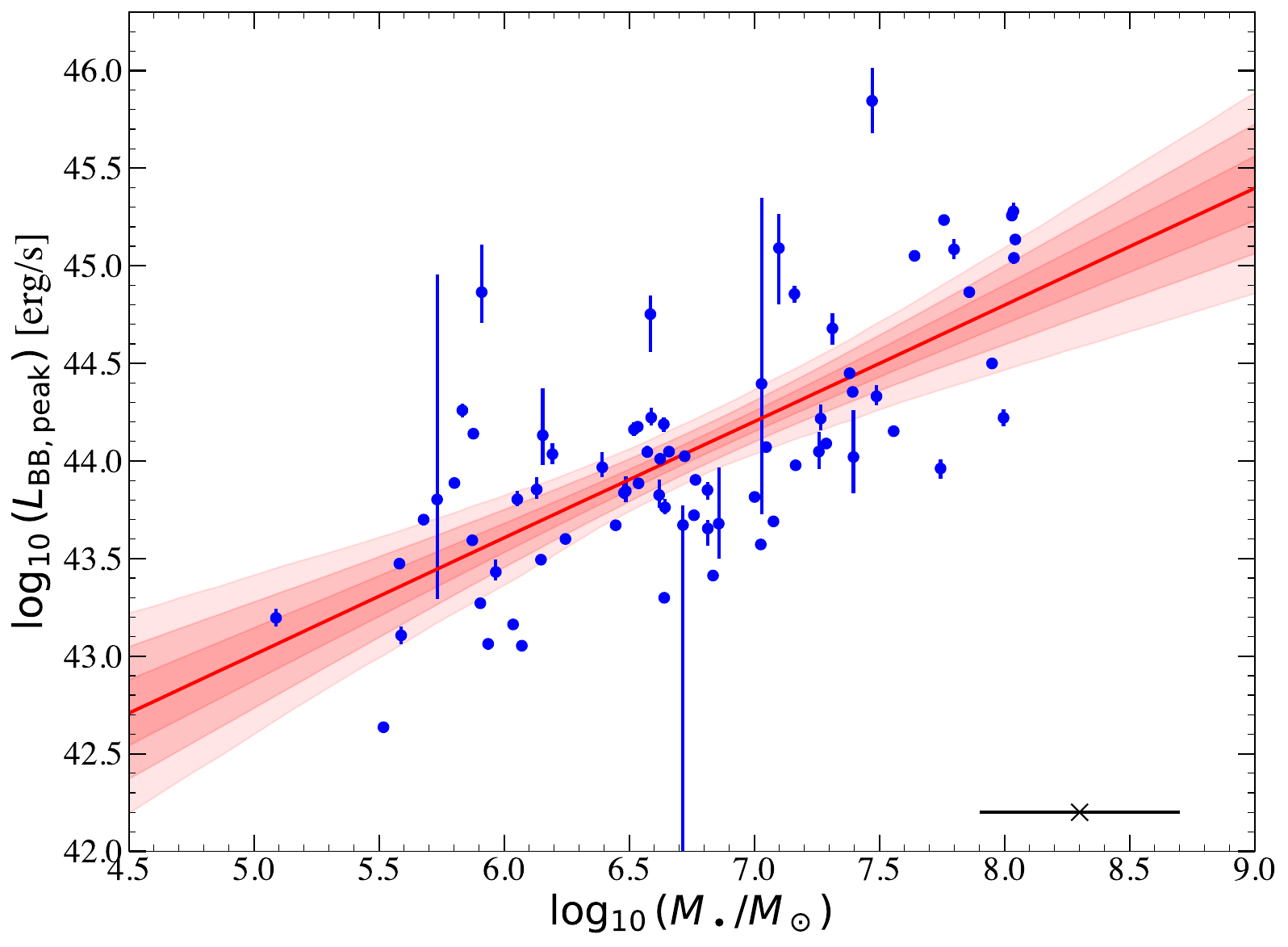}
    \includegraphics[width=0.45\linewidth]{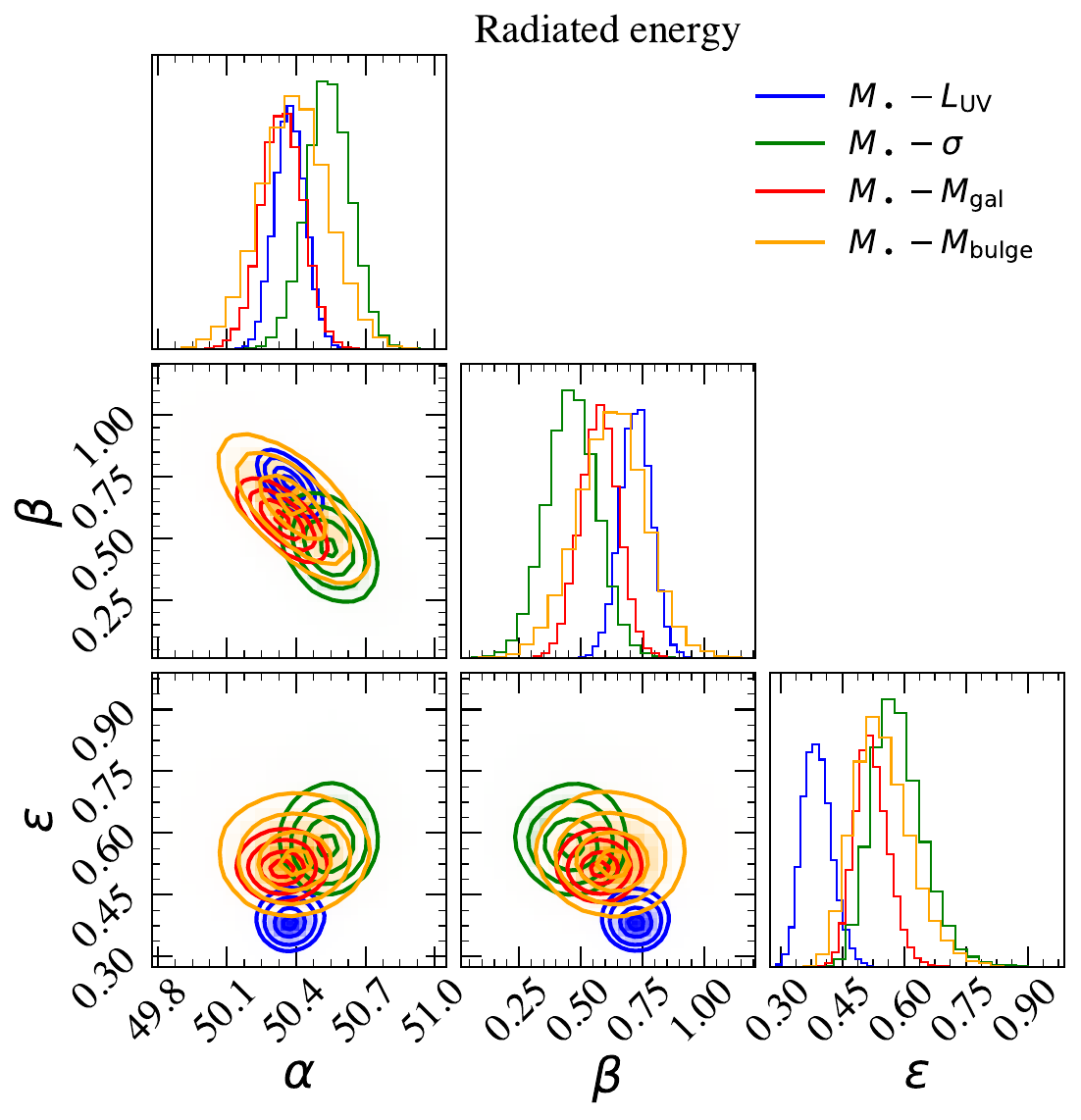}
    \includegraphics[width=0.45\linewidth]{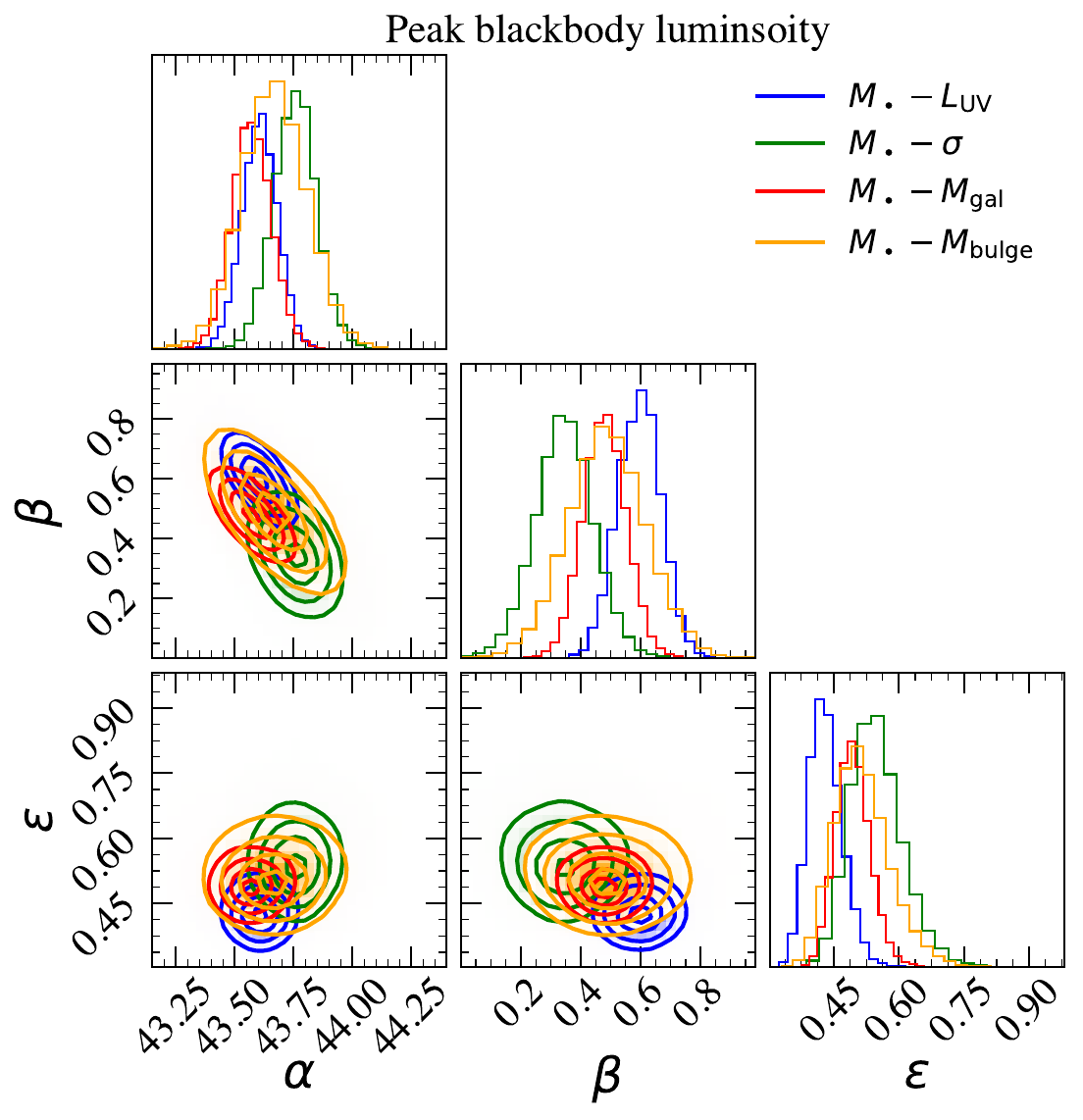}
    \caption{The observed correlation between  (i) the energy radiated in the TDE optical/UV flare and (ii) the peak bolometric luminosity of the TDE optical/UV flare and the black hole mass in the center of the event. The upper panels show the explicit correlations, with the black hole masses inferred from the late-time UV luminosity (which produces the scaling relationships with the least scatter). Typical uncertainties in the black hole masses are shown by the error bar in the lower right hand corner of each plot. The red curve shows the posterior median correlation, while the three shaded regions show $1, 2$ and $3\sigma$ contours respectively. The lower panel shows the posterior distributions of the power-law fits under different assumptions regarding how the black hole mass should be constrained (UV-plateau in blue, $M_\bullet-\sigma$ in green, $M_\bullet-M_{\rm gal}$ in red and $M_\bullet-M_{\rm bulge}$ in orange). All four black hole mass estimation techniques provide consistent results.  }
    \label{fig:fits}
\end{figure*}

In this and following sections we will fit power-law profiles of the following general form 
\begin{equation}\label{power_law_fit}
\log_{10}\left(Y\right) = \alpha + \beta \, \log_{10}\left(X\right) ,
\end{equation}
where 
\begin{equation}
X \equiv {M_\bullet \over 10^6 M_\odot} ,
\end{equation} 
and $Y$ will be some normalised scaling variable. The two parameters $\alpha$ and $\beta$ then have simple physical interpretations, namely the (logarithm of the) magnitude of the typical luminosity/energy of a TDE around a $10^6 M_\odot$ black hole, and the power-law index with which this luminosity/energy then scales with black hole mass, respectively.  

To understand the intrinsic scatter in these scaling relationships, we incorporate an intrinsic scatter $\epsilon$ into the uncertainty of the luminosity/energy measurements 
\begin{equation}
\left(\delta \log_{10} Y\right)^2 \to \left( \delta \log_{10} Y\right)^2 + \epsilon^2 , 
\end{equation}
where $\delta \log_{10} Y$ is the uncertainty in the logarithm of each scaling measurement (we take the mean uncertainty in cases of asymetric errors). We then maximize the likelihood 
\begin{multline}
 {\cal L} = - {1\over 2}\sum_{i}  { \left( \log_{10}(Y_i) - \alpha - \beta \log_{10}\left(X_i\right) \right)^2 \over \left(\delta \log_{10}\left(Y_i\right) \right)^2 + \epsilon^2  } \\ + \ln \Big[ 2\pi \left( \left(\delta \log_{10}\left(Y_i \right)  \right)^2 + \epsilon^2 \right)\Big] ,
\end{multline}
where the summation is over all pairs $(X_i, Y_i)$ of normalised black hole masses and scaling variables. 

We plot the measured peak blackbody luminosities $(L_{\rm BB, pk})$ and radiated energies $E_{\rm rad}$ against the TDE-inferred black hole mass in Figure \ref{fig:fits}. Visual inspection shows that both of these parameters clearly scale positively, and rather strongly, with black hole mass.  

Using the {\tt emcee} code \citep{EMCEE} to run a MCMC and obtain the posterior distribution of the two model parameters (adopting flat priors on all parameters), we infer 
\begin{align}
    &\alpha_{\rm BB, pk} = 43.6 \pm 0.1, \\
    &\beta_{\rm BB, pk} = 0.60 \pm 0.08, \label{LpkObs}
\end{align}
for the peak blackbody luminosity, and 
\begin{align}
    &\alpha_{\rm rad} = 50.4 \pm 0.1, \\
    &\beta_{\rm rad} = 0.72 \pm 0.07,  \label{EradObs}
\end{align}
for the radiated energy. These scaling parameters are robust to our choice of black hole mass inference technique (Appendix \ref{app:M}), and are also robust to imposing redshift cuts (Appendix \ref{app:Z}), as we highlight with their posterior distributions in the lower panel of Figure \ref{fig:fits}. In these lower panels we display the posterior distributions of the power-law fits made with galaxy scaling relationship black hole masses and TDE-inferred (plateau) black hole masses. The amplitude and slope of the power-law fits are consistent between the different techniques, and the plateau-mass measurement provides the lowest scatter (see the $\epsilon$ distributions). We therefore use the plateau mass measurements moving forward. 

It is clear then that any model of these early time flares must seek to reproduce the predicted power-law indices $\beta_{\rm BB,pk}, \beta_{\rm rad}$ and must not grossly under- or over-predict $\alpha_{\rm BB,pk}, \alpha_{\rm rad}$. Over-predicting both the energy and luminosity (compared to those observed) is preferable, owing to the possibility that we are not capturing all of the emission in our blackbody modeling. In the following section we collate all previously published models for optical TDE flares from the literature.

\section{Theoretical Energy and Luminosity Scales}\label{sec:models}
\subsection{General considerations and definitions}
In what follows, we consider the tidal disruption of a star of mass $M_{\star} = m_{\star}M_{\odot}$ by a black hole of mass $M_{\bullet} = 10^{6}M_{\bullet,6}M_{\odot}$.  We focus on main-sequence stars, for which the stellar radius is reasonably well-approximated (for $0.15 < m_{\star} < 3$) as as power-law \citep{Ryu+20b}:
\begin{eqnarray}
    R_{\star} = 0.93 m_{\star}^{0.88}R_{\odot}.
\end{eqnarray}
We focus on complete disruptions, i.e. on pericenter radii $r_{\rm p} \equiv R_{\rm t}/\beta$ well within  the canonical tidal radius $R_{\rm t} \approx R_{\star}(M_{\bullet}/M_{\star})^{1/3},$ where $\beta > 1$ is the penetration factor. In general this estimate of the tidal radius can differ from the true (physical) maximum pericenter radius giving rise to a TDE \citep{Phinney89}.  Following \citet{Ryu+20a}, we therefore take:
\begin{align}
    \Psi &\equiv \frac{R_{\rm T}}{R_{\star}(M_{\bullet}/M_{\star})^{1/3}} \\ &\approx \left[0.80 + 0.26M_{\bullet,6}^{2}\right] \left\{ \frac{1.47 + \exp\left[(m_{\star}-0.67)/0.14\right]}{1+2.34\exp\left[(m_{\star}-0.67)/0.14\right]}\right\},\nonumber
\end{align}
where $R_T$ is the ``true'' tidal radius (the pericentre radius at which the star is fully disrupted). The resulting mass fall-back rate at times $t \gg t_{\mathrm{fb}}$ obeys (e.g., \citealt{Rees88,Phinney89,Lodato+09}),
\begin{equation}
\dot{M}_{\mathrm{fb}} = \dot{M}_{\rm p}\left(\frac{t}{t_{\mathrm{fb}}}\right)^{-5/3},
\label{eq:Mdotfb}
\end{equation}
where we follow \citet{Bandopadhyay+24} in taking a single fall-back time
\begin{equation}
t_{\mathrm{fb}} \simeq 30  M_{\bullet, 6}^{1/2}  \mathrm{~d},
\label{eq:tfb}
\end{equation}
roughly independent of stellar type. 
For simplicity, we assume that half of the star falls back, $M_{\rm fb} = M_{\star}/2$.


In the case of a full disruption, the peak fall-back rate,
\begin{equation}
\dot{M}_{\rm p} = \frac{M_\star}{3 t_{\mathrm{fb}}} \approx 2.6 \times 10^{26}  M_{\bullet,6}^{-1/2}m_{\star} \mathrm{~g} \mathrm{~s}^{-1},
\end{equation}
exceeds the Eddington accretion rate $\dot{M}_{\rm Edd} \equiv L_{\rm Edd}/0.1 c^{2} \approx 1.7\times 10^{24}$ g s$^{-1}$ by orders of magnitude, with
\begin{equation}
L_{\rm Edd} \simeq 1.4\times 10^{44}{\rm erg\,s}^{-1}M_{\bullet,6},
\label{eq:Ledd}
\end{equation}
the Eddington luminosity. 
The fall-back will become sub-Eddington ($\dot{M}_{\rm fb}(t_{\rm Edd}) = \dot{M}_{\rm Edd}$) after a time
\begin{equation}
 t_{\rm Edd} \simeq 610 \,M_{\bullet,6}^{-2/5}m_{\star}^{3/5}\,\,{\rm d}.
\label{eq:tEdd}  
\end{equation}
Of course the above statements relate to the rate at which material returns to pericentre following the disruption itself. Potentially more physically meaningful (for models which seek to relate the optical/UV flare to the accretion rate) is the accretion rate through the inner edge of the disk, which will not in general track the fallback rate unless the viscous time is $\ll$ the fallback time. As we shall discuss later, this approximation (that the accretion rate tracks the fallback rate)  is unlikely even to be approximately true (as it only holds in the asymptotic $t_{\rm visc}/t_{\rm fb} \to 0$ limit -- this is a limit which can not be plausibly satisfied in many regions of TDE parameter space). In Appendix \ref{app:acc} we show how to determine the accretion rate through the inner edge of a black hole disk for a given fallback rate, which self consistently handles the ratio of $t_{\rm fb}$ and the viscous time $t_{\rm visc}$, which varies systematically across the TDE parameter space.



\subsection{Radiatively Efficient Fallback-Accretion}
In the most naive model, where the optical flare is simply the result of radiatively efficient accretion with rate equal to the fallback rate, the accreted $\sim$half of the star ultimately emits energy corresponding to \citep{NovikovThorne73}
\begin{equation}\label{Eradfb}
    E_{\rm fb-acc} \simeq \eta M_{\rm fb}c^{2} \simeq 9\times 10^{52}\,{\rm erg}\,\left(\frac{\eta}{0.1}\right)\left(\frac{M_{\rm fb}}{M_{\star}/2}\right)m_{\star}
\end{equation}
where the radiative efficiency $\eta \approx 0.05-0.2$ is that of a thin (radiatively efficient) accretion disk, depending primarily on the black hole spin. This is in $>5\sigma$ tension with observations, a result of the observed $M_\bullet$ dependence of $E_{\rm rad}$.  The above energy represents an absolute upper limit on the radiated energy and its discrepancy with early-time optical/UV TDE observations is sometimes described as the ``missing energy problem'' (e.g., \citealt{Piran+15,Lu&Kumar18}). It is now understood that this total amount of energy is indeed ultimately emitted by a TDE system, but that this energy escapes only on much longer timescales ($t\gg 1$ year) in the extreme UV by a long lived, spreading disk \citep{Mummery_et_al_2024}. 

In general, the fraction of $E_{\rm acc, fb}$ radiated in the optical band depends on the photosphere temperature, but its maximum value, corresponding to prompt circularization and an optimally-placed reprocessing layer, is given by:
\begin{equation}\label{Lpkfb}
L_{\rm fb-acc} \simeq \eta \dot{M}_{\rm p}c^{2} \approx 2\times 10^{46}\,{\rm erg\,s^{-1}}\left(\frac{\eta}{0.1}\right)M_{\bullet,6}^{-1/2}m_{\star}.
\end{equation}
This is the luminosity assumption employed by {\tt MOSFIT}'s TDE model. We immediately note that the $\sim M_\bullet^{-1/2}$ scaling is at $>5\sigma$ tension with observations. 

\subsection{Eddington-Limited Fallback-Accretion}
The above model is often assumed, but neglects simple and important properties which must be satisfied by any accreting system. Indeed even if an accretion flow has an accretion rate $\dot M_{\rm acc} \gg \dot M_{\rm edd}$ it will not produce luminosity proportional to the accretion rate $L_{\rm bol}\neq \eta \dot M_{\rm acc} c^2$, but will instead be effectively Eddington-limited \citep{SS73} – a simple analytical estimate of the impact of this Eddington limit is $L_{\rm Bol} = L_{\rm Edd} (1 + \ln(\dot M_{\rm acc}/\dot M_{\rm Edd}))$. 


For TDEs what this means is that even if the black hole is accreting at or near the fall-back rate, the inwards advection (``trapping'') of radiation in the accretion flow may prevent it from being radiatively efficient. In particular, radiation is trapped on the inflow time interior to the so-called trapping radius \citep{Begelman79}
\begin{equation}
    R_{\rm tr} \equiv 10R_{\rm g}\left(\frac{\dot{M}}{\dot{M}_{\rm Edd}}\right),
\end{equation}
such that $R_{\rm tr} > R_{\rm circ} = 2R_{\rm t}$ until a time:
\begin{equation}
    t_{\rm tr} \simeq \left(\frac{5R_{\rm g}\beta}{R_{\rm T}}\right)^{3/5}t_{\rm Edd} \approx 161\,{\rm d}\,\beta^{3/5}m_{\star}^{0.96}\Psi^{-3/5}.
\end{equation}
Again, the above expression assumes $\dot M_{\rm acc} = \dot M_{\rm fb}$. 

Radiation hydrodynamic simulations show that the radiation from super-Eddington accretion flows is limited to the Eddington luminosity $L_{\rm Edd}$ (Eq.~\eqref{eq:Ledd}) to within a factor $\lesssim 2$ (e.g., \citealt{Yoshioka+24, Lizhong25} and references therein).
 
The radiated energy up to the time $t_{\rm tr}$ is thus limited to
\begin{equation}\label{EradEddTr}
    E_{\rm Edd,tr} \simeq L_{\rm Edd}t_{\rm tr} \simeq 2.0\times 10^{51}\,{\rm erg}\,\beta^{3/5}m_{\star}^{0.96}M_{\bullet,6}\Psi^{-3/5}.
\end{equation}
The trapping radius remains outside the ISCO radius until $t \simeq t_{\rm Edd},$ resulting in a somewhat larger radiated energy,
\begin{equation}\label{EradEddEdd}
    E_{\rm Edd} \simeq L_{\rm Edd}t_{\rm Edd} \approx 8.0\times 10^{51}\,{\rm erg}\,M_{\bullet,6}^{3/5}m_{\star}^{3/5}
\end{equation}
As in the radiatively efficient case discussed above, the fraction of $E_{\rm acc}$ radiated in the optical band depends on the temperature of the emission and hence radius of the photosphere, $R_{\rm ph}$, something which has not been explicitly computed as far as the authors are aware.  

\subsection{Wind Reprocessed Fallback-Accretion}


Another possibility is that TDE optical/UV emission is powered by reprocessing of thermal energy released in an outflow powered either by accretion or the circularization process (e.g., \citealt{Strubbe&Quataert09,MetzgerStone16,Piro&Lu20}). The escaping radiation field in a dynamical outflow is not Eddington-limited per se, but it remains subject to adiabatic losses in the outflow.

As an explicit example of such a ``wind reprocessing'' scenario, \citet{MetzgerStone16} consider that that a fraction $f_{\rm in} = 0.1f_{\rm in,-1}$ of the stellar fall-back debris from the TDE is accreted, while the majority $f_{\rm out} = 1-f_{\rm in}$ is instead lost to a wind.  The velocity of the wind is assumed to scale with the escape speed at the circularization radius, since this represents the characteristic energy scale required to ``incorporate'' fallback into a Keplerian disk.   

Prior to a critical ``wind trapping time'' $t_{\rm w}$, adiabatic losses in the wind are important and the escaping luminosity obeys 
\begin{equation} 
L = L_{\rm pk}(t/t_{\rm fb})^{-5/9}, \quad t < t_{\rm w},
\end{equation}
where
\begin{equation}
t_{\rm w}/t_{\rm fb} \approx 2.6\beta^{3/5}M_{\bullet,6}^{-1/2}m_{\star}^{1/5}R_{\rm in,6}^{-3/5}.
\label{eq:ttr}
\end{equation}
At times $t > t_{\rm w}$ trapping losses are unimportant and the luminosity is assumed to scale with the fallback rate $L \propto f_{\rm in}\dot{M}_{\rm fb}c^{2} \propto t^{-5/3}$.  The peak luminosity from the wind is given by,
\begin{eqnarray}
 L_{\rm pk} \approx 
\left\{
\begin{array}{lr}
   6\times 10^{44}\,{\rm erg\,s^{-1}}\beta^{-2/3}\eta_{-1}f_{\rm in,-1}M_{\bullet,6}^{0.06}m_{\star}^{0.58}R_{\rm in,6}^{2/3}, \label{LpkWindFb1} \\  (M_{\bullet} < M_{\bullet,\rm w})
 \\\\
2\times 10^{45}\,{\rm erg\,s^{-1}}\,\eta_{-1}f_{\rm in,-1}M_{\bullet,6}^{-1/2}m_{\star}^{4/5}, \\  (M_{\bullet} > M_{\bullet,\rm w})
\\
\end{array}
\right.
\label{eq:Lpk}
\end{eqnarray} 
where $R_{\rm in} = 6 R_{\rm g}R_{\rm in,6}$ is the inner edge of the disk and
\begin{equation} M_{\bullet,\rm w} \equiv 6.6\times 10^{6}M_{\odot}\beta^{6/5}m_{\star}^{2/5}R_{\rm in,6}^{-6/5}.
\label{eq:Mtr}
\end{equation}
is the critical SMBH mass below which adiabatic trapping losses in the wind are important (i.e., for which $t_{\rm w} > t_{\rm fb}$). 

The total energy radiated can thus be approximated as:
\begin{eqnarray}
 E_{\rm rp} \approx 
\left\{
\begin{array}{lr}
   L(t_{\rm w})t_{\rm w} \approx L_{\rm pk}t_{\rm fb}(t_{\rm w}/t_{\rm fb})^{4/9}, \quad\quad (M_{\bullet} < M_{\bullet,\rm w}),  \\ \approx 2\times 10^{51}\,{\rm erg}\,\beta^{-2/5}\eta_{-1}f_{\rm in,-1}M_{\bullet,6}^{0.34}m_{\star}^{0.67}R_{\rm in,6}^{-0.27}  & 
 \\ \\
\eta f_{\rm in} (M_{\star}/2)c^{2},\quad\quad\quad\quad\quad\quad\quad\quad ( M_{\bullet} > M_{\bullet,\rm w})  \\ \approx 9\times 10^{51}\,{\rm erg}\,\eta_{-1}f_{\rm in,-1}m_{\star} \label{EradWindFb1}
\\
\end{array}
\right.
\label{eq:Lpk}
\end{eqnarray}

\subsection{Cooling Envelope (Gravitational Contraction)}


The stellar mass that returns to the SMBH following a TDE is very weakly bound to it.
For this material to form a compact Keplerian accretion disk of size $R_{\rm circ} = 2R_{\rm p} = 2R_{\rm T}/\beta$ commensurate with the angular momentum of the debris, a minimum energy must be released, given by:
\begin{eqnarray}
\label{EradEnv}
    E_{\rm circ} \equiv \frac{GM_{\bullet}M_{\rm env}}{2R_{\rm circ}} &\simeq \frac{GM_{\bullet}^{2/3}M_{\star}^{4/3}}{8R_{\star}}\frac{M_{\rm env}}{(M_{\star}/2)}\frac{\beta}{\Psi(M_{\bullet},M_{\star})} \nonumber \\
    &\approx 5.3\times 10^{51}\,{\rm erg}\,\beta M_{\bullet,6}^{2/3}m_{\star}^{0.40}.
\end{eqnarray}
In the cooling envelope scenario for TDEs \citep{Metzger22}, the UV/optical emission is powered by $E_{\rm circ}$, released via gradual contraction of a quasi-spherical hydrostatic envelope, itself created through rapid dissipation of the kinetic energy of the returning debris streams. In other words, the optical radiation is powered by the gravitational energy that must be liberated in the process of forming a centrifugally supported (Keplerian) disk.  As discussed in \cite{Mummery_et_al_2024}, this energy scaling is in good accord with observations. 


A radiation-dominated envelope in hydrostatic equilibrium radiates at the Eddington luminosity $L_{\rm Edd}$ (Eq.~\eqref{eq:Ledd}). The maximum optical luminosity is thus roughly that of a blackbody of luminosity $L_{\rm Edd} \propto M_{\bullet}$. A version of this model was implemented in the open source software package \texttt{Redback} \citep{Sarin+24} and applied to TDE data in \citet{Sarin&Metzger24}.

\subsection{Colliding Streams}
The maximum heating rate due to collisions between the debris streams at a radius $r_{\rm col}$ is given by (e.g., \citealt{Piran+15})
\begin{equation}
    \dot{E}_{\rm sh} \simeq \frac{GM_{\bullet}\dot{M}_{\rm fb}}{r_{\rm col}}.
\end{equation}
Clearly then, with $\dot M_{\rm fb}$ well constrained by basic arguments, the energetics/luminosity in the colliding streams scenario is determined by the choice of collisional radius $r_{\rm col}$, for which various scenarios could be considered. 

A characteristic maximum scale for the collision radius is the apocenter radius of the most tightly bound debris, $r_{\rm a} \simeq (R_{\star}/2)(M_{\bullet}/M_{\star})^{2/3}$ is the apocenter radius of the most tightly bound debris streams \citep{Piran+15}.  The minimal radial scale for the collision radius is the pericentre radius $r_p$, or effectively the tidal radius $R_T$. Insofar that $r_{\rm a} \propto t^{2/3}$, and $R_T \propto t^0$, and hence $\dot{E}_{\rm sh} \propto t^{-7/3}$ ($t^{-5/3}$) at times $t \gtrsim t_{\rm fb}$ for shocks at the maximum (minimum) radial scale, one can estimate the total shock-heating power as that released on timescales $\sim t_{\rm fb}$ (for any $r_{\rm col}$),   viz.~
\begin{multline}
    E_{\rm sh} \approx \dot{E}_{\rm sh}t_{\rm fb} \approx \frac{GM_{\star}M_{\bullet}}{3r_{\rm col}} \\ \approx 4.6\times 10^{50}{\rm erg}\, M_{\bullet,6}^{1/3}m_{\star}^{0.74}\left(\frac{r_{\rm col}}{r_{\rm a}}\right)^{-1} ,
\end{multline}
which allows us to compute the energy radiated at a given assumed collisional scale. Taking the apocenter radius as the collisional scale gives 
\begin{align}
    L_{\rm pk, a} &\simeq \frac{2 GM_{\bullet}^{1/3} M_\star^{2/3}\dot{M}_{\rm fb}}{R_{\star}} , \nonumber \\
    &\approx 2.7\times10^{43}\,{\rm erg\, s^{-1}}\, M_{\bullet, 6}^{-1/6} m_\star^{0.79} , \label{LpkColA} \\
    E_{\rm rad, a} &\approx  4.6\times 10^{50}{\rm erg}\, M_{\bullet,6}^{1/3}m_{\star}^{0.74} ,\label{EradColA}
\end{align}
while if we take the tidal radius as the collisional scale we find 
\begin{align}
    L_{\rm pk, T} &\simeq \frac{GM_{\bullet}^{2/3} M_\star^{1/3}\dot{M}_{\rm fb}\beta }{\Psi R_{\star}} , \nonumber  \\
    &\approx 1.7\times10^{46} \, {\rm erg/s}\,M_{\bullet, 6}^{1/6} \beta m_\star^{0.45} \Psi^{-1}, \label{LpkColT} \\
    E_{\rm rad, T} &\approx  2.8\times10^{52}\, {\rm erg}\, M_{\bullet,6}^{2/3}m_{\star}^{0.74}\beta \Psi^{-1} , \label{EradColT}
\end{align}

Alternatively, \citet{Dai+15} predict that apsidal precession will cause the most tightly bound streams to intercept at a radius:
\begin{equation}
    R_{\rm int} \simeq \frac{a_{\rm mb}(1-e_{\rm mb}^{2})}{1-e_{\rm mb}\cos(\phi/2)},
\end{equation}
where $\phi = 6\pi R_{\rm g}/a_{\rm mb}(1-e_{\rm mb}^{2})$ is the apsidal precession angle, and $a_{\rm mb} = R_{\rm t}^{2}/2R_{\star}$ and $e_{\rm mb} \approx 1-(2/\beta)(M_{\star}/M_{\bullet})^{1/3}$ are the semi-major axis and eccentricity of the most tightly bound debris stream. The dependence on parameter values here is non-linear, and we simply compute the luminosity/energy scales numerically. Broadly speaking, the scalings with parameters here is similar to the shocks at apocenter case, with a stronger stellar mass dependence. 

A final class of shock-based models (naturally related to those discussed above) is those which make up the {\tt TDEmass} package \citep{Ryu+20a, Krolik24}. A recent discussion of the assumptions involved in this model is presented in \cite{Krolik24}, where the authors also claim that their model reproduces the scalings of TDE observables with parameters (without providing any testing or validation of this claim). We do not reproduce their derivations here, but simply quote their key results, namely a peak luminosity which scales as 
\begin{eqnarray}
    L_{\rm pk} \simeq 8\times 10^{43}\, {\rm erg\, s^{-1}} \, {\Xi ^{5/2} M_{\bullet, 6}^{-1/6} m_\star^{0.47} \over 1+0.26\,  \Xi^{5/2} M_{\bullet, 6}^{-7/6} m_\star^{0.47} }, \label{LpkColK}
\end{eqnarray}
where we have set various nuisance parameters to their default values assumed in \cite{Krolik24}. Note that at the very lowest masses $M_\bullet \lesssim 10^6M_\odot$ this model returns $\sim$ the Eddington luminosity. The function 
\begin{multline}
    \Xi \equiv \left[1.27 - 0.3M_{\bullet, 6}^{0.24}\right]\times \\ \left\{ \frac{0.62 + \exp\left[(m_{\star}-0.67)/0.21\right]}{1+0.55\exp\left[(m_{\star}-0.67)/0.21\right]}\right\},
\end{multline}
is a moderately strongly decreasing function of black hole mass. Combining the dependence of $\Xi$ on $M_\bullet$ with that of $L_{\rm pk}$ results in the approximate scaling of $L_{\rm pk} \propto M_\bullet^{-3/8}$, in $>5\sigma$ tension with observations.  This is likely why {\tt TDEmass} does not recover known galactic scaling relationships from TDE data \citep{Hammerstein23, Guolo25c}.

Similarly, the total energy released in this model is 
\begin{equation}
    E_{\rm rad} \approx  4 \times 10^{50}\, {\rm erg}\,\,   \Xi M_{\bullet, 6}^{1/3} m_\star^{0.79} ,\label{EradColK}
\end{equation}
or an approximately $E_{\rm rad}\propto M_\bullet^{1/4}$ scaling.

The above authors also provide an estimate for the late-time luminosity from their model, distinct from that of compact thin-disk theory (which scales as $L_{\rm late, disk}\propto M_\bullet^{2/3}$; \citealt{Mummery_et_al_2024}), which scales as 
\begin{eqnarray}
    L_{\rm late} \simeq 10^{43} \, {\rm erg\, s^{-1}}\, M_{\bullet, 6}^{-1/2} \, \Xi^{3/2} m_\star^{-0.26} , 
\end{eqnarray}
or an approximately $L_{\rm late} \propto M_\bullet^{-5/8}$ scaling, ruled out at $>5\sigma$ by observations (see Figure \ref{fig:late}). This result is independent of any redshift cuts one places on the population of TDEs (Appendix \ref{app:Z}).  

\subsection{Reprocessed ``viscous'' accretion}
None of the above models incorporate the properties of an accretion flow into their analysis, and typically assume that the instantaneous accretion rate equals the instantaneous fallback rate. This is somewhat surprising  as accretion theory has had a number of successes in reproducing TDE data, such as correctly describing the late time UV luminosity  observed in TDEs \citep{Mummery_et_al_2024}, but also (perhaps more relatively for the initial optical flare) reproducing the peak X-ray luminosity function of TDEs from first principles \citep{MummeryVV25}. This implies that the accretion disks we observe at late times are always present, and can appear at early times in the soft X-rays \citep[see also][for further discussion of this result]{Guolo25c}. It is worthwhile to ask what this early time emission from accretion flows might result in, if reprocessed, into the optical/UV. 

The model we outline here differs strongly from approaches like {\tt MOSFIT}, which effectively set the accretion rate equal to the fallback rate despite this violating angular momentum conservation and implicitly assuming that material moves from the tidal radius to the black hole's event horizon infinitely quickly (indeed, the inferred accretion velocities from {\tt MOSFIT} $v_{\rm acc} \simeq R_T/T_{\rm visc}$, two parameters which can be extracted post-fit, regularly exceed the speed of light, e.g., \citealt{Mockler19, Alexander2025}. This is an unphysical result). 

In reality the returning (or circularised) material has angular momentum, which cannot be neglected. The specific angular momentum of circularised material is approximately equal to that of the incoming star 
\begin{equation}
    j_{\rm circ} \approx \sqrt{GM_\bullet R_{\rm circ}} \approx 10 {GM_\bullet \over c} M_{\bullet, 6}^{-1/3} m_\star^{0.13} \beta^{-1/2}  .
\end{equation}
To be accreted, a fluid element must drop its specific angular momentum to that of the ISCO, or 
\begin{equation}
    j_{\rm acc} = 2\sqrt{3 }\, {GM_\bullet \over c} \, \left(1 - {2a\over 3\sqrt{GM_\bullet r_I}}\right),
\end{equation}
requiring a drop by factor 
\begin{equation}
       {j_{\rm circ} \over j_{\rm acc}} \approx 3 \, \times \, \left({M_{\bullet, 6}^{-1/3} m_\star^{0.13} \over \beta^{1/2} (1-2a_\bullet/3x_I^{1/2})}\right) ,
\end{equation}
where $a_\bullet, x_I$ are the dimensionless spin and ISCO radius of the black hole. Note that this ratio neglects relativistic effects on the circularisation angular momentum which become important when the ratio is close to one. 

As a conserved quantity, angular momentum is difficult to remove. This is in effect the entire physics governing an accretion flow: how does one remove angular momentum from a flow so as to get it onto the central object? The answer is that MHD instabilities instigate turbulence within the flow \citep{BalbusHawley91}, which over the so-called ``viscous'' timescale redistribute this angular momentum within the fluid allowing some of the material to accrete (with some material flowing outwards to compensate for this inward motion). 

The viscous timescale is, within the \cite{SS73} framework, equal to 
\begin{eqnarray}
    t_{\rm visc} \approx \alpha^{-1} \, \theta^{-2} \, t_{\rm orb}, 
\end{eqnarray}
where $\alpha \sim 10^{-2}-10^{-1}$ is a dimensionless fudge factor, $\theta \equiv h/r$ is the opening angle of the disk, and for a TDE 
\begin{equation}
    t_{\rm orb} \approx  \sqrt{R_{\rm circ}^3 \over GM_\bullet} \approx \sqrt{8 R_\star^3 \over \beta^3 GM_\star} \approx 0.05\, {\rm d} \, \beta^{-3/2} \, m_\star^{0.9} ,
\end{equation}
is the orbital time at the disk-circularisation radius, a time-scale which is independent of black hole properties. The viscous timescale can therefore comfortably exceed the fallback time for a large fraction of TDE parameter space 
\begin{eqnarray}
    t_{\rm visc} \approx 50 \, {\rm d} \, \alpha_{-1}^{-1}\theta_{-1}^{-2} \, \beta^{-3/2} \, m_\star^{0.9} .
\end{eqnarray}
Here we have normalised $\alpha_{-1} \equiv \alpha / 0.1$ and $\theta_{-1} \equiv \theta/0.1$. The range of values spanned for plausible changes in the parameters on the right hand side of this expression comfortably includes the range seen in TDEs \citep{Guolo25b}, which span from the very short $t_{\rm visc} \simeq 15\, {\rm d}$ (AT2019dsg), to moderate $t_{\rm visc} \simeq 470\, {\rm d}$ (ASASSN-14li) out to very large $t_{\rm visc} \simeq 2300\, {\rm d}$ (GSN 069). The ratio of the viscous to fallback timescales 
\begin{eqnarray}
    {t_{\rm visc} \over t_{\rm fb}} \approx {5\over 3} \, \alpha_{-1}^{-1}\theta_{-1}^{-2} \, \beta^{-3/2} \, m_\star^{0.9} M_{\bullet, 6}^{-1/2} , 
\end{eqnarray}
implies that it is the turbulent redistribution of angular momentum which is the limiting timescale in the majority of low black hole mass TDE accretion flows (which therefore determines the accretion rate), not the fallback rate. In the limit in which the viscous time is longer than the fallback time, the peak accretion rate depends only on stellar properties and nuisance parameters 
\begin{eqnarray}
    \dot M_{\rm acc, pk} \sim M_{\rm fb}/t_{\rm visc} \sim (M_\star/R_\star)^{3/2} \beta^{3/2} \alpha \, \theta^2 .
\end{eqnarray}
We see that if the optical emission was powered purely by $L_{\rm pk} \simeq \eta \dot M_{\rm acc, pk} c^2$, this would predict no dependence on black hole mass (in stark contrast with observations). However, the peak accretion rate predicted in this framework can be super-Eddington, in which case the resultant luminosity would be $L_{\rm pk} \simeq L_{\rm edd}$. 

Recently, Mummery (in prep.) derived a general expression for the time taken for a viscously evolving TDE disk to drop to a fraction $f_{\rm edd}$ of the Eddington luminosity. If we take that result, rewrite it in the notation of this paper, and take $f_{\rm edd}=1$, we find 
\begin{eqnarray}
    t_{\rm SE} \approx 166\, {\rm d} \, \alpha_{-1}^{-1/4}\theta_{-1}^{-1/2} f_{d, -1}^{3/4}\, \eta_{-1}^{3/4}\beta^{-2/3}  m_\star M_{\bullet, 6}^{-3/4}, \, 
\end{eqnarray}
which implies that for the above timescale a TDE disk is super-Eddington. We have defined $f_{d, -1}$ which is the fraction of the returning 
stellar material which forms a disk (divided by 0.1), and $\eta_{-1}$ the efficiency of the disk (divided by 0.1). If this super-Eddington phase of accretion is reprocessed into optical/UV emission then its energy content would be 
\begin{align}
    E_{\rm rad, SE} & \approx L_{\rm Edd} t_{\rm SE} \label{EradV}\\
    & \approx 2\times 10^{51}\, {\rm erg}\, \nonumber \\
    &\quad\quad \times M_{\bullet, 6}^{1/4}\alpha_{-1}^{-1/4}\theta_{-1}^{-1/2} f_{d, -1}^{3/4}\, \eta_{-1}^{3/4}\beta^{-2/3}  m_\star \nonumber 
\end{align}
During this phase of evolution -- depending once again on the physics of the optical/UV photosphere -- the optical/UV luminosity would be at most the Eddington (equation \ref{eq:Ledd}). We note that detailed spectral modeling of TDE disks in the X-ray typically find Eddington limited bolometric luminosities $L_{\rm Bol}\sim L_{\rm Edd}$ at early times \citep{Mummery_Wevers_23, Guolo25c}. 

\subsection{Summary of the mass-scaling of flare models}
In Table \ref{tab:summary} we collate the various power-law indices of the 10 models discussed here for both the bolometric luminosity, and the radiated energy, of the early time TDE optical/UV emission. We see immediately that this data provides a robust test that many models fail to pass, something which shall be discussed in detail in the next section. Many models display a more complex behavior with black hole mass than can be summarized by a single power law index (mainly due to corrections to e.g., the simplest estimate of the tidal radius, $\Psi$), and so in the following section we numerically evaluate each profile explicitly. 

\begin{table*}[]
    \centering
    \begin{tabular}{|p{155pt}|p{100pt}|p{110pt}|}
    \hline 
       \textcolor{white}{.}\newline Model name \newline & \textcolor{white}{.}\newline Index $\beta_{\rm pk}$ in $L_{\rm pk}\propto M_\bullet^{\beta_{\rm pk}}$\newline & \textcolor{white}{.}\newline Index $\beta_{\rm rad}$ in $E_{\rm rad}\propto M_\bullet^{\beta_{\rm rad}}$ \newline\\
       \hline \hline 
        Efficient Fallback accretion ({\tt MOSFIT}) \newline (eqs. \ref{Lpkfb}, \ref{Eradfb}) & $-1/2$ & $0$ \\
        \hline 
        Eddington-limited Fallback accretion (Eddington-time) \newline (eqs. \ref{eq:Ledd}, \ref{EradEddEdd})  & $1$ & $3/5$ \\
        \hline 
        Eddington-limited Fallback accretion (Trapping-time) \newline (eqs. \ref{eq:Ledd}, \ref{EradEddTr})  & $1$ & $1$ \\
        \hline 
        Eddington-limited ``viscous'' accretion\newline (eqs. \ref{eq:Ledd}, \ref{EradV}) & $1$ & $1/4$ \\
        \hline 
        Wind-reprocessed Fallback-accretion (low black hole masses) \newline (eqs. \ref{LpkWindFb1}, \ref{EradWindFb1}) & $0.06$ & $0.34$ \\
        \hline 
        Wind-reprocessed Fallback-accretion (high black hole masses) \newline (eqs. \ref{LpkWindFb1}, \ref{EradWindFb1}) & $-1/2$ & $0$ \\
        \hline 
        Cooling-envelope \newline (eqs. \ref{eq:Ledd}, \ref{EradEnv})& $1$ & $2/3$ \\
        \hline 
        Colliding streams (apocentre) \newline (eqs. \ref{LpkColA}, \ref{EradColA})& $-1/6$ & $1/3$ \\
        \hline 
        Colliding streams (pericentre)\newline (eqs. \ref{LpkColT}, \ref{EradColT}) & $1/6$ & $2/3$ \\ 
        \hline 
        Colliding streams ({\tt TDEmass}) \newline (eqs. \ref{LpkColK}, \ref{EradColK}) & $-3/8$ & $1/4$ \\ 
        \hline 
        Observed \newline (eqs. \ref{LpkObs}, \ref{EradObs}) & $0.60 \pm 0.08$& $0.72\pm 0.07$ \\
        \hline 
    \end{tabular}
    \caption{The rough power-law slopes with which the peak blackbody luminosity of the optical/UV flare and the radiated energy in the optical/UV flare should scale with black hole mass, for the 10 models collated in this work. We also reproduce the observed scaling indices in the lower row. Popular models such as {\tt MOSFIT} and {\tt TDEmass} have luminosity scalings in $>5\sigma$ tension with observations. Promising models include Eddington-limited accretion and the cooling envelope framework.   }
    \label{tab:summary}
\end{table*}

\section{Comparison to data and discussion}\label{sec:comp}
Figures \ref{fig:lum_comp} and \ref{fig:en_comp} present the key results of this paper, namely a comparison between existing models of the luminosity and energy in the early-time optical flare and those which are observed in TDE samples. We repeat this analysis for different black hole mass inference techniques in Appendix \ref{app:M}, and with different redshift cuts in \ref{app:Z}. Neither make any difference to the results. 

\subsection{The peak optical/UV luminosity}

\begin{figure*}
    \centering
    \includegraphics[width=0.32\linewidth]{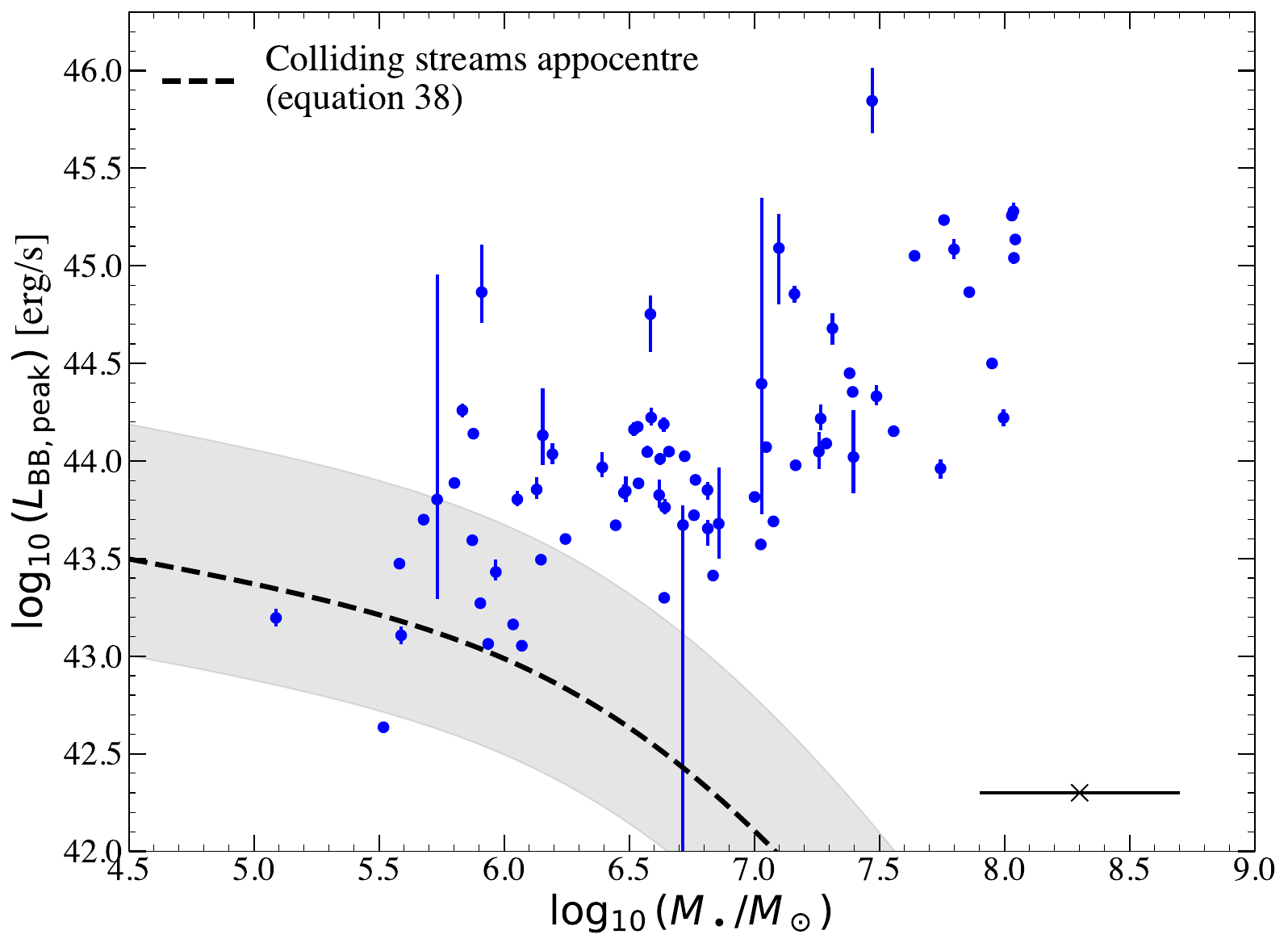}
    \includegraphics[width=0.32\linewidth]{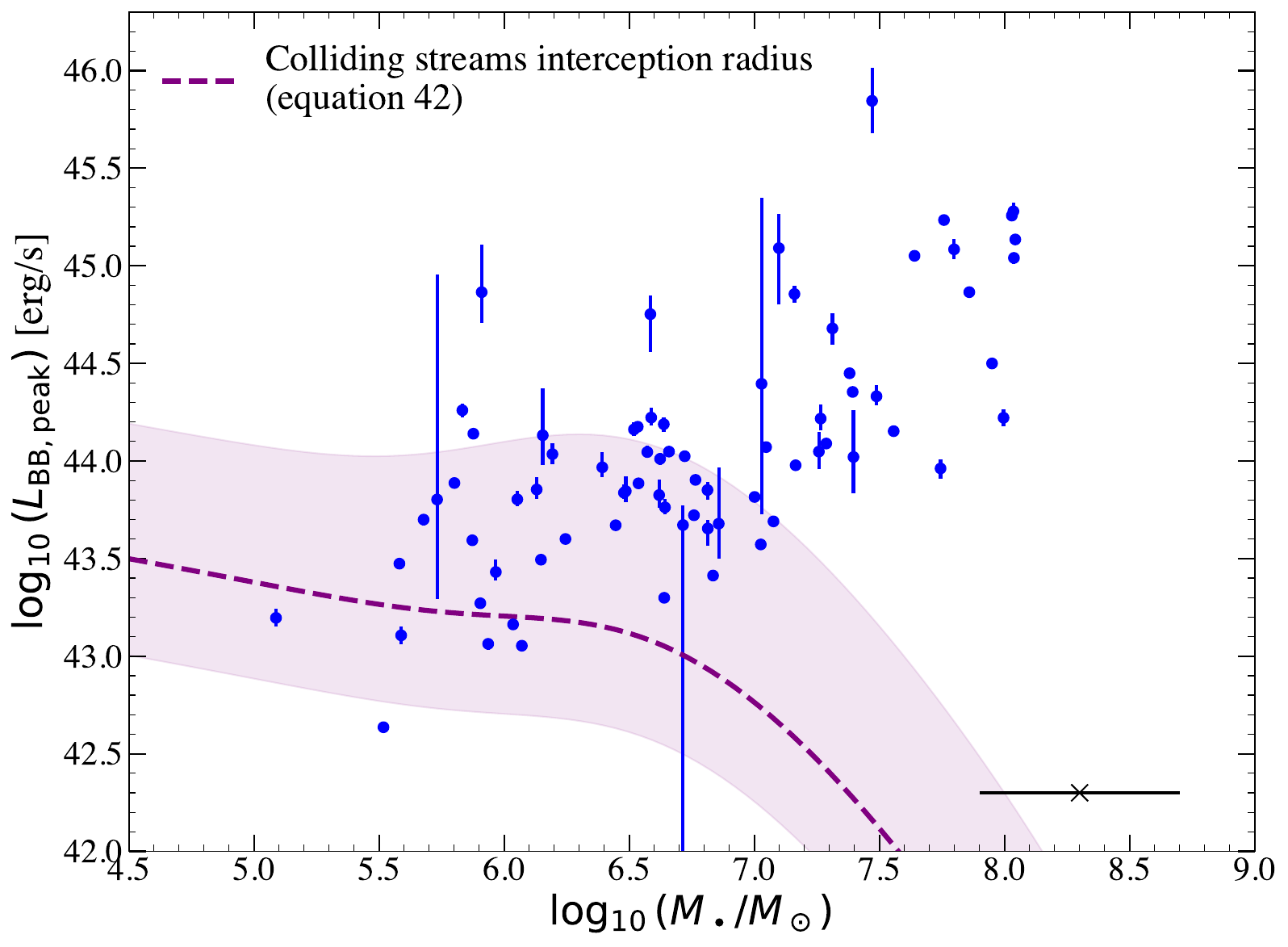}
    \includegraphics[width=0.32\linewidth]{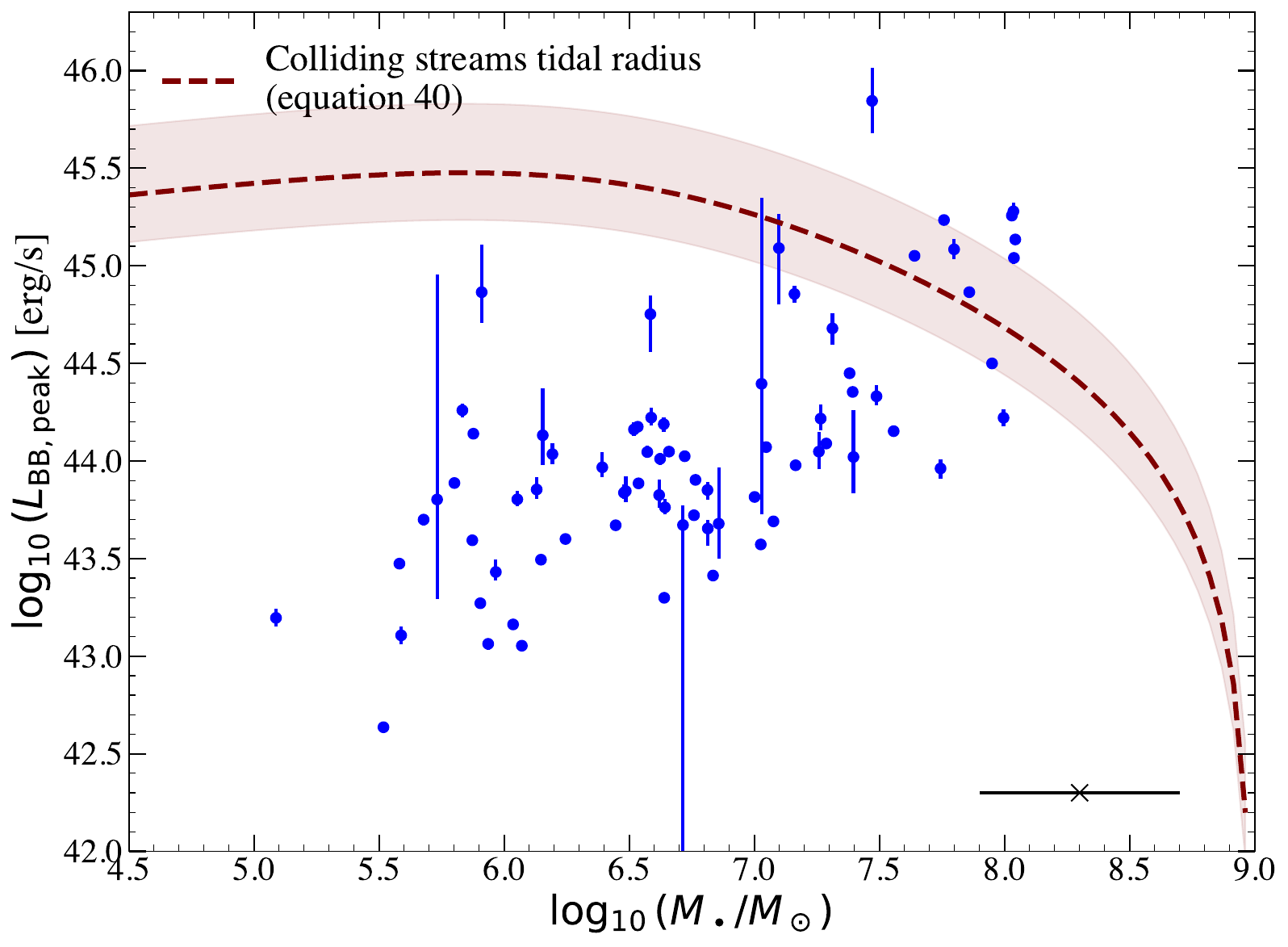}
    \includegraphics[width=0.32\linewidth]{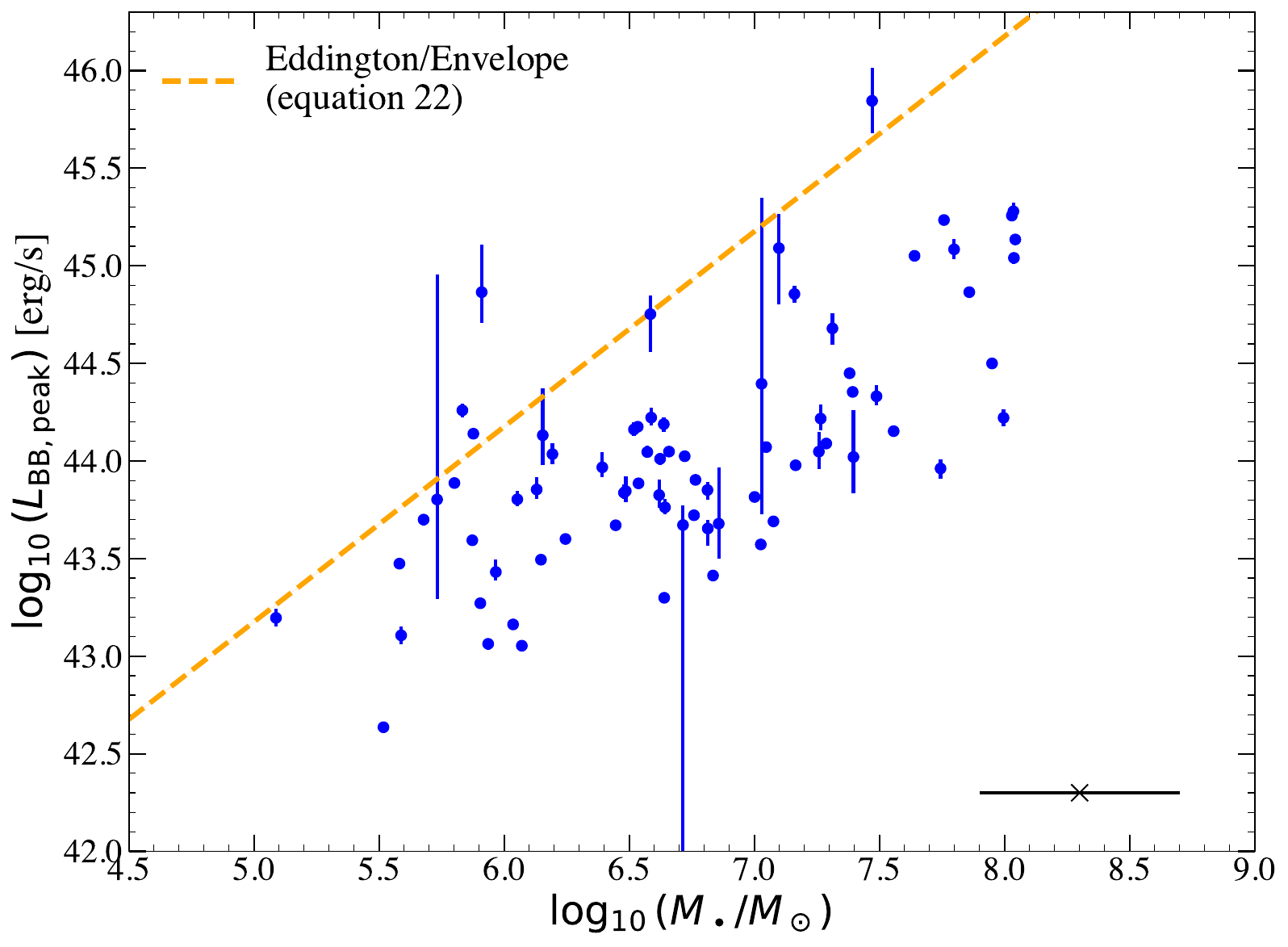}
    \includegraphics[width=0.32\linewidth]{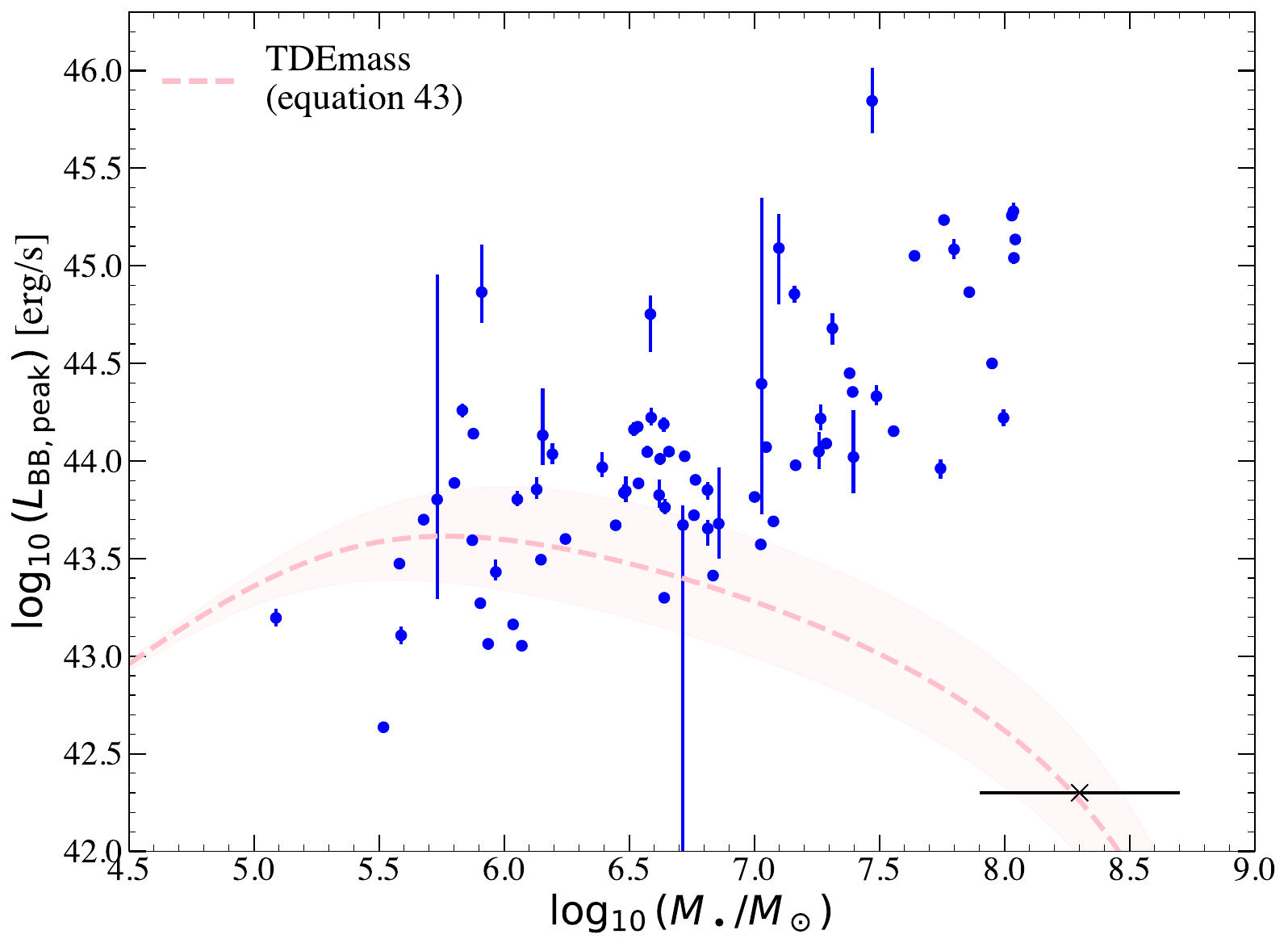}
    \includegraphics[width=0.32\linewidth]{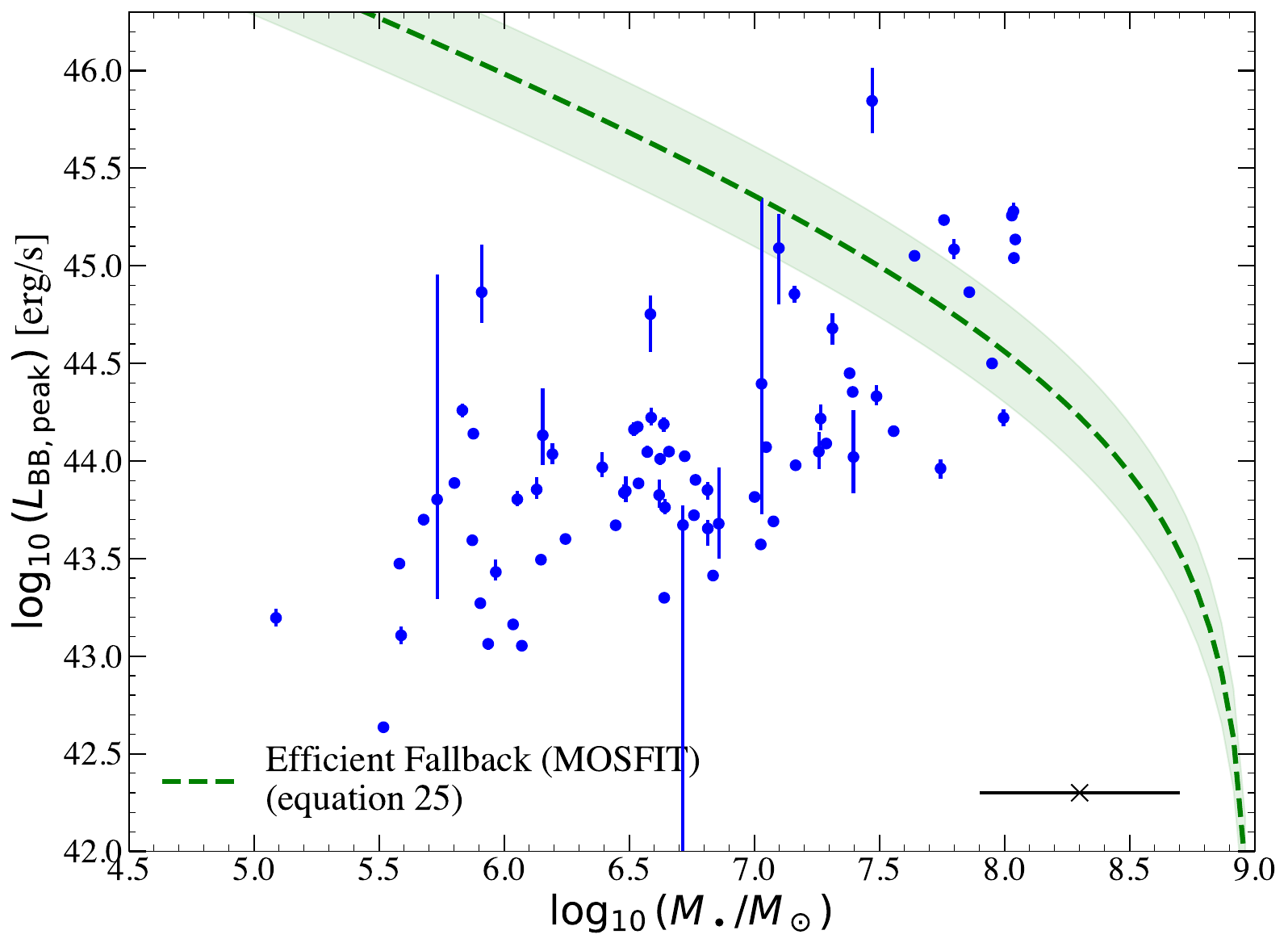}
    \includegraphics[width=0.32\linewidth]{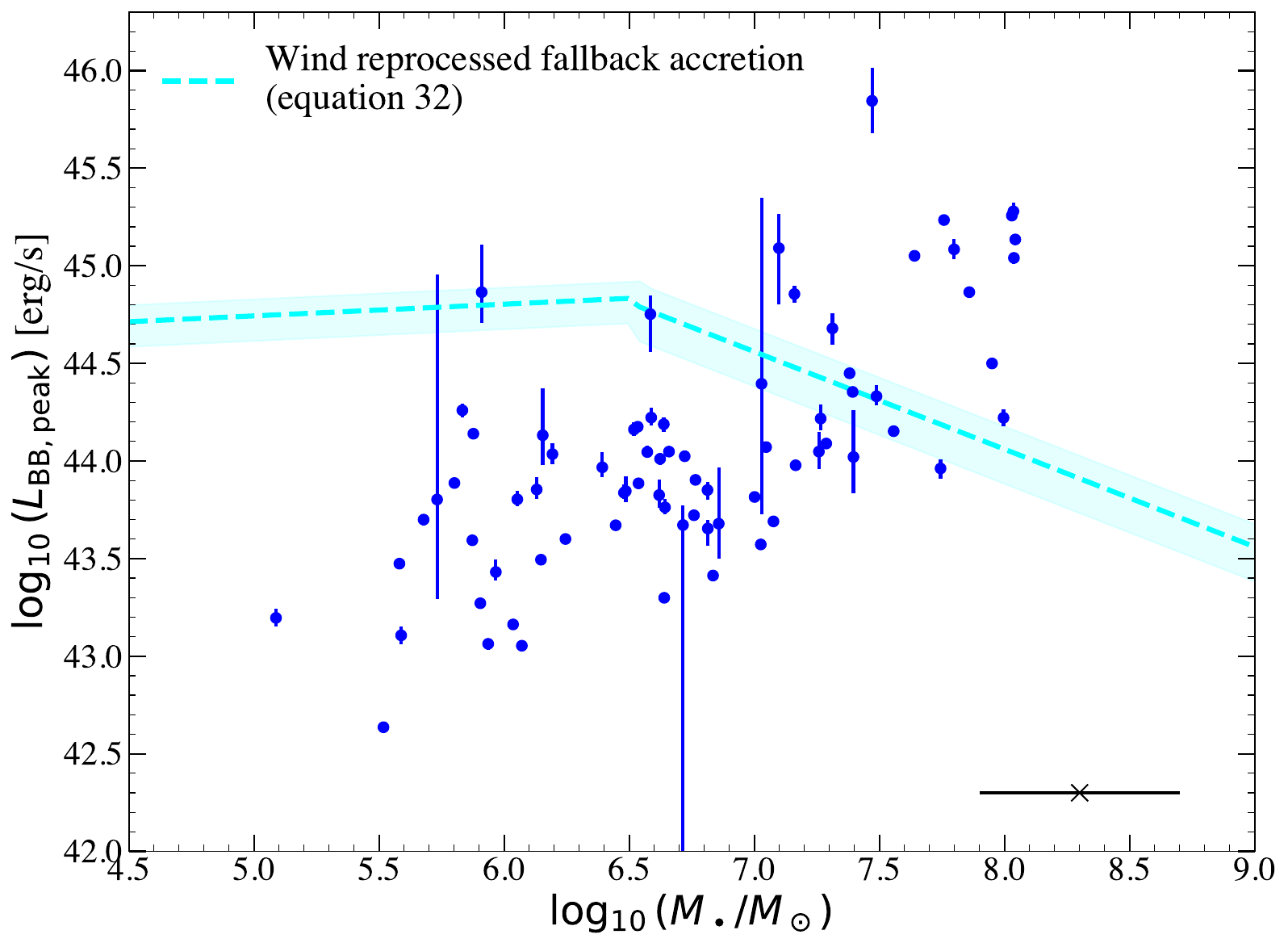}
    \caption{The scaling with black hole mass of the peak blackbody luminosity expected to be observed from the optical/UV flare in a TDE, for the seven models collated in this work. Shown by a dashed curve is the result for $m_\star = 0.5$, and the degree of expected variance in the models is shown by the shaded regions which denote the range for stellar mass $m_\star = 0.3-0.7$. By points we show the observed TDE values. Different models are labeled on each plot.  Typical uncertainties in the black hole masses are shown by the error bar in the lower right hand corner of each plot.  }
    \label{fig:lum_comp}
\end{figure*}

In Figure \ref{fig:lum_comp} we display, as seven panels, the seven peak luminosity model predictions summarized in this paper, as a function of black hole mass. We remind the reader that the black hole masses in this plot come from late-time UV observations of the system, but that this choice does not impact the scaling of the luminosity with mass (see Figure \ref{fig:fits} and Appendix \ref{app:M}).  Typical uncertainties in the black hole masses are shown by the error bar in the lower right hand corner of each plot. To display some of the expected variance in the models we vary the assumed incoming stellar masses, with a dashed curve showing the result for $m_\star = 0.5$, and a shaded region denoting the predicted range for stellar mass $m_\star = 0.3-0.7$.  

We see, immediately, that all models except Eddington-limited accretion, a cooling Eddington-limited envelope, and colliding streams at the tidal radius show a decreasing optical luminosity with black hole mass, in strong tension with what is observed. This immediately rules out many existing models in the literature. 

We can further rule out the model of colliding streams at the tidal radius as it grossly over-predicts the luminosity which should be observed around black holes at the low mass end of the TDE distribution $M_\bullet \sim 10^5-10^{6.5} M_\odot$. This is problematic as, if such a scaling was physical, it should be extremely easy to detect  events around these low mass black holes, as the number of low mass galaxies (and therefore low mass black holes) outnumbers their higher mass companions, and therefore the detected TDE rate should be dominated by these low-mass high-luminosity events. 

It is for this same reason that one cannot invoke, e.g., a high fraction of partial TDEs with lower fallback rates in an attempt to salvage the efficient fallback accretion model. While it is true that a partial TDE results in a significantly lower peak fallback rate, every such partial TDE model still scales as $L_{\rm pk} \sim M_\bullet^{-1/2}$, and one would need to invoke a conspiracy whereby the partial TDE fraction scales strongly with black hole mass. It is clear that full disruptions of solar mass stars by $10^6 M_\odot$ black holes do happen, and if a model suggests that these should be the brightest (and therefore easiest to detect) TDEs, this model is ruled out by the data.

\subsection{The radiated energy}
\begin{figure*}
    \centering
    \includegraphics[width=0.32\linewidth]{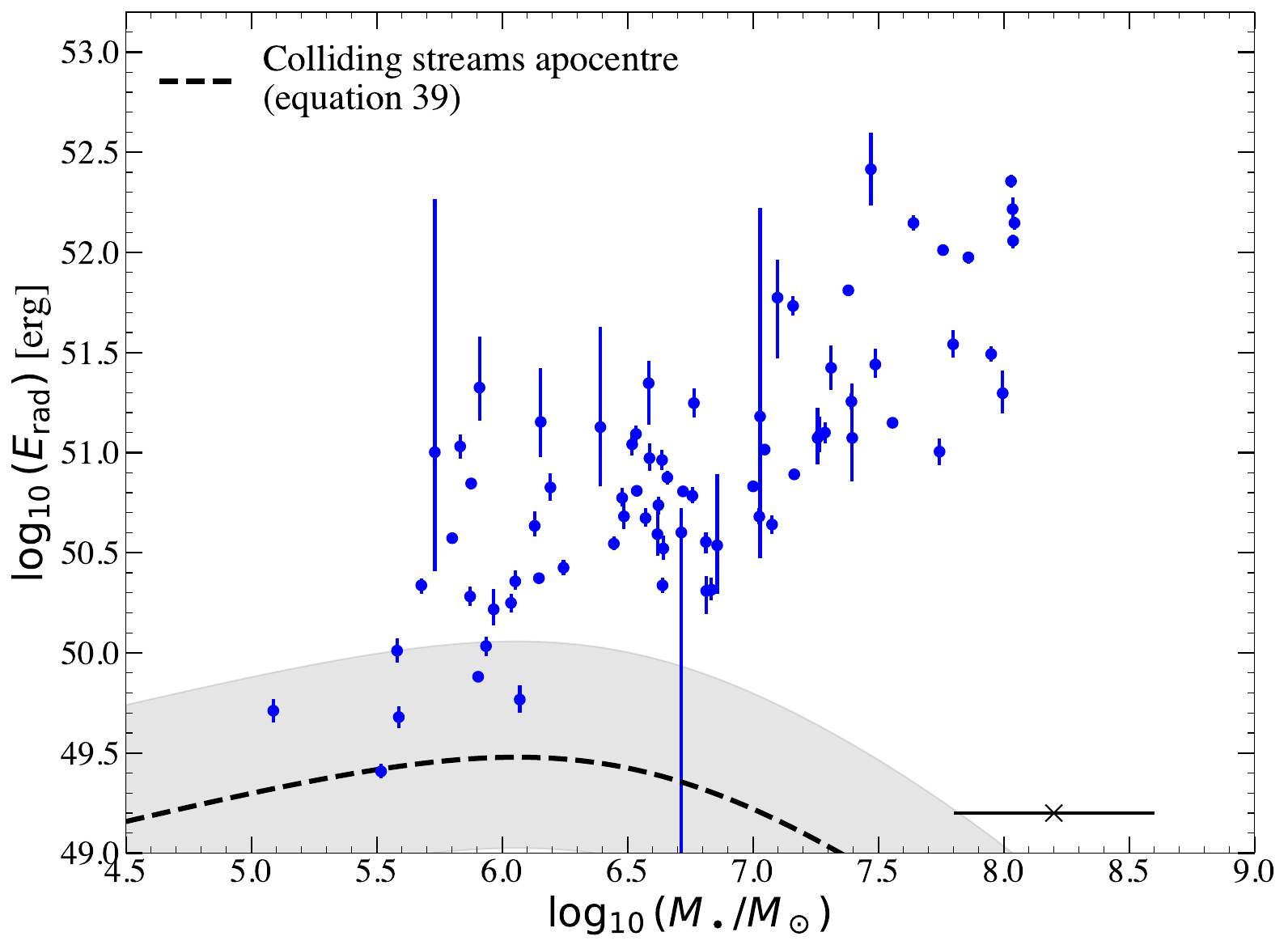}
    \includegraphics[width=0.32\linewidth]{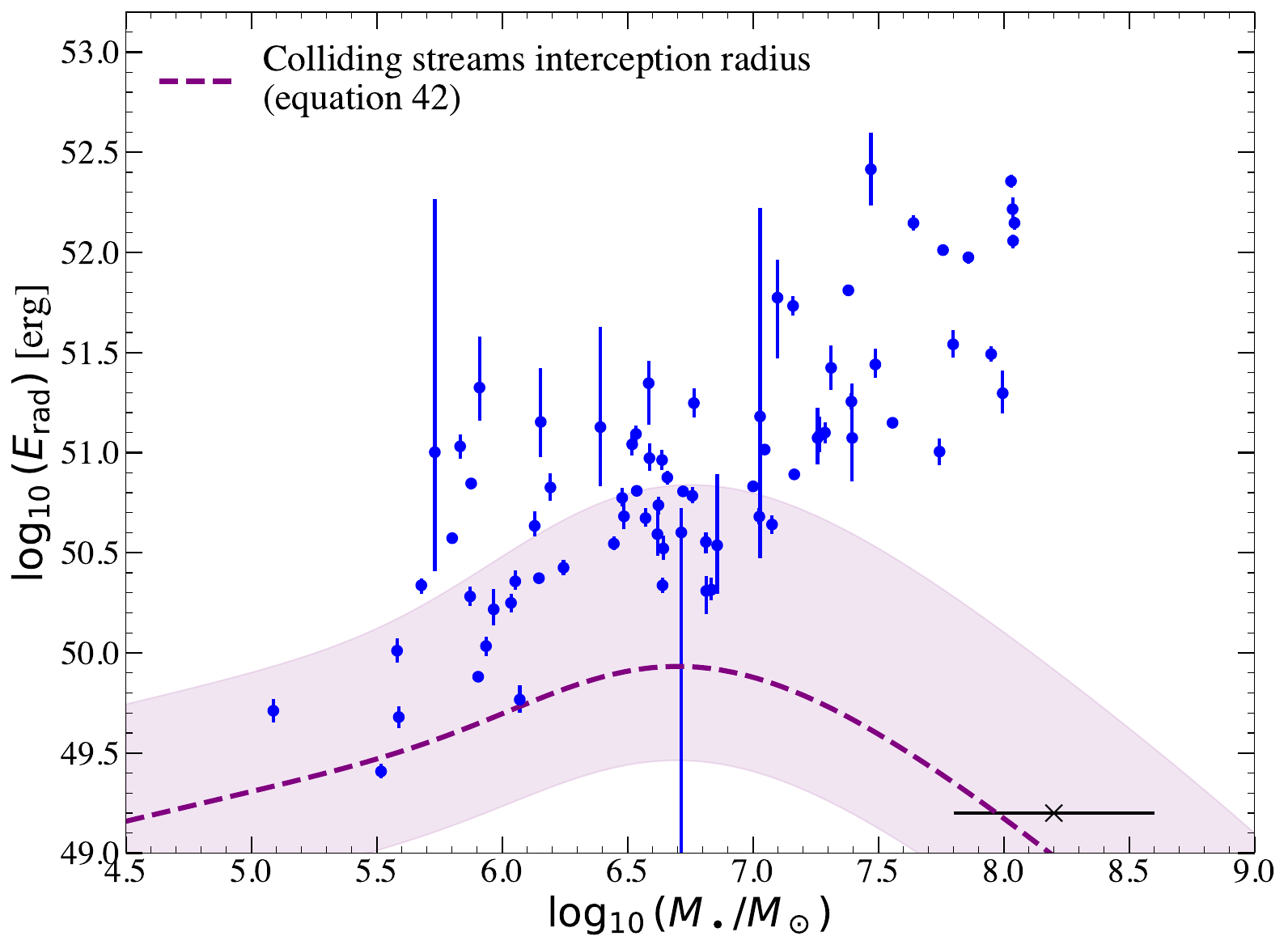}
    \includegraphics[width=0.32\linewidth]{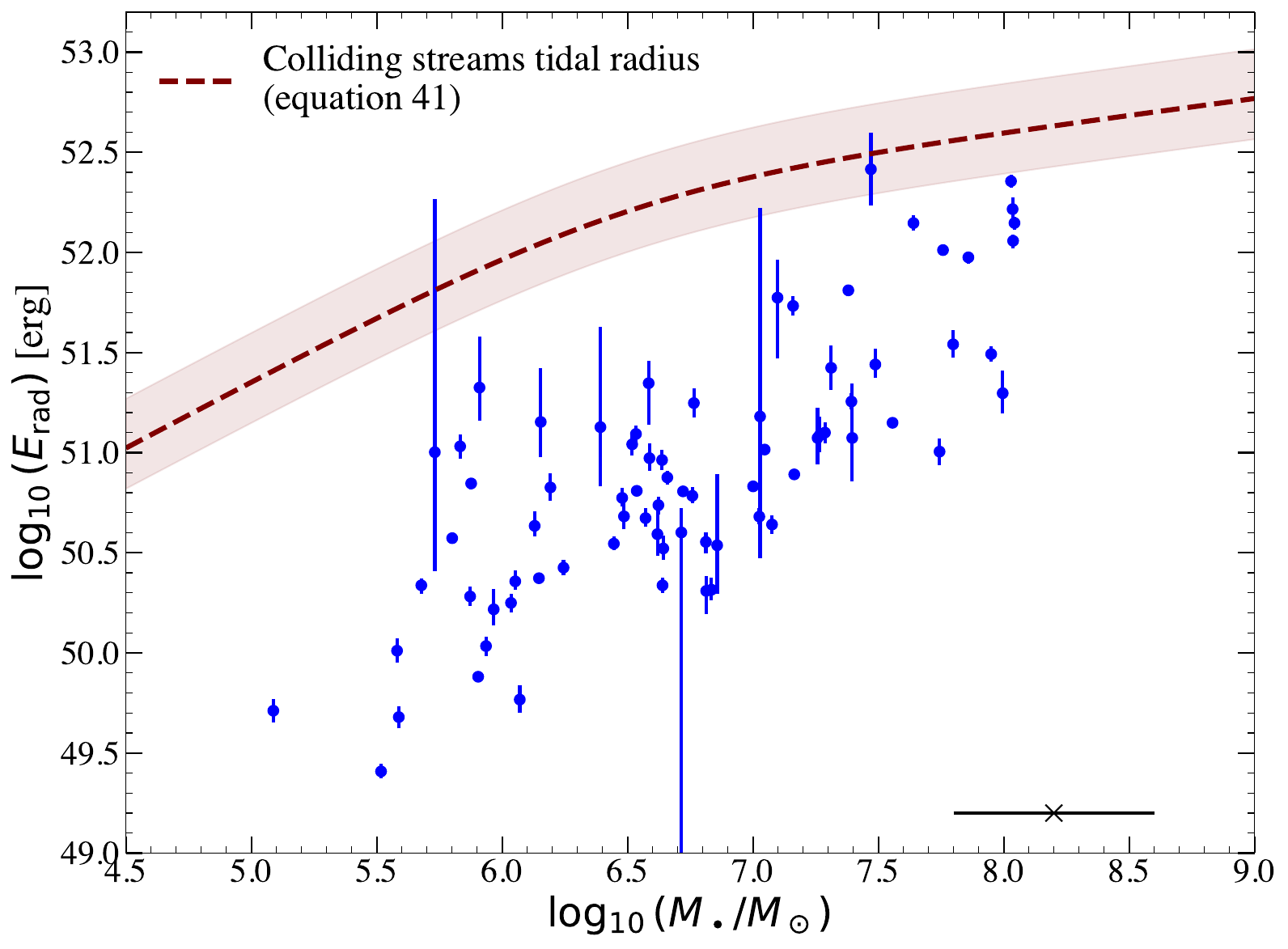}
    \includegraphics[width=0.32\linewidth]{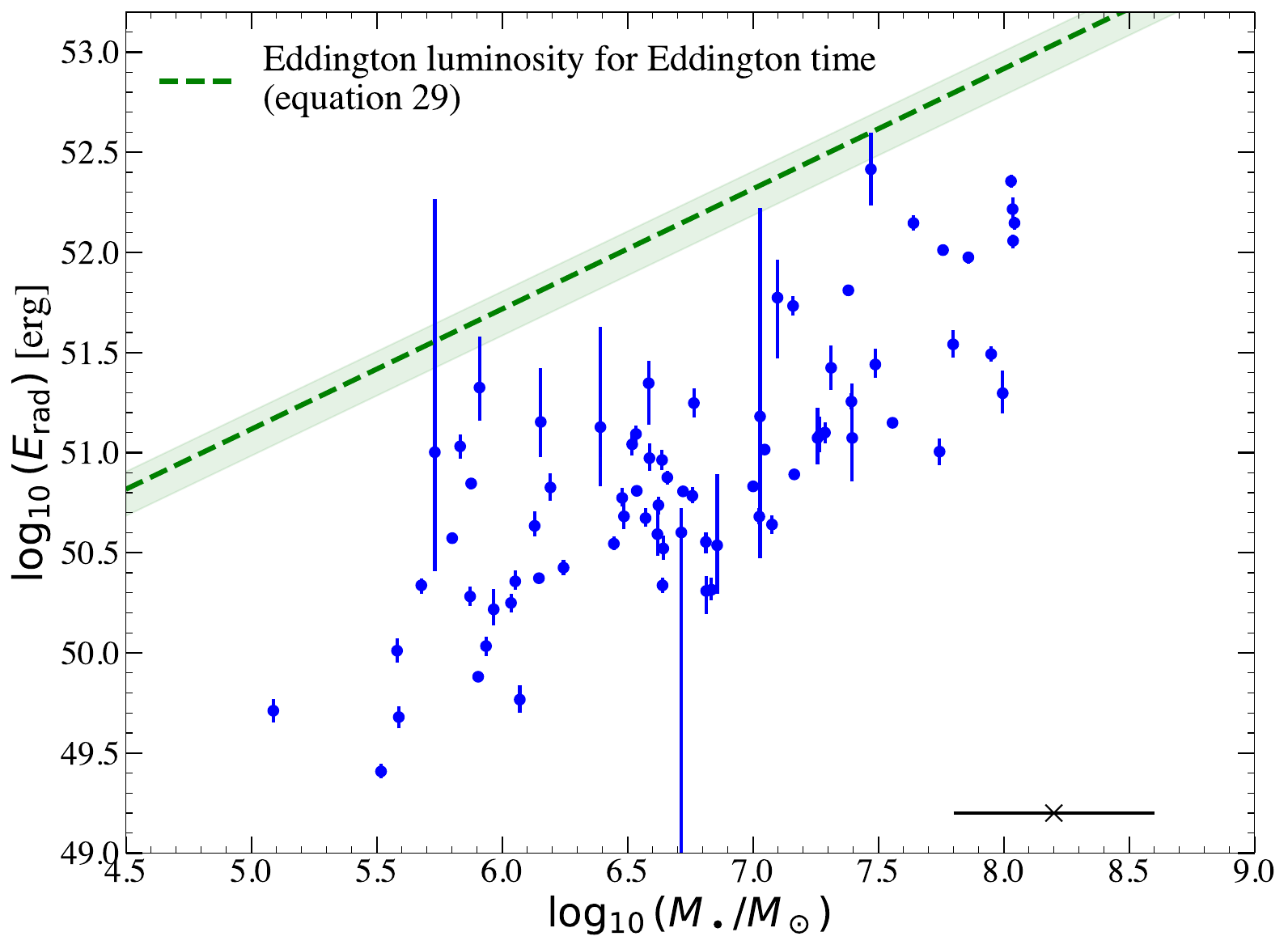}
    \includegraphics[width=0.32\linewidth]{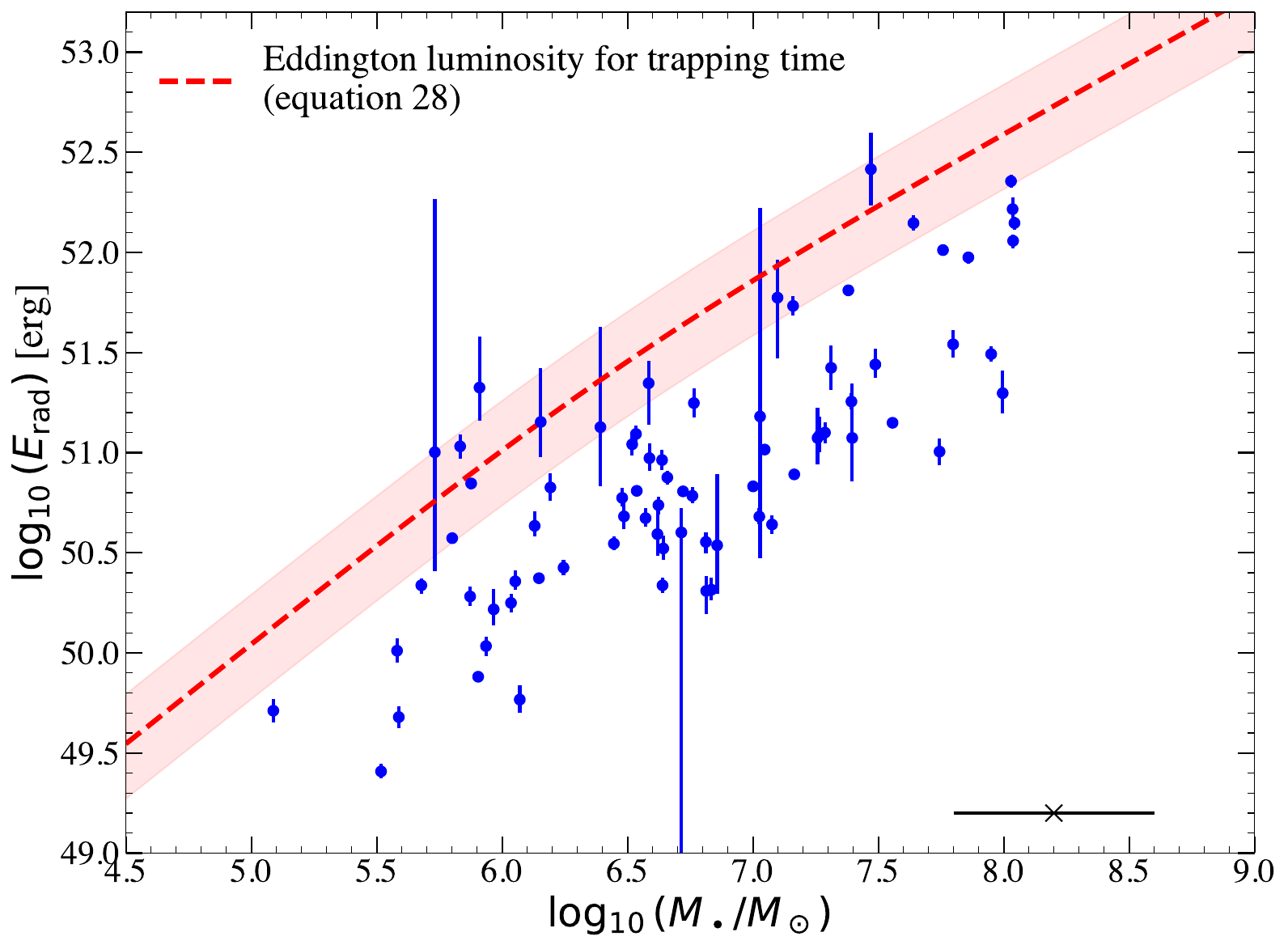}
    \includegraphics[width=0.32\linewidth]{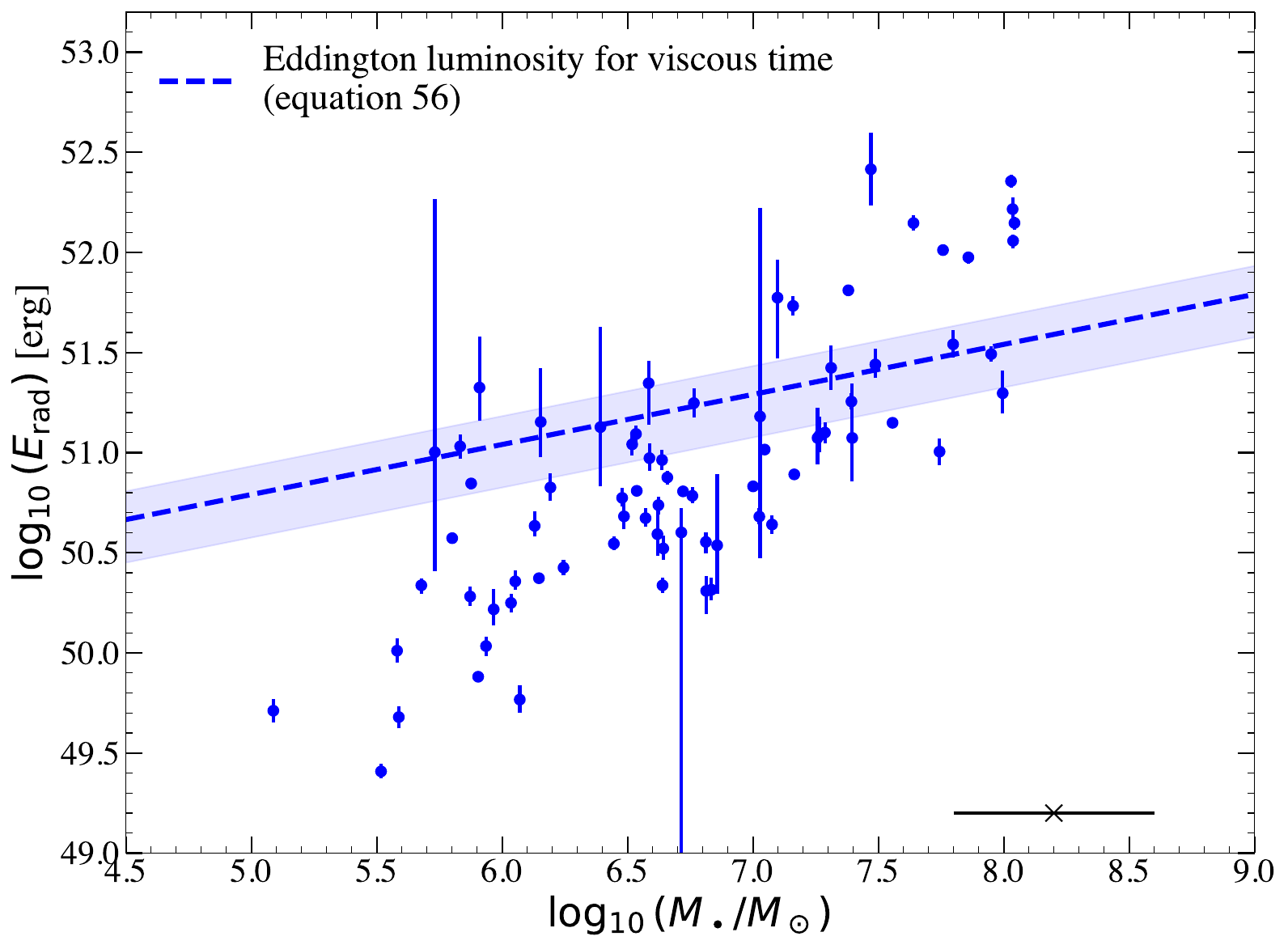}
    \includegraphics[width=0.32\linewidth]{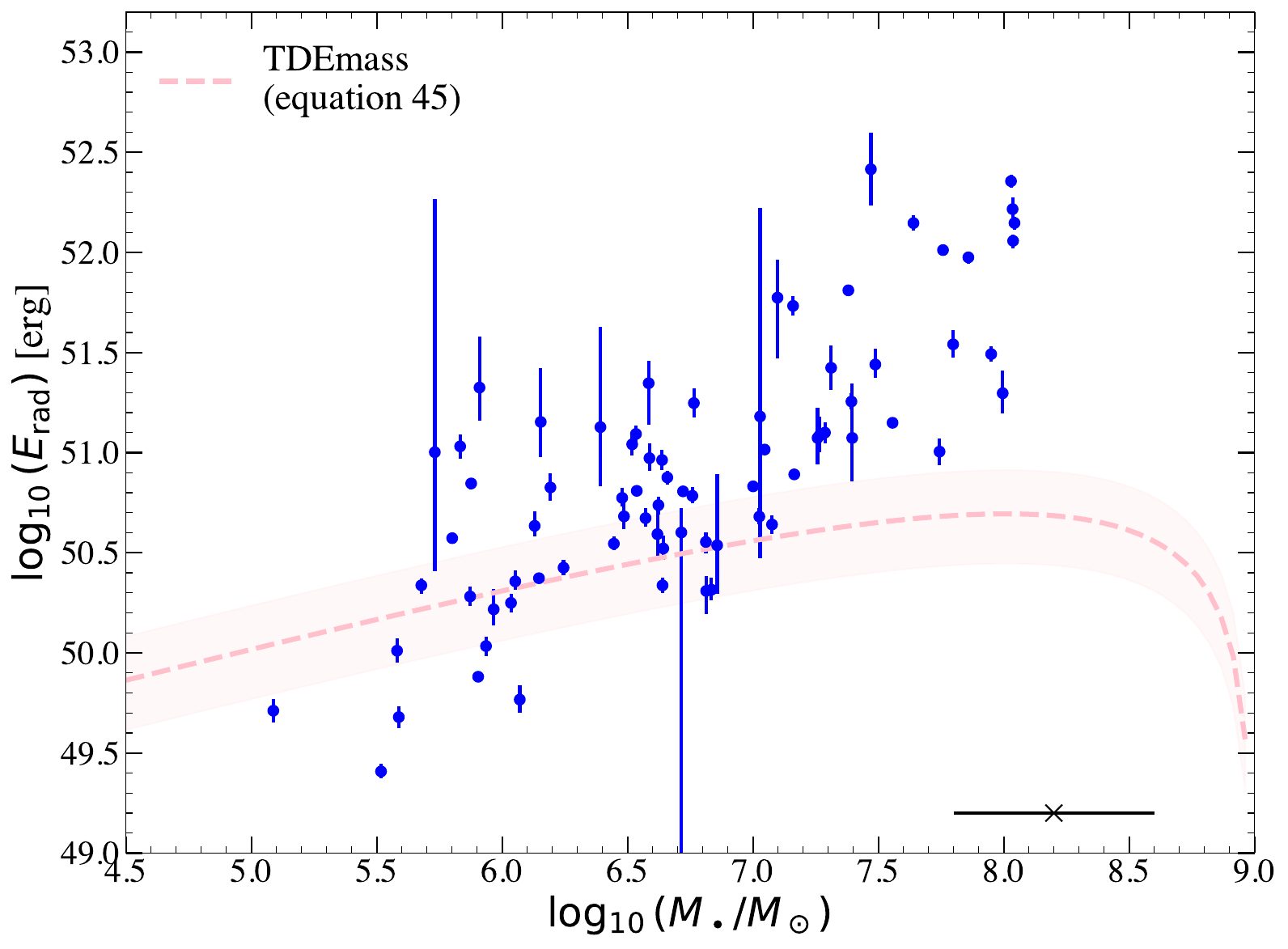}
    \includegraphics[width=0.32\linewidth]{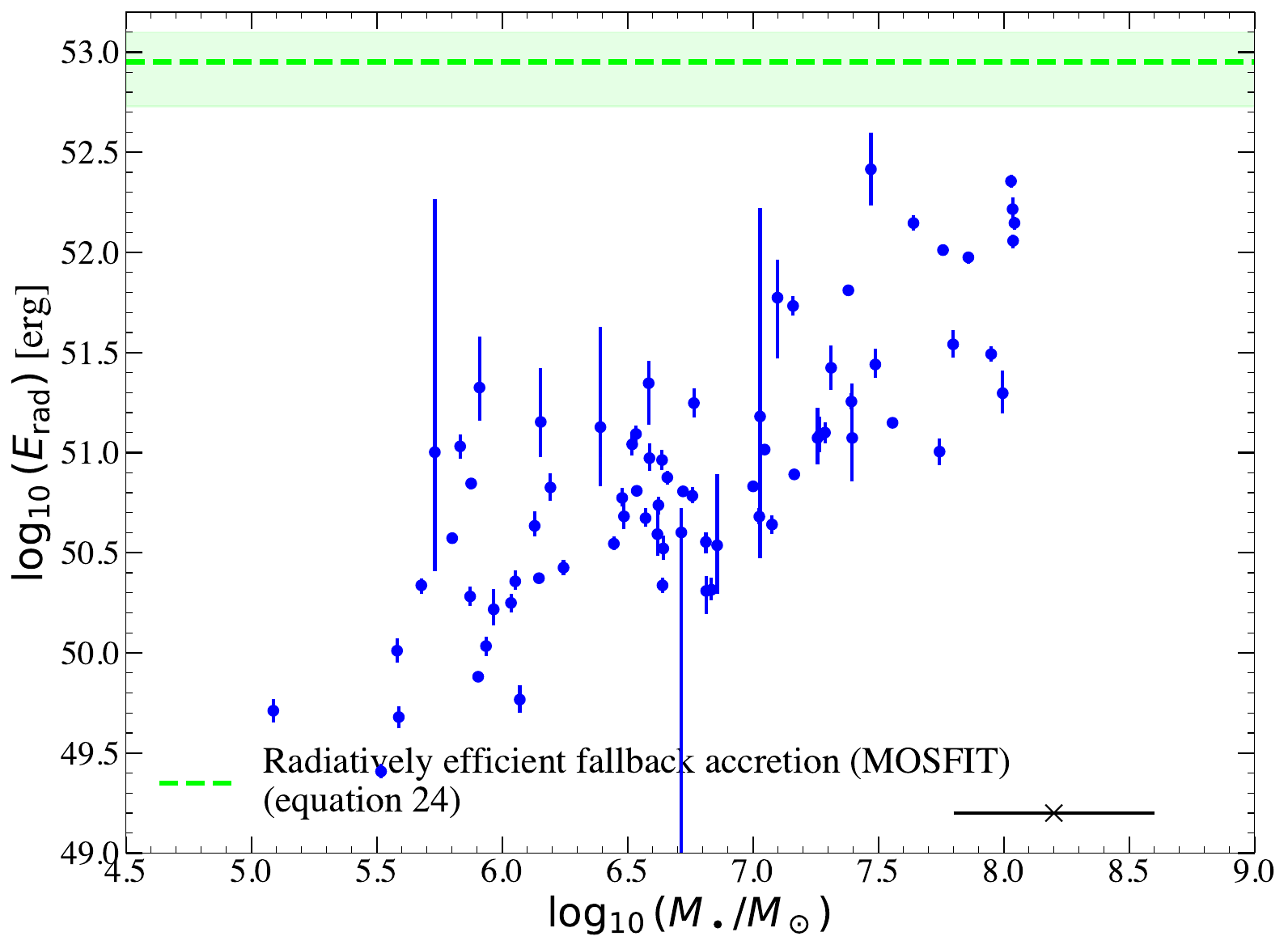}
    \includegraphics[width=0.32\linewidth]{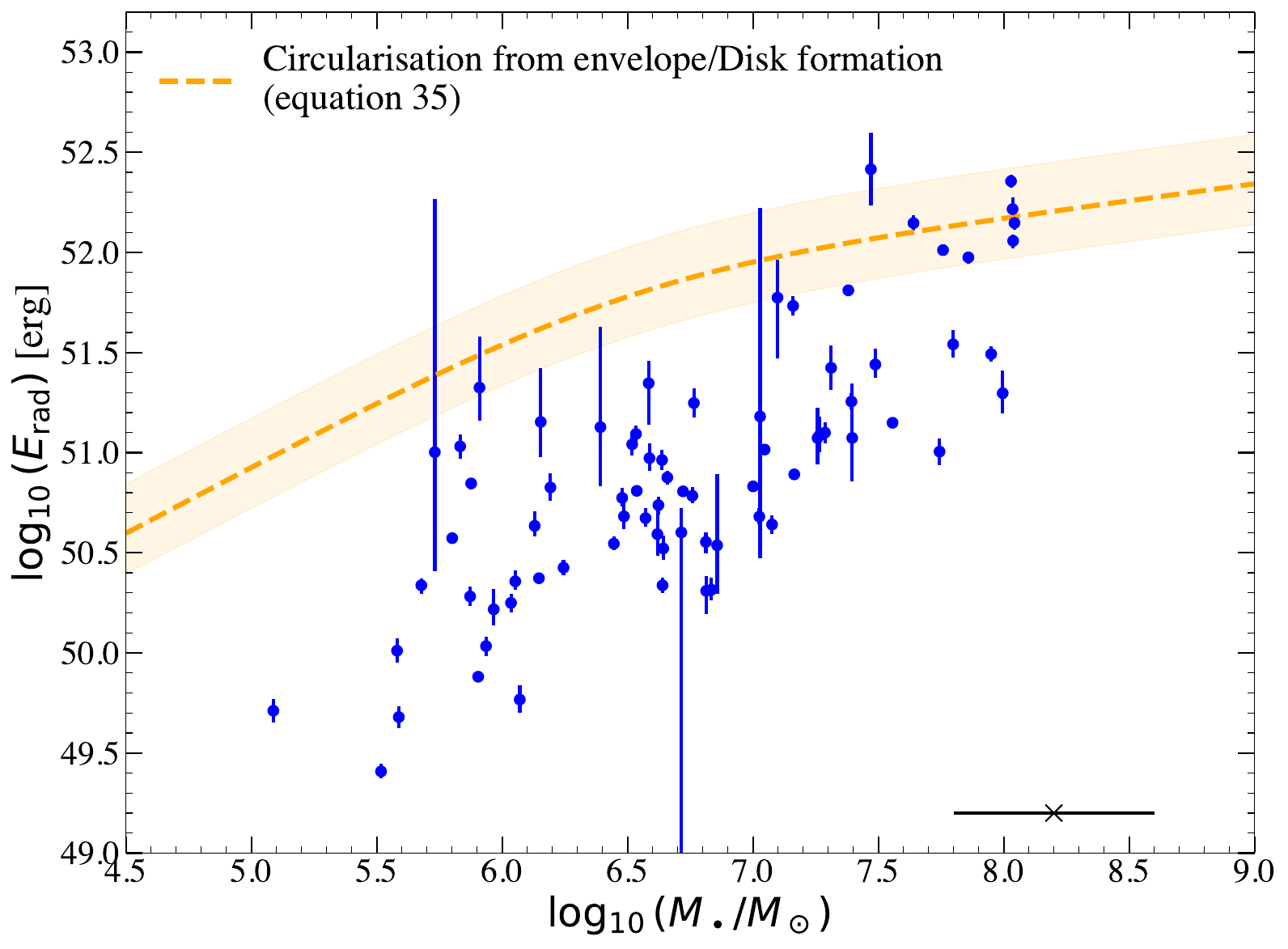}
    \includegraphics[width=0.32\linewidth]{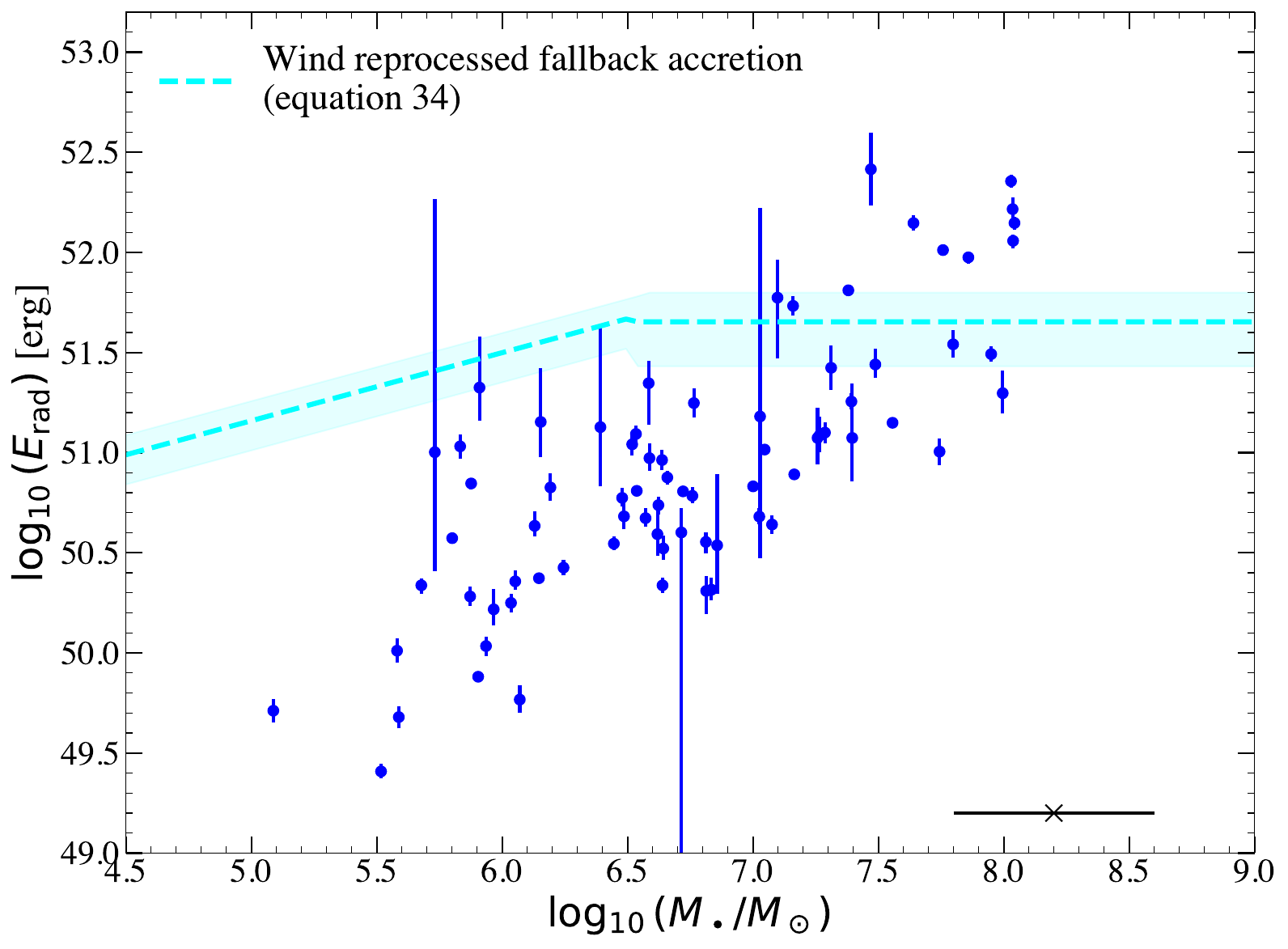}
    \caption{The scaling with black hole mass of the radiated energy expected to be observed in the optical/UV flare in a TDE, for the ten models collated in this work. Shown by a dashed curve is the result for $m_\star = 0.5$, and the degree of expected variance in the models is shown by the shaded regions which denote the range for stellar mass $m_\star = 0.3-0.7$. By points we show the observed TDE values. Different models are labeled on each plot. Typical uncertainties in the black hole masses are shown by the error bar in the lower right hand corner of each plot. }
    \label{fig:en_comp}
\end{figure*}

In Figure \ref{fig:en_comp} we display the scaling with black hole mass of the radiated energy expected to be observed in the optical/UV flare in a TDE, for the ten models collated in this work. Again we show by a dashed curve is the result for $m_\star = 0.5$, and the degree of expected variance in the models is shown by the shaded regions which denote the range for stellar mass $m_\star = 0.3-0.7$. By points we show the observed TDE values. Different models are labeled on each plot. Typical uncertainties in the black hole masses are shown by the error bar in the lower right hand corner of each plot. 

We again find that all models except the various Eddington-limited accretion scenarios and a cooling Eddington-limited envelope/disk formation process are in tension with the data (as the colliding streams at the tidal radius model is ruled out on luminosity grounds).  It appears therefore that both pieces of observational evidence point towards this disk formation/disk reprocessing source of TDE optical/UV flares. 

In the follow section we discuss which frameworks remain plausible given the results of our calorimetric tests in more detail. 

\section{Plausible remaining frameworks}\label{sec:disc}
\subsection{Efficient fallback accretion?}
The radiatively efficient reprocessing of the fallback-rate $\dot M_{\rm fb}$ into optical/UV emission is ruled out by the data on multiple grounds. Firstly, the scaling with black hole mass $\dot M_{\rm fb} \propto M_\bullet^{-1/2}$ is over $5\sigma$ from what is observed, as is the scaling with black hole mass of the radiated energy. 

In addition, for such a scenario to be consistent with the data, it would require that observational surveys are failing to find a large population of the {\it brightest} low black hole mass TDE sources, or that there is some unknown effect which results in an inverse Malmquist bias where only the intrinsically least luminous objects are found in a flux limited survey. 

It is worth reflecting on why models based on this assumption (for example {\tt MOSFIT}) ``work'', in the sense that they produce light curves which pass through observations, given that their governing assumptions are so far removed from observations. We believe this is simply a result of the large number of free parameters that are allowed to vary freely in the fitting process. In particular, {\tt MOSFIT} has a free efficiency $\eta$ in $L=\eta \dot M_{\rm fb}c^2$, and also a free reprocessing radius $(R_{\rm ph, 0})$ for its photosphere. What this means in practice is that {\tt MOSFIT} can reproduce any amplitude of optical emission (because the fallback rate always hugely {\it over predicts} the luminosity), and any temperature of the early flare $T = (L/4\pi\sigma R_{\rm ph}^2)^{1/4}$ (and therefore any SED normalisation and shape), merely by tuning two parameters $(\eta, R_{\rm ph, 0})$ which apriori have no physical grounding. Therefore the only physical parameters which are really being fit to data are those controlling the overall temporal shape of the light curve (i.e., the rise and decay timescales). However, the dynamic range of decay timescales in the TDE data is relatively limited, spanning $\tau \sim 50-300 \, {\rm d}$ \citep{Hammerstein23, Yao23}. In a grid of time-dependent fallback rate curves (which can further be scaled by changing the black hole mass $t_{\rm fb} \propto M_\bullet^{1/2}$) it appears to us that one will always be able to find a time-dependency that reproduces the time dependency of the data. Coupled with the free normalisation and temperature of the emission, there are enough free parameters to reproduce a wide range of light curves. 

To highlight one example of this parameter compensation explicitly, we note that \cite{Nicholl2022} (using a sample of TDEs and \texttt{MOSFIT}), recovered an efficiency  scaling of $\eta \propto M_{\bullet}^{0.97 \pm 0.36}$. This can be traced directly to the integrated energy in the flares, which scales only with $E_{\rm rad} \sim \eta M_\star$ in {\tt MOSFIT}, see Fig.~\ref{fig:en_comp}. Clearly, $\eta$ is simply compensating for the true scaling in the data, and has no physical meaning. This is likely why {\tt MOSFIT} does not recover known galactic scaling relationships from TDE data \citep{Ramsden22, Hammerstein23, Guolo25c}. 

\subsection{A shock conspiracy?}
An interesting result to note is that at low black hole masses $M_\bullet \lesssim 10^{6.5}M_\odot$, shocks at apocentre produce a reasonable description of the luminosity and energetics of TDE optical flares, while at $M_\bullet \gtrsim 10^{7.5} M_\odot$ shocks at pericentre (the tidal radius) provide a reasonable description of the luminosity and energetics. One could therefore imagine a scenario at which the collision radius moves systematically with black hole mass between the outer and inner radial scales, to generate a growing luminosity/energy profile. 

We believe this scenario is unlikely to be physically realistic, and instead represents fine tuning of models aposteri as it was not, as far as we are aware, predicted ahead of time.  It would also require the generation of smoothly growing luminosity and energy profiles from a sequence of profiles that all decline with mass.

\begin{figure*}
    \centering
    \includegraphics[width=0.6\linewidth]{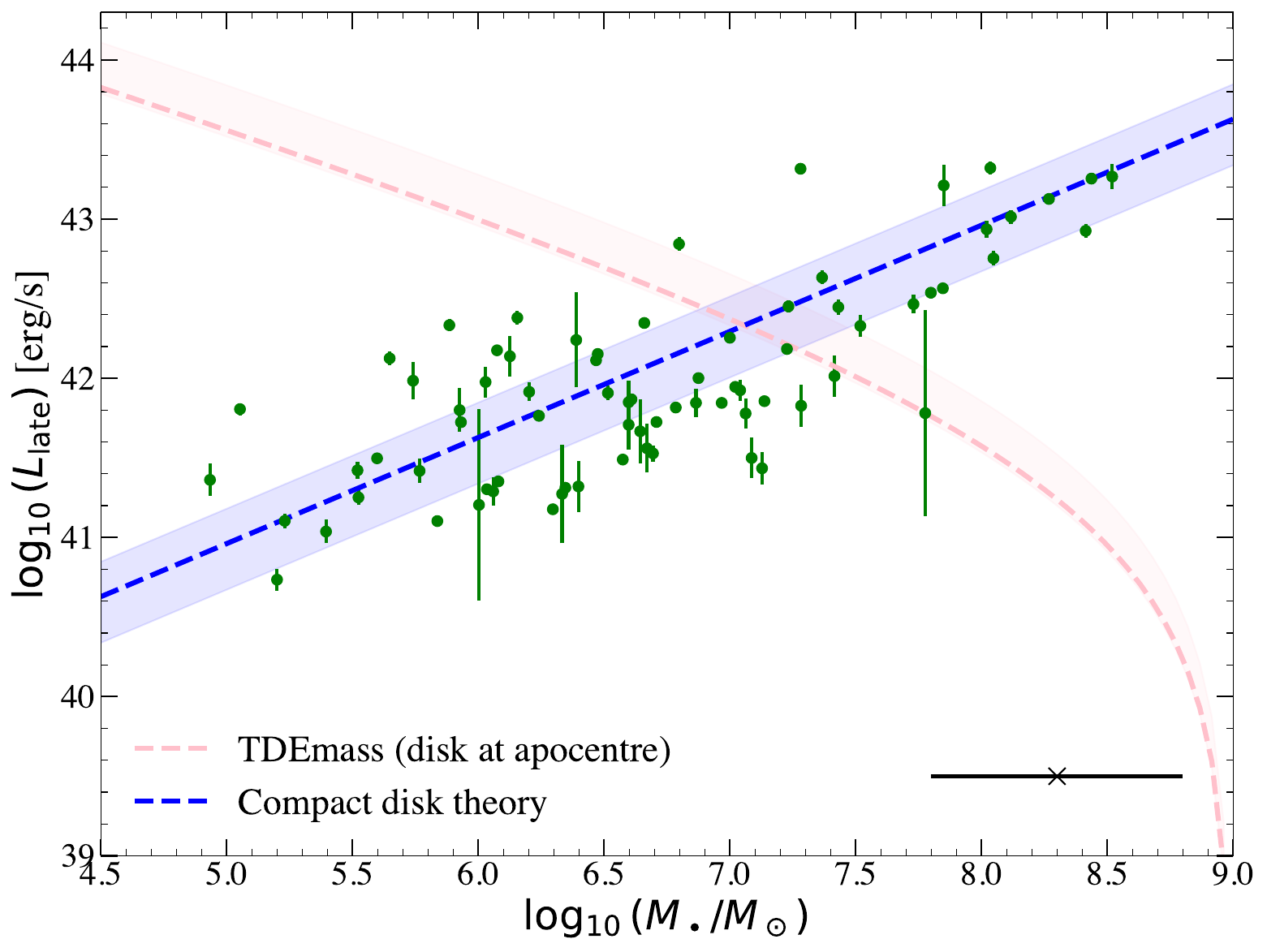}
    \caption{The scaling with black hole mass of the late time ($\Delta t > 1\, {\rm yr}$) optical/UV luminosity expected to be observed after the initial optical/UV flare in a TDE, for disk formation scenarios at different radial scales. Shown by a dashed curve is the result for $m_\star = 0.5$, and the degree of expected variance in the models is shown by the shaded regions which denote the range for stellar mass $m_\star = 0.3-0.7$. By points we show the observed TDE values, with black hole masses inferred from galactic scaling relationships ($M_\bullet-\sigma$ where available, and $M_\bullet-M_{\rm bulge}$ otherwise).  Typical uncertainties in the black hole masses are shown by the error bar in the lower right hand corner of each plot. }
    \label{fig:late}
\end{figure*}

Perhaps most problematic for this framework is that shocks/disk formation at apocentre (which would be required in this framework to power the low black hole mass optical flares) predict a late-time ($\Delta t \gtrsim 1 \,{\rm yr}$) luminosity which is in strong contention with the data at these low masses (Figure \ref{fig:late}). 

In Figure \ref{fig:late} we contrast the predictions of disks which form at apocentre (pink curve and shaded region, \citealt{Krolik24}) with the predictions of compact (i.e., formed at $\sim$ the tidal radius) disk theory \citep[blue curve and shaded region][in particular their equations 1, 2 and 46]{Mummery_et_al_2024}. The green points represent observations of TDEs, with the mass estimated from galactic scaling relationships (we use $M_\bullet-\sigma$ where available, and $M_\bullet-M_{\rm bulge}$ where not). Shown by a dashed curve is the result for $m_\star = 0.5$, and the degree of expected variance in the models is shown by the shaded regions which denote the range for stellar mass $m_\star = 0.3-0.7$. The stellar mass is not the only source of scatter in this relationship, and both should also have additional $\cos i$ scatter from the disk-observer inclination (as disk are two dimensional). We see that disks at apocentre are ruled out owing to both their late time luminosity scaling with black hole mass, but also their systematic over prediction of late time luminosities at low masses (which again should be easiest to find, and are not). This result is robust to any redshift cuts one wishes to place on the data (Appendix \ref{app:Z}), as this makes no difference to the observed scaling, it only reduces the number of points. Plateau detection in TDEs is almost ubiquitous (Appendix \ref{app:Z}), with every TDE within $z<0.08$ and with $>2$ years of data having a detected plateau, making this a very strong observational constraint. 

While shocks are clearly essential in getting light out of a tidal disruption event (without them no disk can form), we believe that the current data rules them out as being directly involved with producing a dominant component of the early time optical/UV emission. 

\subsection{Reprocessed Eddington-limited accretion?}
We now move onto models which appear in better accord with the data. Figures \ref{fig:lum_comp}, \ref{fig:en_comp} and \ref{fig:late} suggest that a framework whereby a compact accretion flow forms and accretes at high rates, with a luminosity which is then reprocessed into optical/UV emission, provides an acceptable description of the gross luminosity/energy scales at early and late times. 

At this stage this of course is by no means a ``model'' for optical flares in TDEs (in the sense that we have made no prediction for the evolution of the light curves from such a system), but this work clearly suggest that this represents an interesting avenue for future study. It is clear from late-time observations in TDEs that compact accretion flows are formed in the vast majority of TDEs (Figure \ref{fig:late}, Appendix \ref{app:Z}), and given their high natural luminosity scales it is also clear that the energetics of this early phase of disk evolution could provide a plausible energy source for the observed optical emission. 

Further evidence for Eddington-limited accretion in TDEs comes from the observed break in the TDE X-ray luminosity function at $L_{X, {\rm br}} \sim 10^{44}$ erg/s \citep{Guolo24, Grotova25}. This was predicted ahead of time by \cite{Mum21} as a clear signature of Eddington-limited accretion in TDE disks \citep[a result confirmed by more recent simulations][]{MummeryVV25}. Any model of the early evolution in TDEs has to allow for Eddington limited accretion in a compact disk to explain this X-ray observational result. 

For this interpretation of optical flares to be considered plausible a number of questions must be answered/clarified. Firstly, what is the correct timescale over which this reprocessing should act? The fallback, trapping and viscous timescale can all be expected to play an important role, possibly in different regions of parameter space. Secondly, is it plausible to produce a time-evolving optical/UV flare from a $\sim $ constant (Eddington) input luminosity? It is clear that a time-evolving photosphere (similar in spirit to that in the cooling envelope model, \citealt{Metzger22}) would be required to generate time-evolving optical emission. 

The main flaws with this interpretation are the following. Firstly, the fact that we see prompt (coincident with the optical flare) and extremely luminous ($L_X\sim 10^{44}$ erg/s) X-ray emission from some TDEs means that all of the luminosity emitted from these compact disks cannot possibly be being reprocessed. This could be circumvented by some geometric/viewing angle argument \citep[a la][]{Dai18}, a possibility certainly worth exploring (but note that \citealt{Dai18} set the accretion rate equal to the fallback rate, and so this is a framework which must be modified). 

The most consequential flaw in this framework, however, is the simple question of how it is possible to sustain Eddington-limited accretion at the highest ($M_\bullet \gtrsim 10^{7.5} M_\odot$) black hole masses, when not even the fallback rate is super-Eddington (eq. \ref{eq:Mdotfb})? Resolving this (i.e., producing a high enough luminosity in these systems) appears to require fine-tuning, as it must invoke high mass stars and high black hole spins to generate enough luminosity in this limit.  This may not be completely unreasonable (both high mass stars and high spins are required to get a tidal disruption near to the Hills mass limit \citealt{Hills75,Kesden12, Mummery24} when the fallback rate starts to become sub-Eddington at peak for solar-type stars), but appears unsatisfactory. More generally, if the viscous timescale is long (compared to the fallback/circularisation timescale) it is not difficult for many TDEs to produce sub-Eddington  disks at peak (see Appendix \ref{app:acc}). 



\subsection{Cooling envelope/disk formation?}
The cooling envelope/compact disk formation scenario appears to provide a natural framework to interpret many of the observational properties seen on the population level in TDEs. The energy budget which must be released to form a compact disk
\begin{equation}
    E_{\rm circ} \approx {GM_\bullet M_{\rm disk} \over 2 R_{\rm circ}} \propto M_\bullet^{2/3} ,
\end{equation}
provides the best description of the observations. As it is clear that a compact disk does form in TDEs, with evidence from a wide range of observational fronts \citep[e.g.,][]{Guolo24, Mummery_et_al_2024, Grotova25, MummeryVV25, Guolo25c}, it is clear that ultimately this energy budget must be released in the process. If, as discussed in \cite{Metzger22}, this energy is released by radiative cooling at the Eddington luminosity, then this provides a natural resolution of the growing optical/UV luminosity with black hole mass. 

As this framework naturally produces compact accretion disks at $\sim$ the tidal radius, it is also consistent with observations at late times (Figure \ref{fig:late}). 

However, it is not clear how this and the previously discussed framework (of Eddington limited accretion)  interact. It is clear observationally that Eddington-limited accretion flows do form in TDEs, and at least some of them form promptly (during the optical flare). This provides an additional energy source (plausibly larger than the disk formation energy budget), which will  inevitably be deposited into the envelope, modifying the picture presented in \cite{Metzger22}. A spherically symmetric envelope which blocks all X-ray emission at early times (during the optical flare) also cannot reproduce the properties of many TDE systems, requiring at least some sight lines where X-ray radiation can escape unimpeded. 

The cooling envelope/disk formation picture similarly (as Eddington-limited accretion) struggles to reproduce observations at the highest black hole mass scales, where the Eddington luminosity begins to exceed plausible luminosity scales in the problem. In this limit the cooling time of the envelope (unless the star is very massive) should be very short, with the material forming a disk on $\sim$ the dynamical timescale. 



\section{Conclusions}\label{sec:conc}
In this paper we have collated all available  data of  optical/UV flares seen at early times in TDEs, and compared the black hole mass dependence of the peak luminosity and radiated energy observed in the  TDEs early time optical/UV flare with that predicted from a comprehensive survey of published models from the literature. 

No one existing framework provides a truly satisfactory description of the data at all black hole mass scales, and we believe that the precise origin of early-time optical/UV emission in TDEs remains an open question. This work has not been without utility however, as many popular frameworks in the literature can be shown to be in strong contention with observations, and are therefore ruled out. 

In particular, the efficient reprocessing of the fallback rate into an optical luminosity is completely ruled out by the data (as is any model which sets the optical/UV luminosity proportional to the fallback rate). Both the peak luminosity and radiated energy seen in TDEs are of fundamentally distinct amplitude to that assumed within the efficient-fallback-accretion framework, and have the opposite scaling with black hole mass. While models invoking this framework can produce light curves consistent with the data, this is only because of unconstrained free parameters such as the radiative efficiency $\eta$ and photosphere radius $R_{\rm ph, 0}$, which allow the amplitude and shape of the resulting model SED to vary freely (i.e., at fixed fallback rate). 

Similarly, models invoking luminosity from shocks at apocentre cannot generate enough luminosity or energy to explain observations for higher mass black holes, and shocks at pericentre predict an abundance of high-luminosity low-mass TDE systems which are never observed. A model whereby the shock radius moves systematically to compensate for these two flaws is, in a strictly data-reproduction sense, impossible to rule out, but requires a high degree of fine tuning.

Shocking debris streams are, of course, an essential component of getting luminosity out of a TDE system (without them no compact disk can form). It appears however that these shocks do not produce the bulk of the observed emission themselves, but simply act to initiate other luminosity-generating processes. 

Frameworks that broadly pass the calorimetric tests we perform in this paper can be summarized into two (likely interacting) classes, those associated with the formation of a compact disk (a cooling envelope), and those associated with the reprocessing of the Eddington-limited emission from these compact disks by a photosphere. The natural luminosity scale of both frameworks (the Eddington) provides a perfectly plausible resolution to the growing luminosity of TDEs with black hole mass, while the energy associated with disk formation/a period of Eddington limited accretion lasting $\sim 100$ days, is comparable to the observed energy in optical TDE flares.  

A picture which is supported by the data therefore appears to be emerging. The optical/UV emission observed during the initial flare in a TDE is powered by a process involving compact disk formation and early (possibly super-Eddington) disk evolution. 

The precise dominant energy source/physical process in such a scenario depends on the interaction between three timescales in the problem, the fallback timescale $t_{\rm fb} \propto M_\bullet^{1/2}$ (which feeds the system and initiates disk formation), the disk formation/envelope cooling timescale $t_{\rm cool} \propto E_{\rm circ}/L_{\rm Edd} \propto M_\bullet^{-1/3}$ and the viscous $t_{\rm visc} \propto M_\bullet^0$ timescale of the formed disk. 

Current models invoking disk reprocessing/disk formation are in some sense unsatisfactory for two reasons. The first is that no published model allows for the interaction between the two scenarios (i.e., does not allow for the accretion disk luminosity to feed back onto the disk formation process), in a sense they assume that only one of the above three timescales is dominant, and then seek to extend these scalings into regimes where other timescales can be expected to dominate. The need for interplay between accretion and disk formation is a hard observational requirement owing to sources which display bright (Eddington-limited) X-ray emission during the TDEs optical flare. 

Secondly, and potentially more pertinently, both frameworks invoke the Eddington limit, which seems apriori to be a hard constraint to satisfy in TDE systems at the highest black hole masses. This work leaves this as an open problem. Plausible resolutions may be the systematically higher stellar masses/black hole spins required at these high masses to overcome the Hills mass limit (and permit a disruption at all, \citealt{Hills75}), or some feedback/interplay between the high (but likely sub-Eddington) luminosity of the formed disk and the contracting disk formation process itself. 

We conclude that while no one framework appears to currently have all the answers, the parameter space of viable theories is steadily being cut down by wide-field optical surveys of TDEs. A promising future avenue for theoretical study appears to be optical/UV flares sourced during the disk formation and early evolution process.

\section*{Acknowledgments}
This research benefited from discussions at the Kavli Institute for Theoretical Physics (KITP) program on Tidal Disruption Events in April 2024 that was supported in part by the National Science Foundation under PHY1748958. MG acknowledges support from NASA through XMM-Newton grant 80NSSC24K1885.

\bibliography{andy}{}

\begin{thebibliography}{}
\expandafter\ifx\csname natexlab\endcsname\relax\def\natexlab#1{#1}\fi
\providecommand{\url}[1]{\href{#1}{#1}}
\providecommand{\dodoi}[1]{doi:~\href{http://doi.org/#1}{\nolinkurl{#1}}}
\providecommand{\doeprint}[1]{\href{http://ascl.net/#1}{\nolinkurl{http://ascl.net/#1}}}
\providecommand{\doarXiv}[1]{\href{https://arxiv.org/abs/#1}{\nolinkurl{https://arxiv.org/abs/#1}}}

\bibitem[{{Alexander} {et~al.}(2025){Alexander}, {Margutti}, {Gomez}, {Stroh}, {Chornock}, {Laskar}, {Cendes}, {Berger}, {Eftekhari}, {Franz}, {Hajela}, {Metzger}, {Terreran}, {Bietenholz}, {Christy}, {de Colle}, {Komossa}, {Nicholl}, {Ramirez-Ruiz}, {Saxton}, {Schroeder}, {Williams}, \& {Wu}}]{Alexander2025}
{Alexander}, K.~D., {Margutti}, R., {Gomez}, S., {et~al.} 2025, arXiv e-prints, arXiv:2506.12729, \dodoi{10.48550/arXiv.2506.12729}

\bibitem[{{Balbus} \& {Hawley}(1991)}]{BalbusHawley91}
{Balbus}, S.~A., \& {Hawley}, J.~F. 1991, \apj, 376, 214, \dodoi{10.1086/170270}

\bibitem[{{Bandopadhyay} {et~al.}(2024){Bandopadhyay}, {Fancher}, {Athian}, {Indelicato}, {Kapalanga}, {Kumah}, {Paradiso}, {Todd}, {Coughlin}, \& {Nixon}}]{Bandopadhyay+24}
{Bandopadhyay}, A., {Fancher}, J., {Athian}, A., {et~al.} 2024, \apjl, 961, L2, \dodoi{10.3847/2041-8213/ad0388}

\bibitem[{{Begelman}(1979)}]{Begelman79}
{Begelman}, M.~C. 1979, \mnras, 187, 237, \dodoi{10.1093/mnras/187.2.237}

\bibitem[{{Bricman} \& {Gomboc}(2020)}]{Bricman20}
{Bricman}, K., \& {Gomboc}, A. 2020, \apj, 890, 73, \dodoi{10.3847/1538-4357/ab6989}

\bibitem[{{Dai} {et~al.}(2015){Dai}, {McKinney}, \& {Miller}}]{Dai+15}
{Dai}, L., {McKinney}, J.~C., \& {Miller}, M.~C. 2015, ArXiv e-prints.
\newblock \doarXiv{1507.04333}

\bibitem[{{Dai} {et~al.}(2018){Dai}, {McKinney}, {Roth}, {Ramirez-Ruiz}, \& {Miller}}]{Dai18}
{Dai}, L., {McKinney}, J.~C., {Roth}, N., {Ramirez-Ruiz}, E., \& {Miller}, M.~C. 2018, \apjl, 859, L20, \dodoi{10.3847/2041-8213/aab429}

\bibitem[{{Foreman-Mackey} {et~al.}(2013){Foreman-Mackey}, {Hogg}, {Lang}, \& {Goodman}}]{EMCEE}
{Foreman-Mackey}, D., {Hogg}, D.~W., {Lang}, D., \& {Goodman}, J. 2013, \pasp, 125, 306, \dodoi{10.1086/670067}

\bibitem[{{Gezari} {et~al.}(2009){Gezari}, {Heckman}, {Cenko}, {Eracleous}, {Forster}, {Gon{\c{c}}alves}, {Martin}, {Morrissey}, {Neff}, {Seibert}, {Schiminovich}, \& {Wyder}}]{Gezari+09}
{Gezari}, S., {Heckman}, T., {Cenko}, S.~B., {et~al.} 2009, \apj, 698, 1367, \dodoi{10.1088/0004-637X/698/2/1367}

\bibitem[{{Greene} {et~al.}(2020){Greene}, {Strader}, \& {Ho}}]{Greene20}
{Greene}, J.~E., {Strader}, J., \& {Ho}, L.~C. 2020, \araa, 58, 257, \dodoi{10.1146/annurev-astro-032620-021835}

\bibitem[{{Grotova} {et~al.}(2025){Grotova}, {Rau}, {Baldini}, {Goodwin}, {Liu}, {Merloni}, {Salvato}, {Anderson}, {Arcodia}, {Buchner}, {Krumpe}, {Malyali}, {Masterson}, {Miller-Jones}, {Nandra}, \& {Shirley}}]{Grotova25}
{Grotova}, I., {Rau}, A., {Baldini}, P., {et~al.} 2025, arXiv e-prints, arXiv:2504.08424, \dodoi{10.48550/arXiv.2504.08424}

\bibitem[{{Guolo} {et~al.}(2024){Guolo}, {Gezari}, {Yao}, {van Velzen}, {Hammerstein}, {Cenko}, \& {Tokayer}}]{Guolo24}
{Guolo}, M., {Gezari}, S., {Yao}, Y., {et~al.} 2024, \apj, 966, 160, \dodoi{10.3847/1538-4357/ad2f9f}

\bibitem[{{Guolo} {et~al.}(2025{\natexlab{a}}){Guolo}, {Mummery}, {Ingram}, {Nicholl}, {Gezari}, \& {Nathan}}]{Guolo25b}
{Guolo}, M., {Mummery}, A., {Ingram}, A., {et~al.} 2025{\natexlab{a}}, arXiv e-prints, arXiv:2504.20148, \dodoi{10.48550/arXiv.2504.20148}

\bibitem[{{Guolo} {et~al.}(2025{\natexlab{b}})}]{Guolo25c}
{Guolo}, M., {et~al.} 2025{\natexlab{b}}, arXiv e-prints.
\newblock \doarXiv{arXiv:2510.26774}

\bibitem[{{Hammerstein} {et~al.}(2023){Hammerstein}, {van Velzen}, {Gezari}, {Cenko}, {Yao}, {Ward}, {Frederick}, {Villanueva}, {Somalwar}, {Graham}, {Kulkarni}, {Stern}, {Andreoni}, {Bellm}, {Dekany}, {Dhawan}, {Drake}, {Fremling}, {Gatkine}, {Groom}, {Ho}, {Kasliwal}, {Karambelkar}, {Kool}, {Masci}, {Medford}, {Perley}, {Purdum}, {van Roestel}, {Sharma}, {Sollerman}, {Taggart}, \& {Yan}}]{Hammerstein23}
{Hammerstein}, E., {van Velzen}, S., {Gezari}, S., {et~al.} 2023, \apj, 942, 9, \dodoi{10.3847/1538-4357/aca283}

\bibitem[{{Hills}(1975)}]{Hills75}
{Hills}, J.~G. 1975, \nat, 254, 295, \dodoi{10.1038/254295a0}

\bibitem[{{Kesden}(2012)}]{Kesden12}
{Kesden}, M. 2012, \prd, 85, 024037, \dodoi{10.1103/PhysRevD.85.024037}

\bibitem[{Kormendy \& Ho(2013)}]{Kormendy13}
Kormendy, J., \& Ho, L.~C. 2013, Annual Review of Astronomy and Astrophysics, 51, 511, \dodoi{https://doi.org/10.1146/annurev-astro-082708-101811}

\bibitem[{{Krolik} {et~al.}(2024){Krolik}, {Piran}, \& {Ryu}}]{Krolik24}
{Krolik}, J., {Piran}, T., \& {Ryu}, T. 2024, arXiv e-prints, arXiv:2409.02894, \dodoi{10.48550/arXiv.2409.02894}

\bibitem[{{Lodato} {et~al.}(2009){Lodato}, {King}, \& {Pringle}}]{Lodato+09}
{Lodato}, G., {King}, A.~R., \& {Pringle}, J.~E. 2009, \mnras, 392, 332, \dodoi{10.1111/j.1365-2966.2008.14049.x}

\bibitem[{{Lu} \& {Kumar}(2018)}]{Lu&Kumar18}
{Lu}, W., \& {Kumar}, P. 2018, \apj, 865, 128, \dodoi{10.3847/1538-4357/aad54a}

\bibitem[{{Metzger}(2022)}]{Metzger22}
{Metzger}, B.~D. 2022, \apjl, 937, L12, \dodoi{10.3847/2041-8213/ac90ba}

\bibitem[{{Metzger} \& {Stone}(2016)}]{MetzgerStone16}
{Metzger}, B.~D., \& {Stone}, N.~C. 2016, \mnras, 461, 948, \dodoi{10.1093/mnras/stw1394}

\bibitem[{{Mockler} {et~al.}(2019){Mockler}, {Guillochon}, \& {Ramirez-Ruiz}}]{Mockler19}
{Mockler}, B., {Guillochon}, J., \& {Ramirez-Ruiz}, E. 2019, \apj, 872, 151, \dodoi{10.3847/1538-4357/ab010f}

\bibitem[{{Mummery}(2021)}]{Mum21}
{Mummery}, A. 2021, \mnras, 504, 5144, \dodoi{10.1093/mnras/stab1187}

\bibitem[{{Mummery}(2023)}]{Mummery23a}
---. 2023, \mnras, 518, 1905, \dodoi{10.1093/mnras/stac2846}

\bibitem[{{Mummery}(2024)}]{Mummery24}
---. 2024, \mnras, 527, 6233, \dodoi{10.1093/mnras/stad3636}

\bibitem[{{Mummery} \& {Balbus}(2020)}]{MumBalb20a}
{Mummery}, A., \& {Balbus}, S.~A. 2020, \mnras, 492, 5655, \dodoi{10.1093/mnras/staa192}

\bibitem[{{Mummery} \& {van Velzen}(2025)}]{MummeryVV25}
{Mummery}, A., \& {van Velzen}, S. 2025, \mnras, 541, 429, \dodoi{10.1093/mnras/staf938}

\bibitem[{{Mummery} {et~al.}(2024){Mummery}, {van Velzen}, {Nathan}, {Ingram}, {Hammerstein}, {Fraser-Taliente}, \& {Balbus}}]{Mummery_et_al_2024}
{Mummery}, A., {van Velzen}, S., {Nathan}, E., {et~al.} 2024, \mnras, 527, 2452, \dodoi{10.1093/mnras/stad3001}

\bibitem[{{Mummery} {et~al.}(2023){Mummery}, {Wevers}, {Saxton}, \& {Pasham}}]{Mummery_Wevers_23}
{Mummery}, A., {Wevers}, T., {Saxton}, R., \& {Pasham}, D. 2023, \mnras, 519, 5828, \dodoi{10.1093/mnras/stac3798}

\bibitem[{{Nicholl} {et~al.}(2022){Nicholl}, {Lanning}, {Ramsden}, {Mockler}, {Lawrence}, {Short}, \& {Ridley}}]{Nicholl2022}
{Nicholl}, M., {Lanning}, D., {Ramsden}, P., {et~al.} 2022, \mnras, 515, 5604, \dodoi{10.1093/mnras/stac2206}

\bibitem[{{Novikov} \& {Thorne}(1973)}]{NovikovThorne73}
{Novikov}, I.~D., \& {Thorne}. 1973, in Black Holes (Les Astres Occlus), 343--450

\bibitem[{{Phinney}(1989)}]{Phinney89}
{Phinney}, E.~S. 1989, in The Center of the Galaxy, ed. M.~{Morris}, Vol. 136, 543

\bibitem[{{Piran} {et~al.}(2015){Piran}, {Svirski}, {Krolik}, {Cheng}, \& {Shiokawa}}]{Piran+15}
{Piran}, T., {Svirski}, G., {Krolik}, J., {Cheng}, R.~M., \& {Shiokawa}, H. 2015, \apj, 806, 164, \dodoi{10.1088/0004-637X/806/2/164}

\bibitem[{{Piro} \& {Lu}(2020)}]{Piro&Lu20}
{Piro}, A.~L., \& {Lu}, W. 2020, \apj, 894, 2, \dodoi{10.3847/1538-4357/ab83f6}

\bibitem[{{Ramsden} {et~al.}(2022){Ramsden}, {Lanning}, {Nicholl}, \& {McGee}}]{Ramsden22}
{Ramsden}, P., {Lanning}, D., {Nicholl}, M., \& {McGee}, S.~L. 2022, \mnras, 515, 1146, \dodoi{10.1093/mnras/stac1810}

\bibitem[{{Ramsden} {et~al.}(2025){Ramsden}, {Nicholl}, {McGee}, \& {Mummery}}]{Ramsden25}
{Ramsden}, P., {Nicholl}, M., {McGee}, S.~L., \& {Mummery}, A. 2025, \mnras, 541, 1218, \dodoi{10.1093/mnras/staf1059}

\bibitem[{{Rees}(1988)}]{Rees88}
{Rees}, M.~J. 1988, \nat, 333, 523, \dodoi{10.1038/333523a0}

\bibitem[{{Ryu} {et~al.}(2020{\natexlab{a}}){Ryu}, {Krolik}, {Piran}, \& {Noble}}]{Ryu+20a}
{Ryu}, T., {Krolik}, J., {Piran}, T., \& {Noble}, S.~C. 2020{\natexlab{a}}, \apj, 904, 98, \dodoi{10.3847/1538-4357/abb3cf}

\bibitem[{{Ryu} {et~al.}(2020{\natexlab{b}}){Ryu}, {Krolik}, {Piran}, \& {Noble}}]{Ryu+20b}
---. 2020{\natexlab{b}}, \apj, 904, 100, \dodoi{10.3847/1538-4357/abb3ce}

\bibitem[{{Sarin} \& {Metzger}(2024)}]{Sarin&Metzger24}
{Sarin}, N., \& {Metzger}, B.~D. 2024, \apjl, 961, L19, \dodoi{10.3847/2041-8213/ad16d8}

\bibitem[{{Sarin} {et~al.}(2024){Sarin}, {H{\"u}bner}, {Omand}, {Setzer}, {Schulze}, {Adhikari}, {Sagu{\'e}s-Carracedo}, {Galaudage}, {Wallace}, {Lamb}, \& {Lin}}]{Sarin+24}
{Sarin}, N., {H{\"u}bner}, M., {Omand}, C. M.~B., {et~al.} 2024, \mnras, 531, 1203, \dodoi{10.1093/mnras/stae1238}

\bibitem[{{Sazonov} {et~al.}(2021){Sazonov}, {Gilfanov}, {Medvedev}, {Yao}, {Khorunzhev}, {Semena}, {Sunyaev}, {Burenin}, {Lyapin}, {Meshcheryakov}, {Uskov}, {Zaznobin}, {Postnov}, {Dodin}, {Belinski}, {Cherepashchuk}, {Eselevich}, {Dodonov}, {Grokhovskaya}, {Kotov}, {Bikmaev}, {Zhuchkov}, {Gumerov}, {van Velzen}, \& {Kulkarni}}]{Sazonov21}
{Sazonov}, S., {Gilfanov}, M., {Medvedev}, P., {et~al.} 2021, \mnras, 508, 3820, \dodoi{10.1093/mnras/stab2843}

\bibitem[{{Shakura} \& {Sunyaev}(1973)}]{SS73}
{Shakura}, N.~I., \& {Sunyaev}, R.~A. 1973, \aap, 24, 337

\bibitem[{{Strubbe} \& {Quataert}(2009)}]{Strubbe&Quataert09}
{Strubbe}, L.~E., \& {Quataert}, E. 2009, \mnras, 400, 2070, \dodoi{10.1111/j.1365-2966.2009.15599.x}

\bibitem[{{van Velzen} {et~al.}(2019){van Velzen}, {Stone}, {Metzger}, {Gezari}, {Brown}, \& {Fruchter}}]{vanVelzen19}
{van Velzen}, S., {Stone}, N.~C., {Metzger}, B.~D., {et~al.} 2019, \apj, 878, 82, \dodoi{10.3847/1538-4357/ab1844}

\bibitem[{{van Velzen} {et~al.}(2021){van Velzen}, {Gezari}, {Hammerstein}, {Roth}, {Frederick}, {Ward}, {Hung}, {Cenko}, {Stein}, {Perley}, {Taggart}, {Foley}, {Sollerman}, {Blagorodnova}, {Andreoni}, {Bellm}, {Brinnel}, {De}, {Dekany}, {Feeney}, {Fremling}, {Giomi}, {Golkhou}, {Graham}, {Ho}, {Kasliwal}, {Kilpatrick}, {Kulkarni}, {Kupfer}, {Laher}, {Mahabal}, {Masci}, {Miller}, {Nordin}, {Riddle}, {Rusholme}, {van Santen}, {Sharma}, {Shupe}, \& {Soumagnac}}]{vanVelzen21}
{van Velzen}, S., {Gezari}, S., {Hammerstein}, E., {et~al.} 2021, \apj, 908, 4, \dodoi{10.3847/1538-4357/abc258}

\bibitem[{{Yao} {et~al.}(2023){Yao}, {Ravi}, {Gezari}, {van Velzen}, {Lu}, {Schulze}, {Somalwar}, {Kulkarni}, {Hammerstein}, {Nicholl}, {Graham}, {Perley}, {Cenko}, {Stein}, {Ricarte}, {Chadayammuri}, {Quataert}, {Bellm}, {Bloom}, {Dekany}, {Drake}, {Groom}, {Mahabal}, {Prince}, {Riddle}, {Rusholme}, {Sharma}, {Sollerman}, \& {Yan}}]{Yao23}
{Yao}, Y., {Ravi}, V., {Gezari}, S., {et~al.} 2023, \apjl, 955, L6, \dodoi{10.3847/2041-8213/acf216}

\bibitem[{{Yoshioka} {et~al.}(2024){Yoshioka}, {Mineshige}, {Ohsuga}, {Kawashima}, \& {Kitaki}}]{Yoshioka+24}
{Yoshioka}, S., {Mineshige}, S., {Ohsuga}, K., {Kawashima}, T., \& {Kitaki}, T. 2024, arXiv e-prints, arXiv:2407.15927, \dodoi{10.48550/arXiv.2407.15927}

\bibitem[{{Zhang} {et~al.}(2025){Zhang}, {Stone}, {White}, {Davis}, {Jiang}, \& {Mullen}}]{Lizhong25}
{Zhang}, L., {Stone}, J.~M., {White}, C.~J., {et~al.} 2025, arXiv e-prints, arXiv:2509.10638, \dodoi{10.48550/arXiv.2509.10638}

\end{thebibliography}
\bibliographystyle{aasjournal}

\appendix
\section{From fallback rate to accretion rate}\label{app:acc}
Material which falls back post-disruption to the pericentre radius in a TDE has angular momentum, and therefore must be delayed in its ultimate accretion onto the black hole. This delay (while the material redistributes its angular momentum) means that the accretion rate in a TDE can never equal the instantaneous fallback rate, and indeed these two quantities are often not even approximately equal. This is a persistent misconception in the TDE literature, and we discuss here how the accretion rate precisely relates to the fallback rate. 

In complete generality the accretion rate at a radius $r$, and a time $t$, in an accretion flow is equal to 
\begin{equation}    
    \dot M(r, t) = \int_{-\infty}^t \, \int_0^\infty g_{\dot M}(r, t|r', t')\, {\cal S}_{\dot M}(r', t')\, {\rm d}r'\, {\rm d}t'. 
\end{equation}
In this expression ${\cal S}_{\dot M}(r', t')$ is the matter source function (i.e., the fallback rate for a TDE), normalised such that 
\begin{equation}
    \int_{-\infty}^{+\infty} \, \int_0^\infty  {\cal S}_{\dot M}(r', t')\, {\rm d}r'\, {\rm d}t' = M_{\rm fb} .
\end{equation}
In other words, eventually, and somewhere, $M_{\rm fb}$ returns to the disk system. The function $g_{\dot M}$ is the ``Greens function'' of the accretion process, normalised such that 
\begin{equation}
    \int_{-\infty}^{+\infty}   {g}_{\dot M}(r_{\rm in}, t | r', t')\,  {\rm d}t = 1, 
\end{equation}
in other words, eventually $(t\to\infty)$, all of the mass which enters the disk (regardless of where or when it enters) falls through the inner edge of the disk $r_{\rm in}$ onto the black hole. At the moment this is a purely formal mathematical expression, and is not particularly illuminating. The point is, however, that while the fully general relativistic version of $g_{\dot M}(r, t|r', t')$ in unwieldily, the relativistic Greens function for the accretion rate at the inner disk edge (the relevant one in this case) is extremely simple, a result derived in \cite{Mummery23a}. Indeed, it equals 
\begin{equation}\label{eq:g_gen}
    g_{\dot M}(r_I, t|r', t') = {C(r')\over \tau^{n}} \exp\left(-{f(r') \over 4\tau} \right) \, \Theta(\tau), 
\end{equation}
where $C$ is a normalisation (that depends on the feeding radius) chosen so that the integral above equals $1$, $f(r')$ is a relatively simple function of position \citep{Mummery23a}, and $\tau \equiv (t-t')/t_{\rm acc}$, where $t_{\rm acc}$ is a timescale which can be written down explicitly. The function $\Theta(x)$ the Heaviside function ($\Theta(x)=1$ for $ x\geq0$ and $\Theta(x) = 0$ for $x<0$), enforcing causality in this case (material cannot be accreted before it has entered the disk). The index $n$ is apriori unknown, as it depends on the micro-physical processes governing the turbulent redistribution of angular momentum within the flow. In some sense the most ``natural'' value is $n = 4/3$, for which the viscous time at a given radius scales directly proportional to the orbital period. On dimensional grounds we see that $f$ is dimensionless, and $C$ (the normalisation) must have dimensions of time$^{-1}$. Differentiating equation \ref{eq:g_gen} with respect to time allows one to relate $f$ to the time of maximum accretion rate $t_{\rm max}$, an excellent proxy for the viscous time (for this Greens function). Then performing the integral over time we can relate $C$ to $t_{\rm max}$, leading to a more convenient formulation of the Greens function
\begin{equation}\label{eq:g_simp}
    g_{\dot M}(r_I, t|r', t') ={1\over t_{\rm max}}{n^{n-1} \over \Gamma(n-1)} {\tau_{\rm max}^n \over \tau^{n}} \exp\left(-n{\tau_{\rm max} \over \tau} \right) \, \Theta(\tau) ,
\end{equation}
 where $\Gamma$ is the usual Gamma function.

If we follow the standard assumption that material returns at times following the fallback time as $\sim t^{-5/3}$ and promptly circularizes at $r_0 \sim 2r_T$, then the matter source term can be approximated by 
\begin{equation}
    {\cal S}_{\dot M}(r', t') = {M_\star \over 3 t_{\rm fb}} \, \left({t' \over t_{\rm fb}}\right)^{-5/3} \, \delta(r'-2r_T)\, \Theta(t'-t_{\rm fb}),
\end{equation}
where $\delta(x)$ is the usual delta function.  Therefore the accretion rate onto the black hole is given by 
\begin{equation}
    \dot M(r_I, t) = {M_\star \over 3t_{\rm fb}} {n^{n-1} \over \Gamma(n-1)}\int_{t_{\rm fb}}^t \left({ t_{\rm max} \over t-t'}\right)^n \,  \left({t' \over t_{\rm fb}}\right)^{-5/3} \exp\left(-n{t_{\rm max} \over t-t'} \right) \, {{\rm d}t'\over t_{\rm max}}.
\end{equation}
This is a trivial integral to compute numerically.

\begin{figure}
    \centering
    \includegraphics[width=0.45\linewidth]{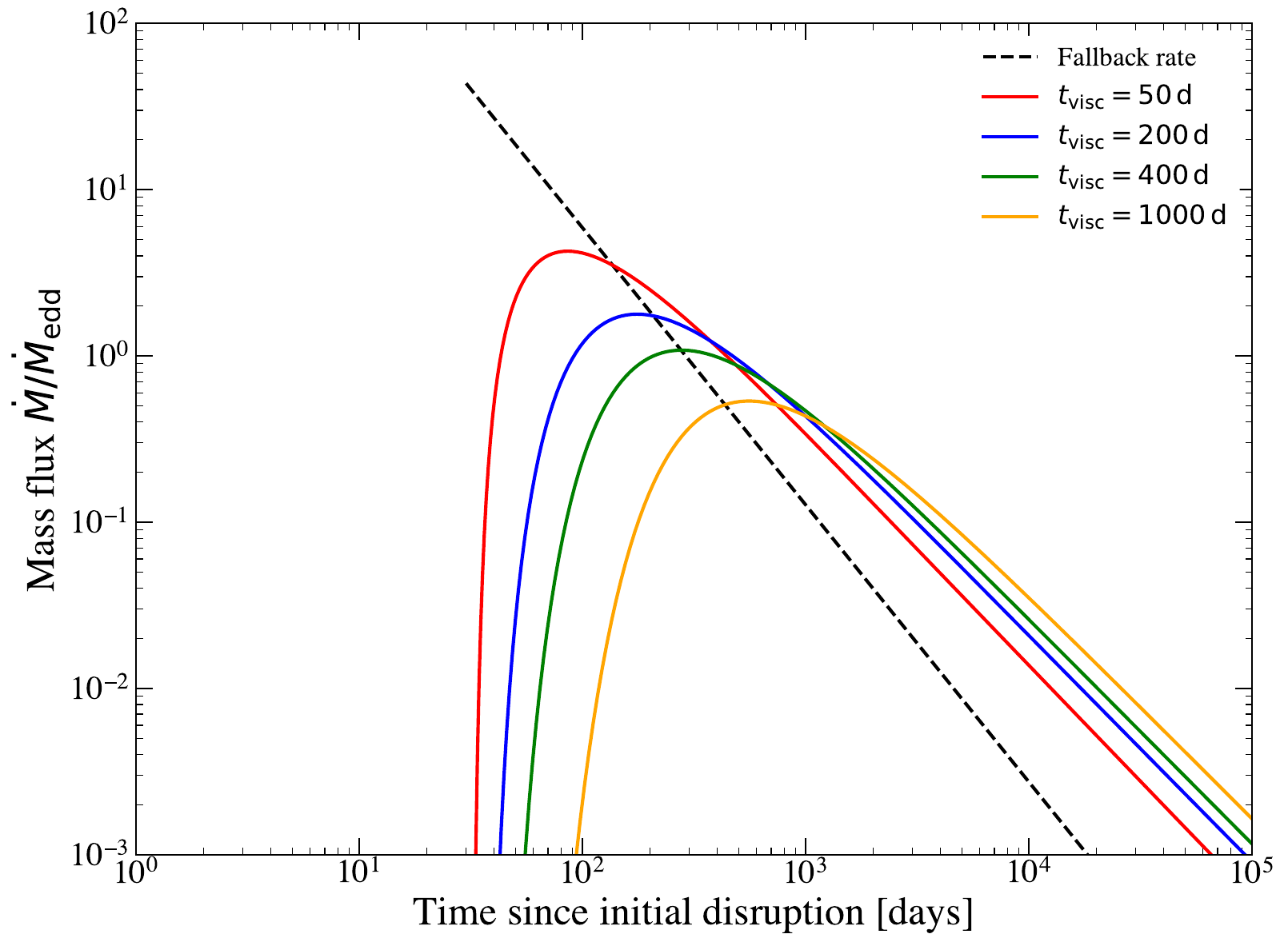}
    \includegraphics[width=0.45\linewidth]{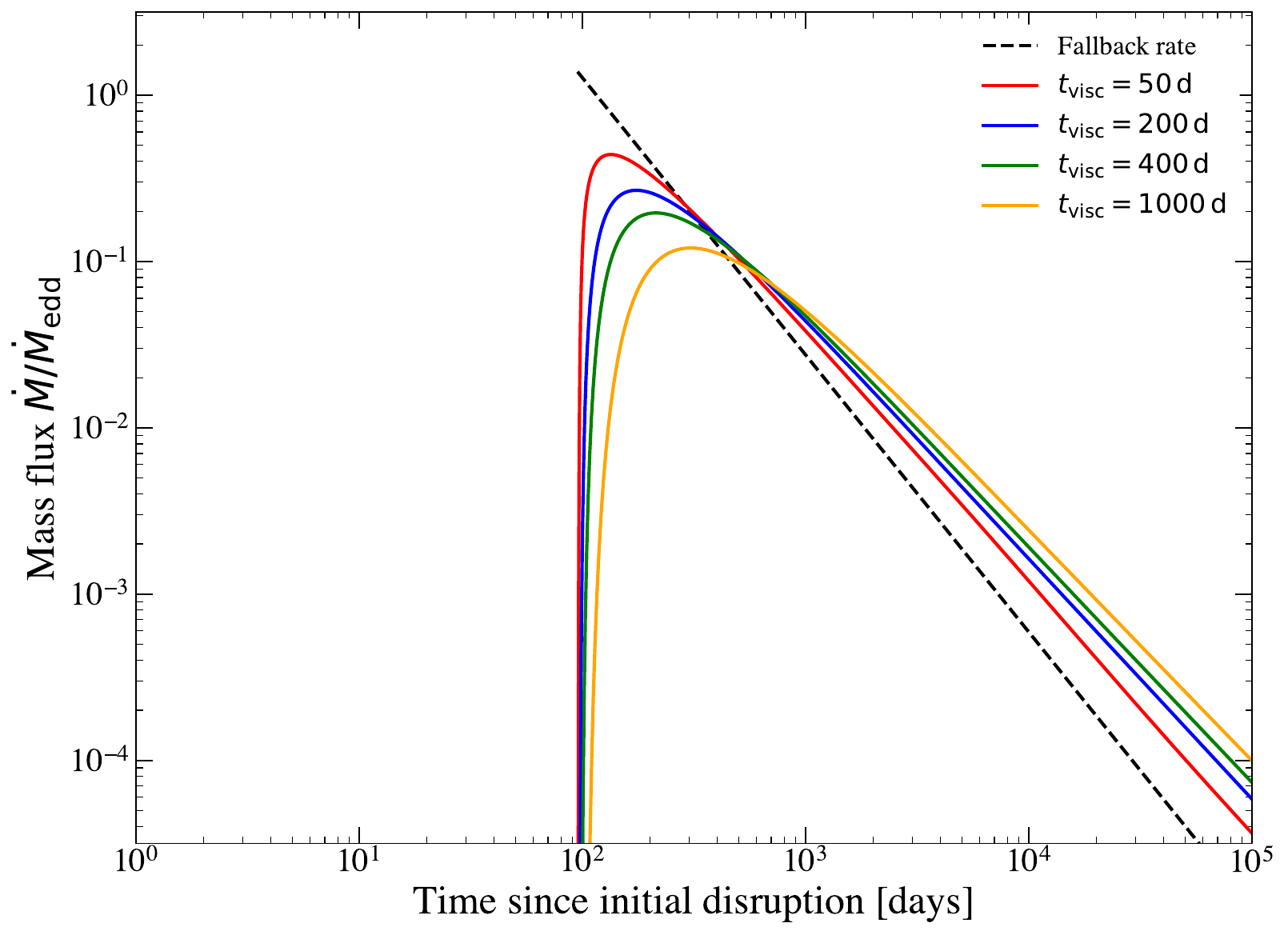}
    \caption{The fallback rate (black dashed curve) and the black hole accretion rate (colored solid lines) for four different viscous timescales (denoted on plot) for the full disruption of a $m_\star = 0.5$ star about a $10^6M_\odot$ black hole (left) and a $10^7M_\odot$ black hole (right). The accretion rate onto the black hole is generically suppressed with respect to the fallback rate, with a degree of suppression which depends on the ratio of $t_{\rm visc}/t_{\rm fb}$. Note that the late time evolution of the accretion rate (the power law fall off at late times) is set by the microphysics of accretion (the index $n$) not the index in the fallback rate.  }
    \label{fig:acc}
\end{figure}

In Figure \ref{fig:acc} we show the accretion rate onto the horizon (i.e., the solution of the above integral), for a fallback rate (i.e., outer disk feeding rate) shown by the black dashed curve. We assume an incoming stellar mass of $m_\star = 0.5$, and a $10^6M_\odot$ black hole (left) and a $10^7M_\odot$ black hole (right), which results in both different peak fallback rates and fallback timescales. We assume a natural disk microphysics parameter, i.e., $n = 4/3$. The viscous times are set to four values which span the observed range in TDE X-ray light curves \citep{Guolo25b}. 

We note immediately that for a $10^6M_\odot$ black hole mass, the accretion rate onto the black hole is strongly suppressed with respect to the peak fallback rate. This is simply because the viscous time is (even for the shortest viscous times seen in TDE X-ray light curves) long compared to the fallback time. In this limit the peak accretion rate $\dot M_{\rm pk} \sim M_{\rm disk}/t_{\rm visc}$ is set by the total mass budget and the disks viscous time, not the fallback rate. We see that even for peak fallback rates of $\dot m \sim 100$, the accretion rate may well be sub-Eddington at all times for sufficiently long viscous timescales. Note that for a $10^6M_\odot$ black hole the fallback rate is a terrible approximation to the accretion rate at all times, and for all plausible values of the viscous time.

\begin{figure}
    \centering
    \includegraphics[width=0.45\linewidth]{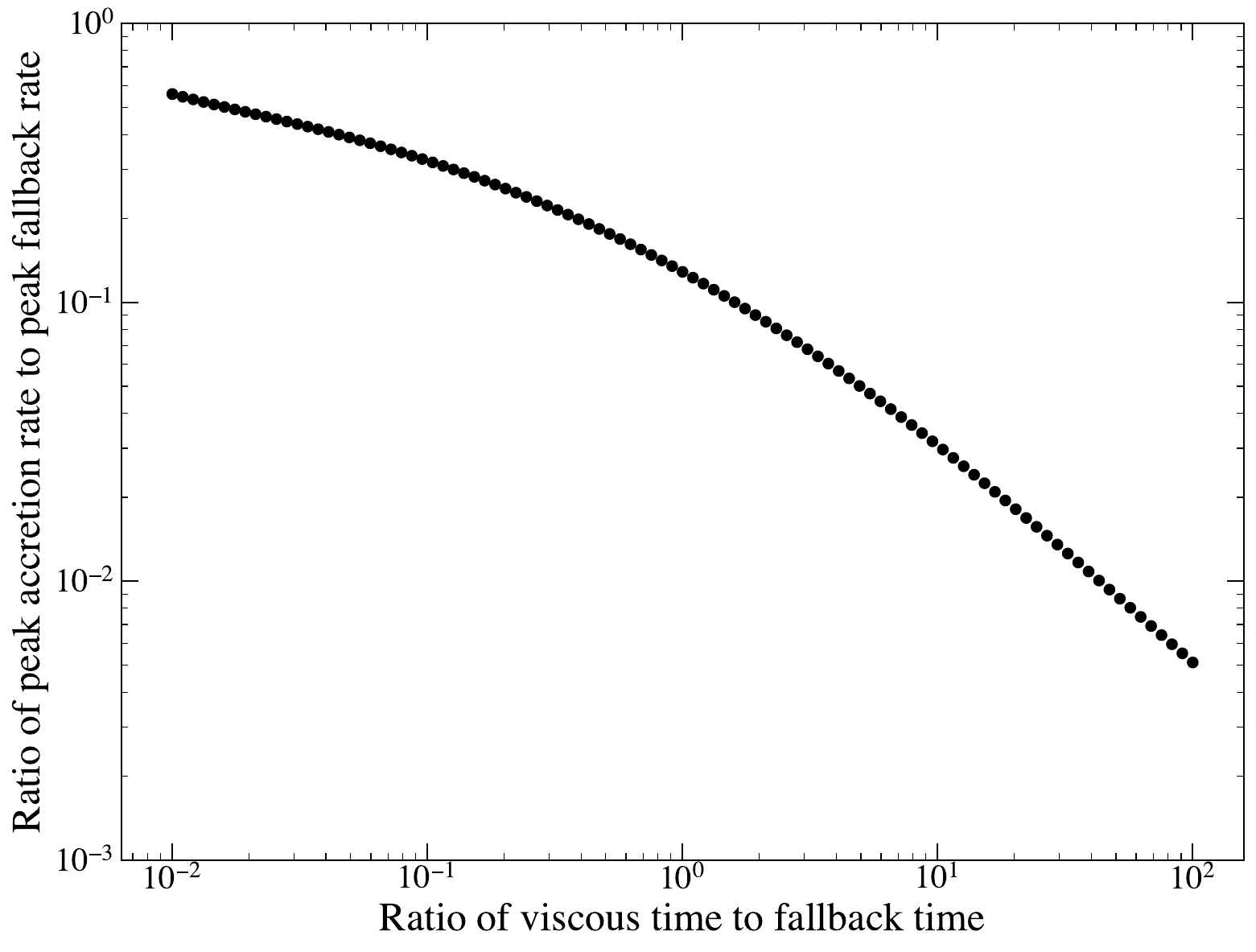}
    \includegraphics[width=0.45\linewidth]{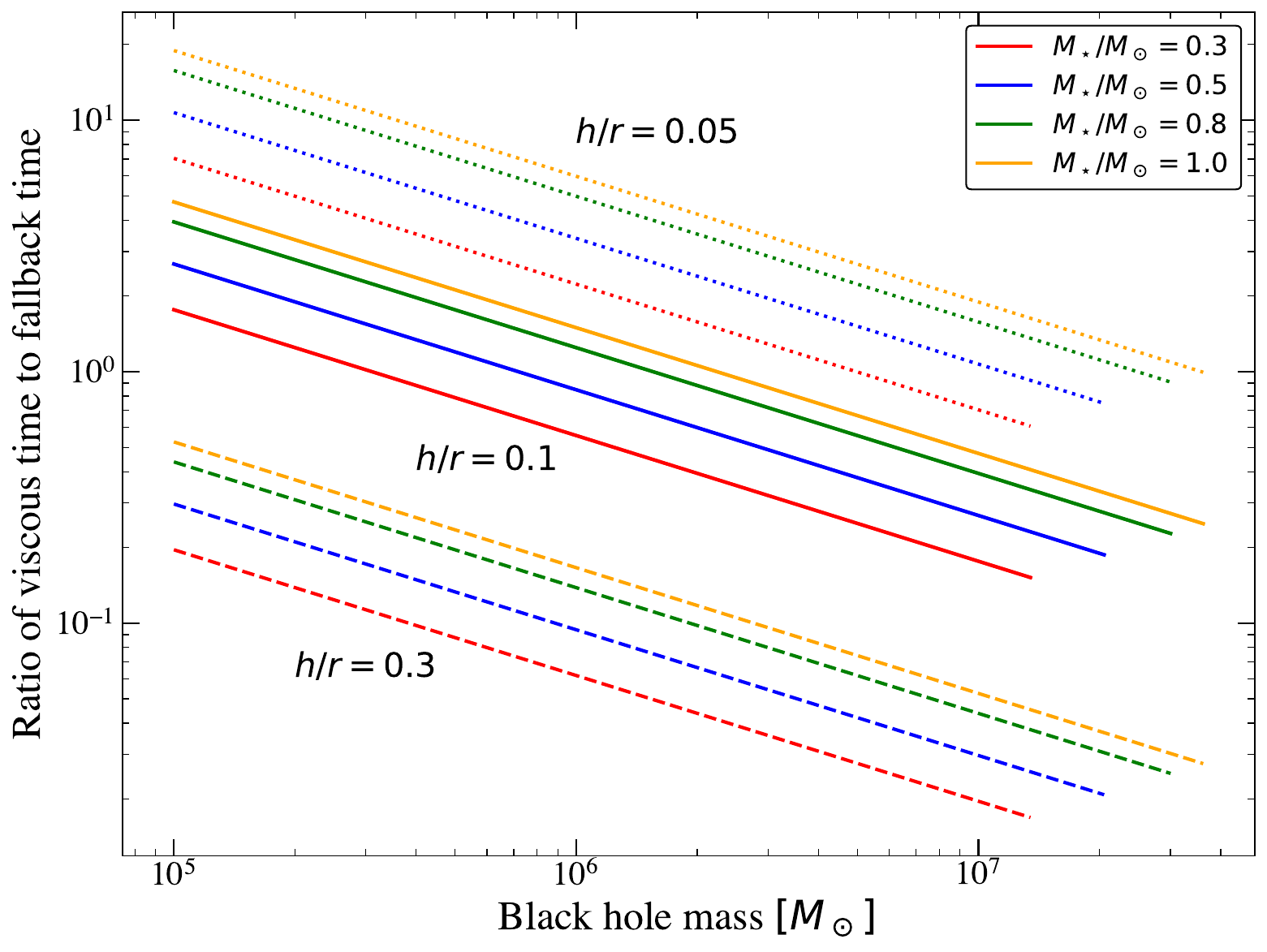}
    \includegraphics[width=0.65\linewidth]{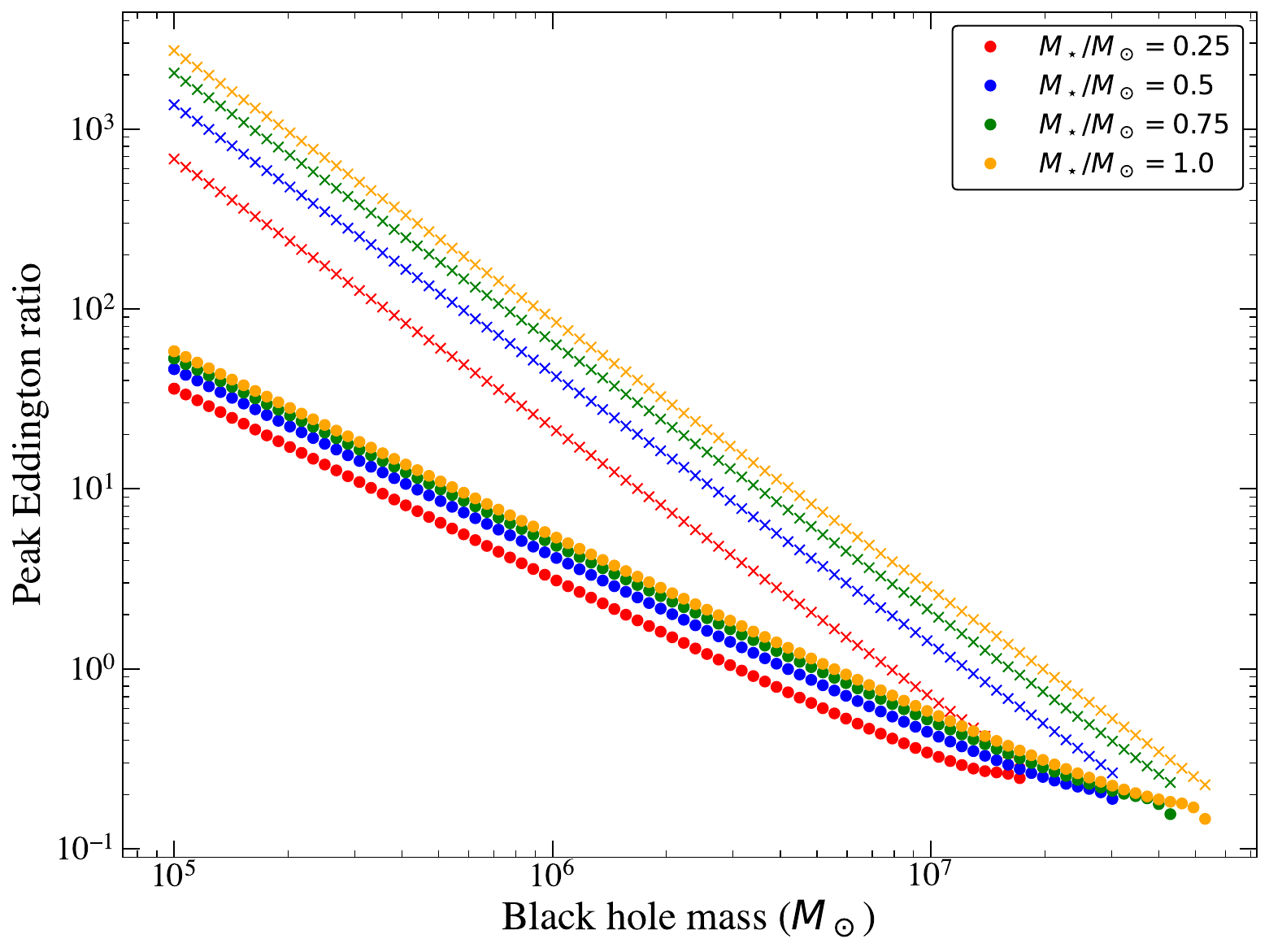}
    \caption{The effects of the systematic variance of the fallback and viscous timescales across a TDE population on the peak accretion rate. In the upper left panel we show the ratio of the peak accretion rate to the peak fallback rate, as a function of the ratio $t_{\rm visc}/t_{\rm fb}$. We see that typically the redistribution of angular momentum in an accretion flow leads to a suppression of $\sim 1-2$ orders of magnitude in the peak accretion rate, except for extremely short viscous times $t_{\rm visc}/t_{\rm fb}\ll 10^{-2}$. In the upper right plot we show the expected range for this ratio, taking the fallback timescale from equation \ref{eq:tfb}, and the viscous time for different stellar masses (denoted by the colors on the plot), and disk thicknesses (different line styles). We see that when the peak fallback rate is largest (low black hole masses), the ratio of $t_{\rm visc}/t_{\rm fb}$ is also largest, leading to the most suppression in the peak accretion rate. This means that across the entire black hole mass space (lower panel), the peak accretion Eddington ratio (dots)  is better described by a $\dot m \sim (M_{\rm disk}/t_{\rm visc}) \times 1/M_\bullet$ profile, than the canonical $\dot m_{\rm fb} \propto 1/M_\bullet^{3/2}$ fallback scaling (crosses).   }
    \label{fig:acc2}
\end{figure}

For a $10^7 M_\odot$ black hole however, the accretion rate more closely follows the fallback rate. This is simply a result of the fallback time growing, and becoming larger than the shorter end of the viscous timescales seen in X-ray observations. This result can be extended to TDEs across the entire black hole mass range, as we do in Figure \ref{fig:acc2}.  In the upper left panel we show the ratio of the peak accretion rate to the peak fallback rate, as a function of the ratio $t_{\rm visc}/t_{\rm fb}$. We see that typically the accretion rate is suppressed by $\sim 1-2$ orders of magnitude from the peak fallback rate, except for extremely short viscous times $t_{\rm visc}/t_{\rm fb}\ll 10^{-2}$. This will lead to systematic effects, as the fallback rate grows for higher mass black holes. In the upper right plot we show the expected range for the $t_{\rm visc}/t_{\rm fb}$ ratio, taking the fallback timescale from equation \ref{eq:tfb}, and the viscous time for different stellar masses (denoted by the colors on the plot), and disk thicknesses (different line styles). We see that when the peak fallback rate is largest (low black hole masses), the ratio of $t_{\rm visc}/t_{\rm fb}$ is also largest, leading to the most suppression in the peak accretion rate. This means that across the entire black hole mass space (lower panel), the peak accretion Eddington ratio (dots)  is better described by a $\dot m \sim (M_{\rm disk}/t_{\rm visc}) \times 1/M_\bullet$ profile, than the canonical $\dot m_{\rm fb} \propto 1/M_\bullet^{3/2}$ fallback scaling (crosses).  

A final point we wish to make is that the asymptotic decay in the accretion rate onto the black hole is equal to $\dot M(t)\sim t^{-n}$, i.e., it is set entirely by the disk turbulent microphysics, and is independent of how the fallback rate may scale with time. To understand this physically, realize that the late time decay in the accretion rate is set entirely by how quickly the flow can redistribute its angular momentum budget (i.e., the angular momentum which was fed in by the material which fell into the disk), and eventually the accretion rate stops caring about any mass being fed in at a given radius. This disk physics result should be contrasted with the ``viscous delay timescale'' in {\tt MOSFIT}, which simply forces (with an exponential convergence) the accretion rate to {\it follow} the fallback rate after a delay of $\Delta t \sim T_{\rm visc}$ ({\tt MOSFIT} notation). This exponential convergence to the fallback rate is not what the physics of an accretion disk leads to, and therefore the viscous delay in {\tt MOSFIT} does not account for the physics of ``viscous'' (really turbulent)  redistribution of angular momentum within an accretion flow. 

\section{Checking for redshift bias}\label{app:Z}
The work in the main body of the paper utilized every observed TDE, and applied no cuts on redshift. Of course, in a flux-limited survey it is obvious that there will be an induced correlation between observed luminosity and redshift, simply because the flux limit
\begin{equation}
    F_{\rm lim} = {L\over 4\pi D_L^2}, 
\end{equation}
results in intrinsically more luminous objects being found at higher redshift (larger luminosity distance $D_L$). 
\begin{figure*}
    \centering
    \includegraphics[width=0.45\linewidth]{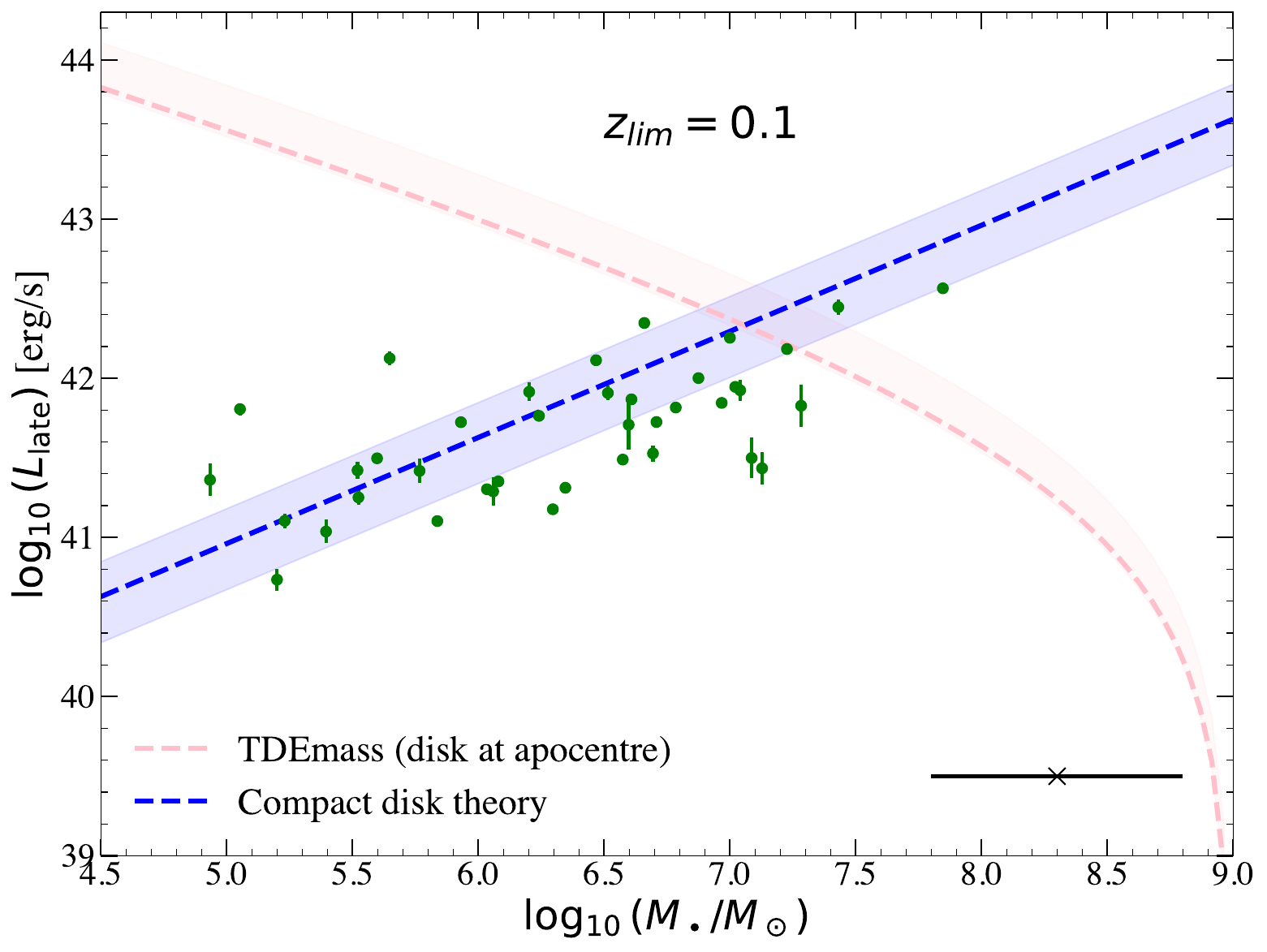}
    \includegraphics[width=0.45\linewidth]{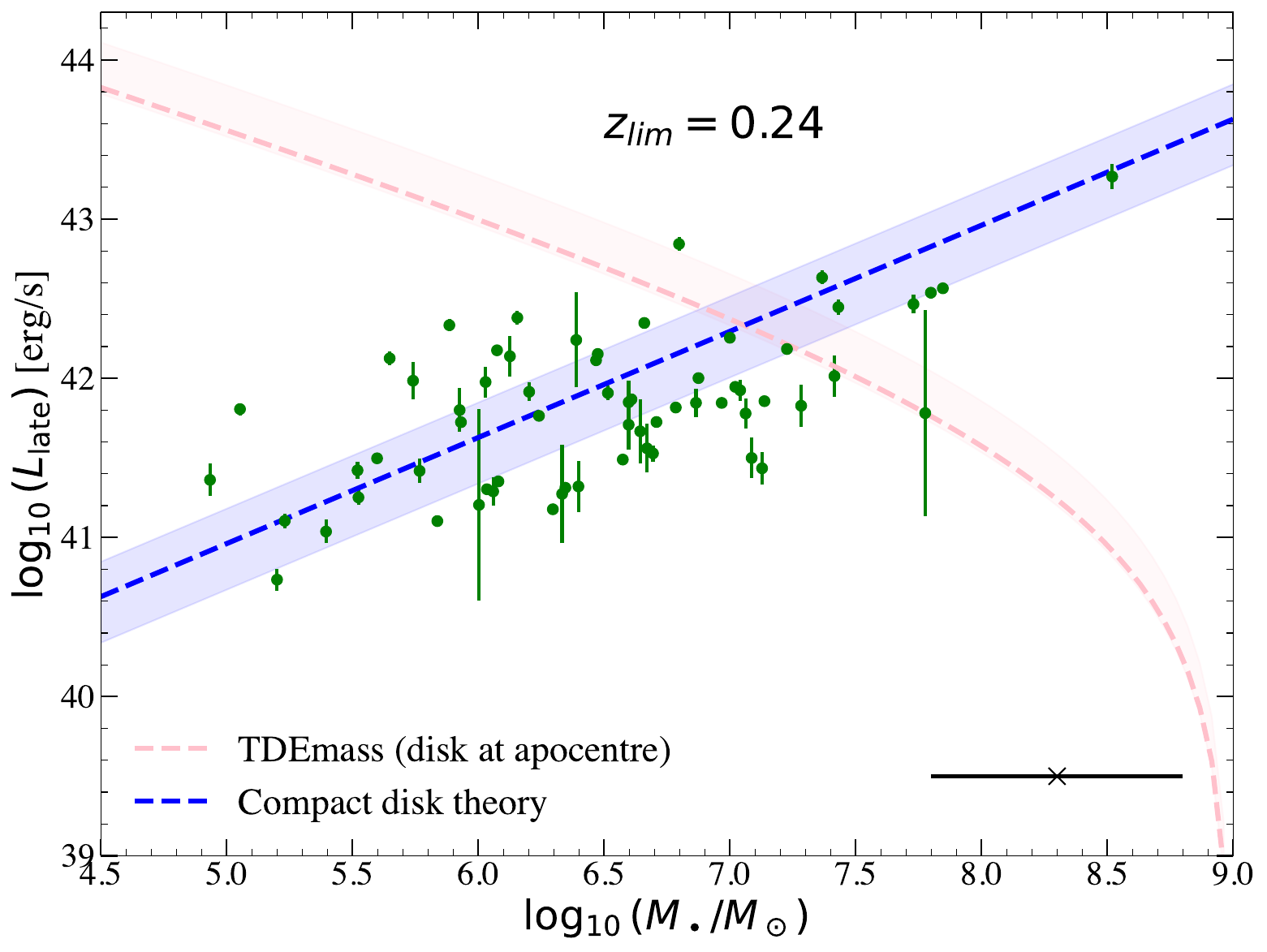}
    \includegraphics[width=0.45\linewidth]{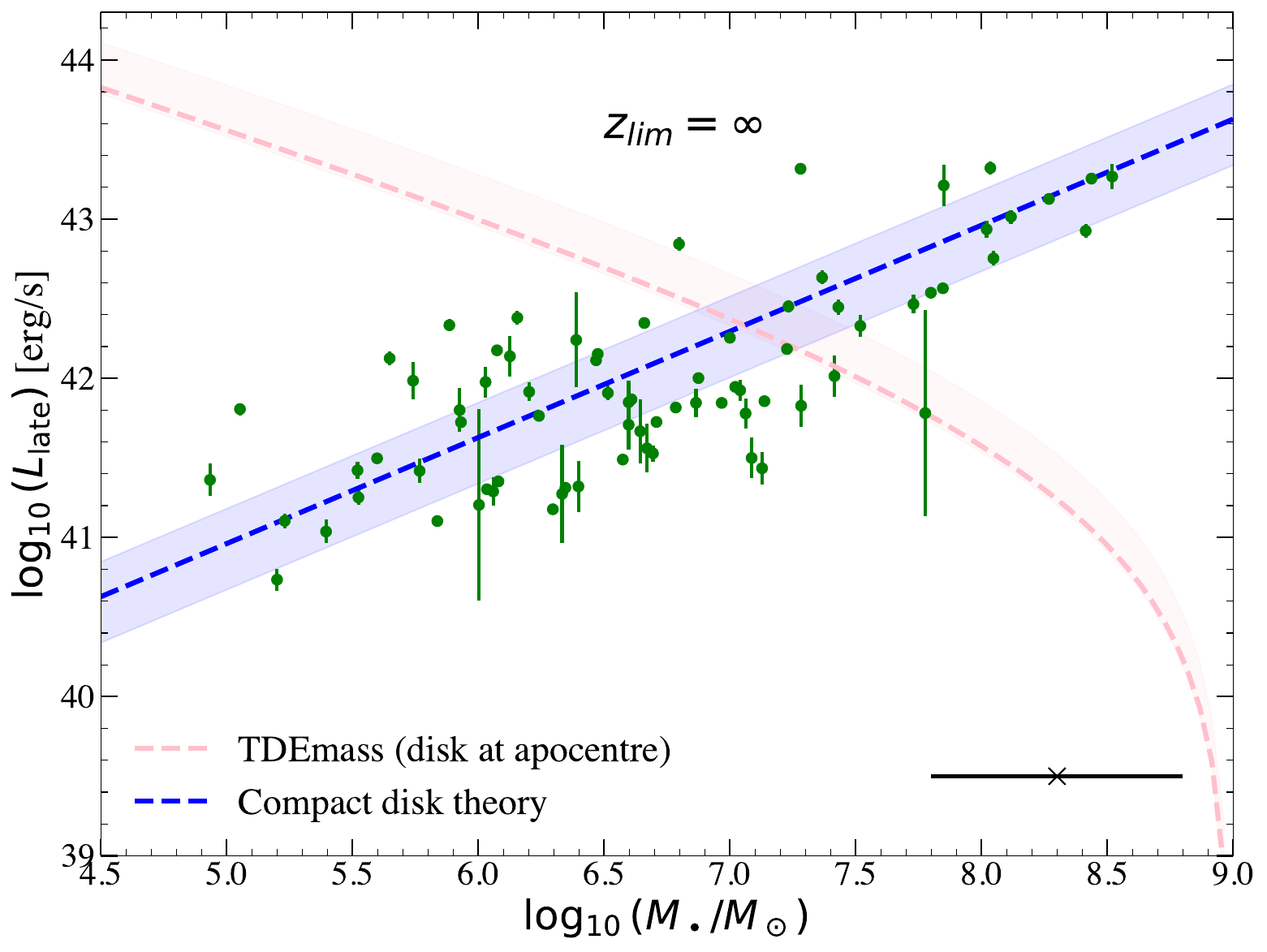}
    \includegraphics[width=0.4\linewidth]{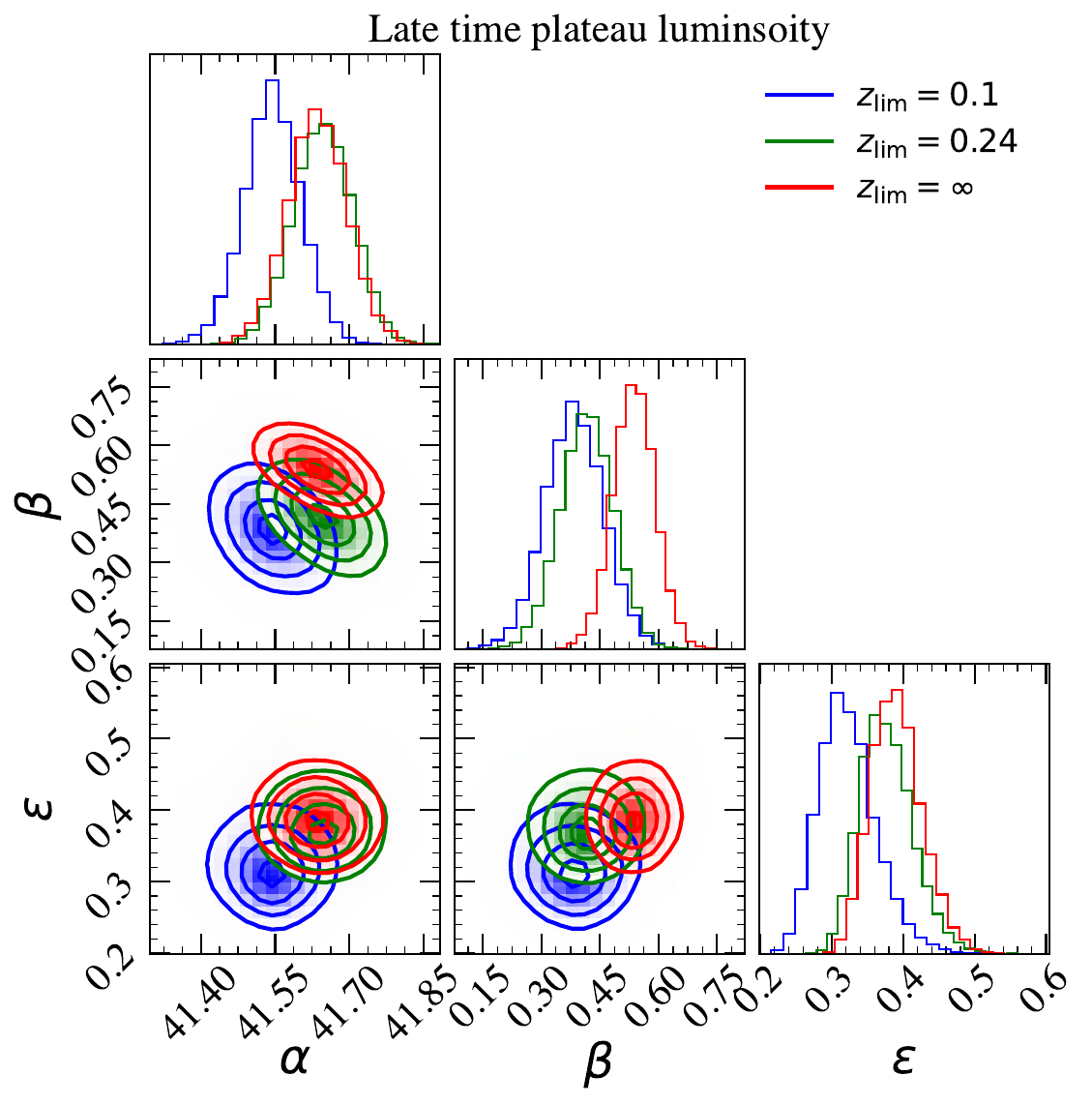}
    \caption{The late time plateau luminosity versus black hole mass (measured from $M_\bullet-\sigma$ where available, and $M_\bullet-M_{\rm bulge}$ otherwise), for different redshift cuts for the TDE population (as denoted on plot). The cut $z_{\rm lim} = 0.24$  represents the redshift out to which TDE searches with ZTF have no redshift bias for the host galaxies, see \citealt{Yao23}. The choice of redshift cut is completely irrelevant for the inferred scaling relationship, as is highlighted in the lower right panel where we show the results of power-law fits to the data for different redshift cuts. All data sets, independent of redshift cut, strongly support compact disk formation over apocentre disk formation. This result holds at extremely high statistical significance.   }
    \label{fig:late_Z}
\end{figure*}

It is worth being careful to check that this induced luminosity-redshift relationship does not introduce any bias into our results. Most important is the late-time plateau luminosity-black hole mass scaling, as this is what we use to constrain the black hole masses in our sample. In Figure \ref{fig:late_Z} we show that any choice of redshift cut makes no difference to the inferred plateau-black hole mass scaling relationship, or the reliability of using compact disk theory as a means to measuring black hole masses. 

In Figure \ref{fig:late_Z} we consider three redshift cuts, $z_{\rm lim} = \infty$ (i.e., using every TDE, the same as Figure \ref{fig:late}), $z_{\rm lim} = 0.24$ (the redshift out to which TDE searches with ZTF have no redshift bias for the host galaxies, see \citealt{Yao23}) and $z_{\rm lim} = 0.1$ (chosen simply because it is a round number). We see by inspection of each plateau luminosity-black hole mass plot (upper left, upper right and lower left panels) that the trend is unchanged, only the number of points is changed\footnote{This is an unsurprising result, as it has been shown that compact disk theory can reproduce the optical peak {\it luminosity function} if a direct coupling between peak luminosity and black hole mass is assumed (\citealt{MummeryVV25}). The luminosity function represents the luminosity scaling when the different observing volumes of different sources are {\it explicitly accounted for}. }. This is confirmed explicitly in the lower right panel were we fit a power-law profile to each redshift-limited set. The amplitude of the plateau luminosity around a $10^6 M_\odot$ black hole $(\alpha)$ and the power-law scaling index with black hole mass $(\beta)$ are consistent with each other at $1\sigma$ independent of redshift cut applied. The power-law index is robustly positive, in strong $(\gg 5\sigma)$ contention with the {\tt TDEmass} prediction of $\beta = -5/8$. 

\begin{figure}
    \centering
    \includegraphics[width=0.65\linewidth]{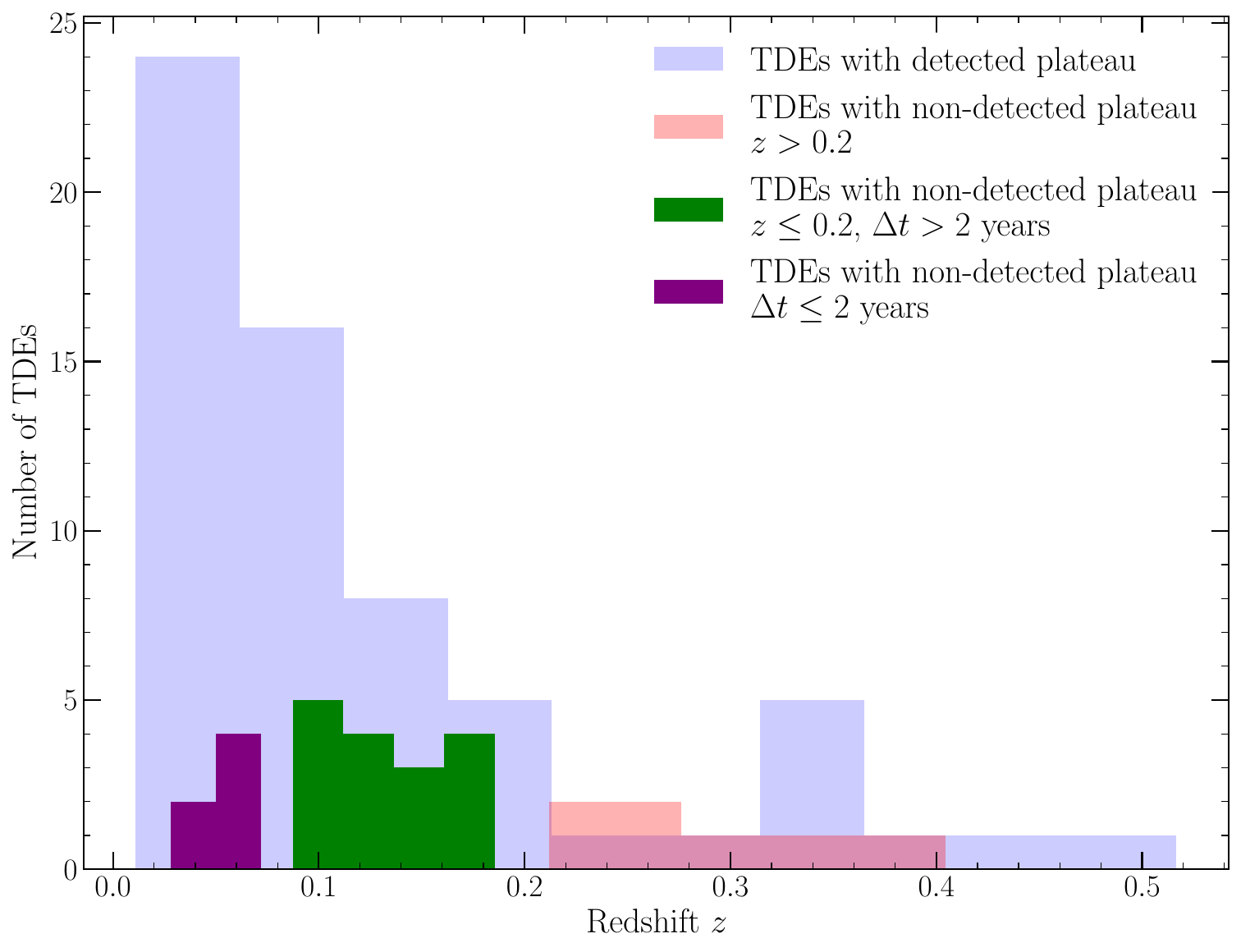}
    \caption{The ubiquity of the late-time UV plateau highlighted by the red-shift and TDE age distribution of those sources with and without robust plateau detections (data and classification from \citealt{Mummery_et_al_2024}). Every TDE with $>2$ years of data which is nearby $z < 0.08$ has a detected plateau, while high redshift $z > 0.2$ TDEs are generally much less likely to be detected (owing to their intrinsic faintness). Only a small fraction of TDEs with $> 2$ years of data and an intermediate redshift $0.1 < z < 0.2$ are non-detected, none of which have deep observations in the UV (e.g., with HST) which would be required to conclusively rule out a plateau.  }
    \label{fig:plat_find}
\end{figure}

This is lack of a redshift dependence is likely because the plateau population is dominated by low redshift sources (see Figure \ref{fig:plat_find}), which highlights the ubiquity of the late-time UV plateau detections. In Figure \ref{fig:plat_find} we show the red-shift and TDE age distribution of those sources with and without robust plateau detections (data and classification from \citealt{Mummery_et_al_2024}).  Every TDE with $>2$ years of data which is nearby $z < 0.08$ has a detected plateau, while high redshift $z > 0.2$ TDEs are generally much less likely to be detected (owing to their intrinsic faintness). Only a small fraction of TDEs with $> 2$ years of data and an intermediate redshift $0.1 < z < 0.2$ are non-detected, none of which have deep observations in the UV (e.g., with HST) which would be required to conclusively rule out a plateau.

\begin{figure*}
    \centering
    \includegraphics[width=0.45\linewidth]{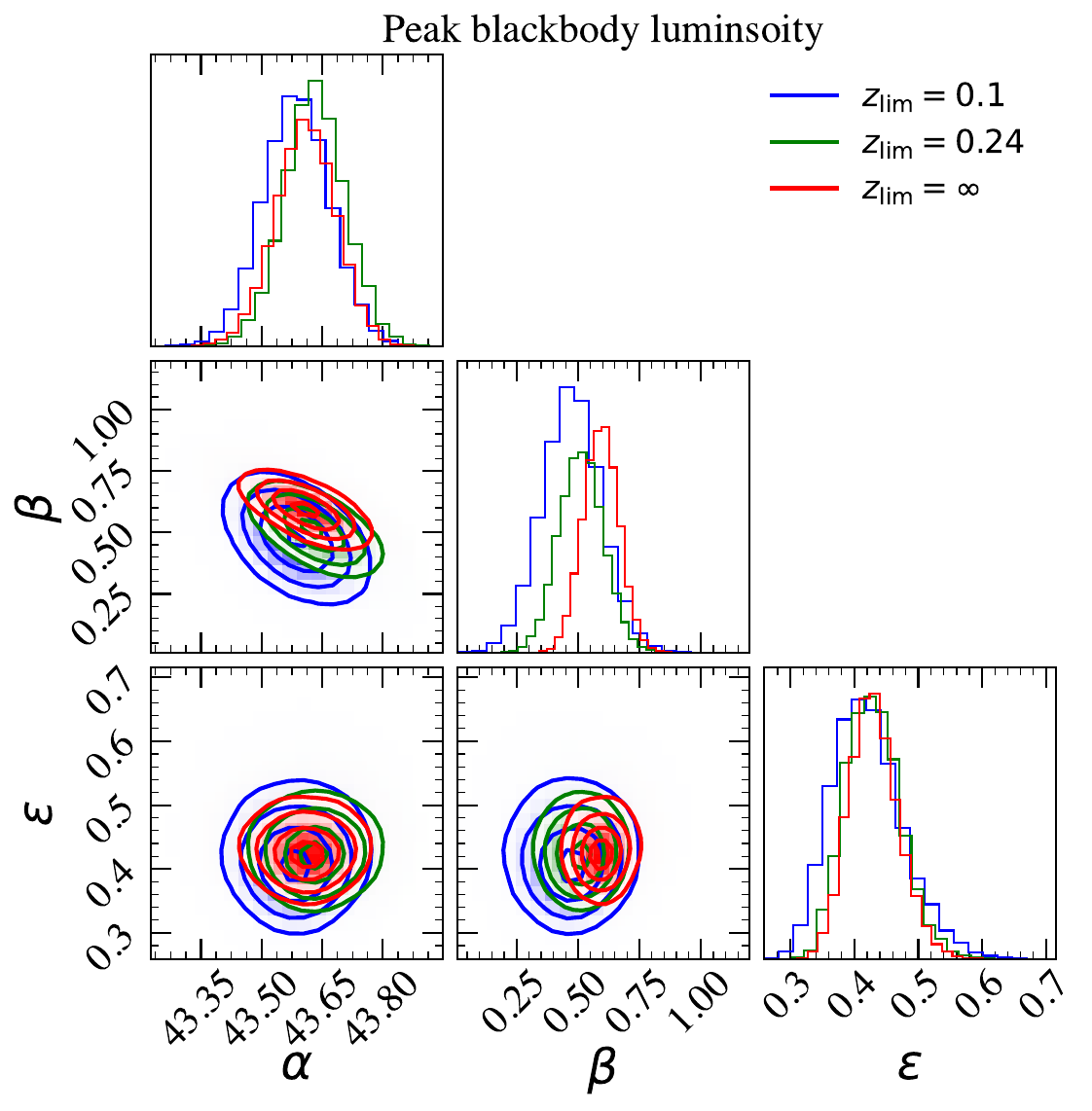}
    \includegraphics[width=0.45\linewidth]{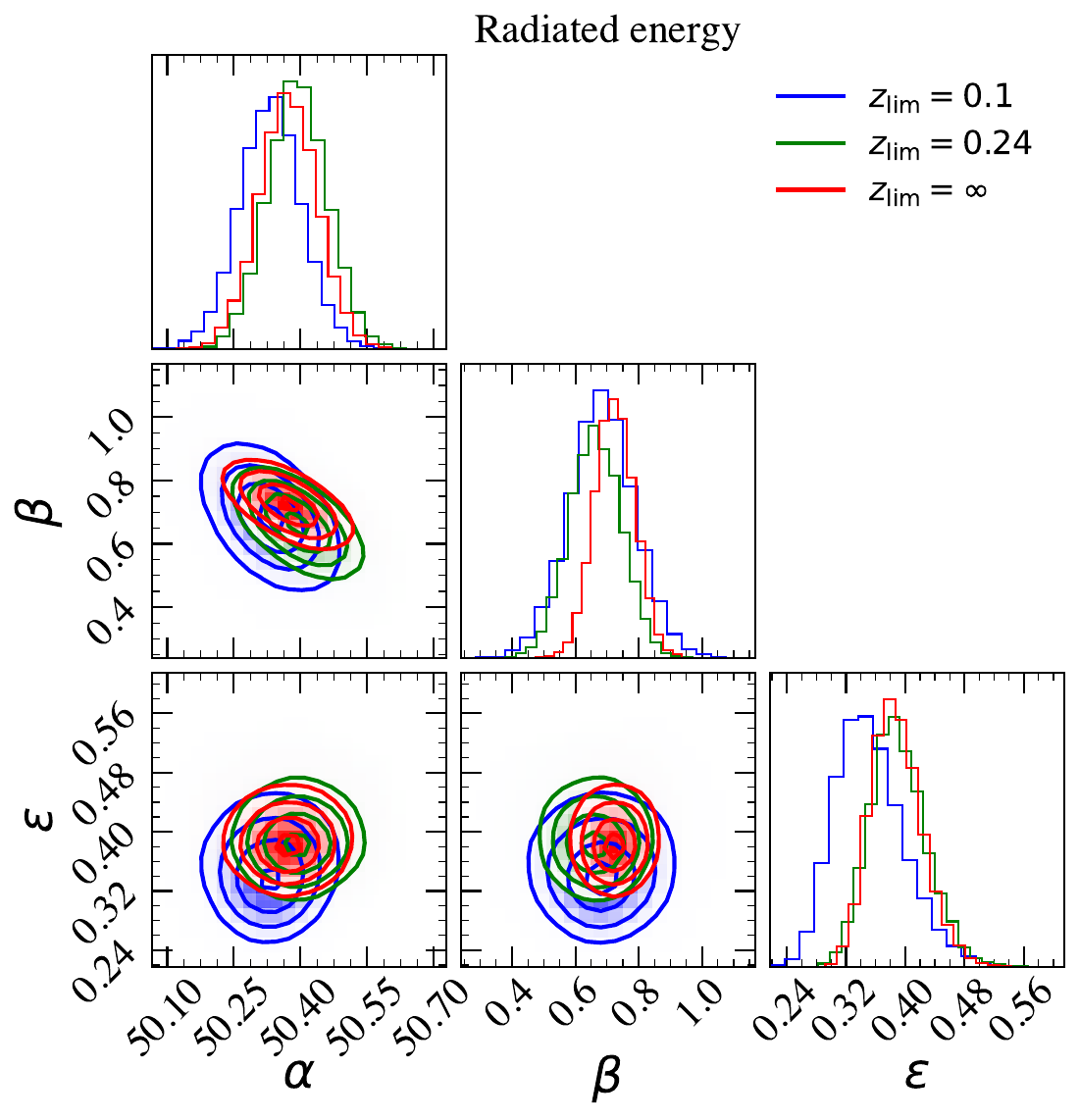}
    \caption{The posterior distributions of the power-law fits to the peak blackbody luminosity (left) and total radiated energy (right) under different redshift cuts for the TDE population. All results are entirely consistent, and the choice of a redshift cut is completely irrelevant to the inferred scaling relationship.  }
    \label{fig:Other_Z}
\end{figure*}

Now that we have confirmed that the plateau luminosity-black hole mass correlation is robust to any choice of redshift cut, it is worth testing that the same applies to the scaling of the peak blackbody luminosity and radiated energy with black hole mass.  In Figure \ref{fig:Other_Z} we show the results of power-law fits to both quantities versus black hole mass, with the same redshift cuts applied as above. Once again, imposing any choice of redshift cut results in no change in the power law profiles. All of the results in this paper are robust to any Malmquist bias. 

\section{Different black hole mass inference techniques}\label{app:M}
As a final precaution against any induced bias, in the following sections we repeat the analysis of the main paper, with black hole masses inferred from three galactic scaling relationships. These are the black hole mass galaxy mass correlation of \cite{Greene20}
\begin{equation}
    \log_{10} M_\bullet/M_\odot = 7.43 + 1.61 \log_{10} M_{\rm gal}/3\times10^{10} M_\odot ,
\end{equation}
the black hole mass-velocity dispersion scaling relationship \citep{Greene20}
\begin{equation}
    \log_{10} M_\bullet/M_\odot = 7.87 + 4.38 \log_{10} \sigma_\star/160\, {\rm km\, s}^{-1} ,
\end{equation}
and the black hole mass-bulge mass scaling relationship \citep{Kormendy13}
\begin{equation}
    \log_{10} M_\bullet/M_\odot = 8.69 + 1.16 \log_{10} M_{\rm bulge}/10^{11} M_\odot .
\end{equation}
In the following sub-sections we show that the choice of scaling relation is irrelevant for the general point made in the manuscript. We remake Figures \ref{fig:lum_comp} and \ref{fig:en_comp} using black hole masses inferred from each different technique, in order of the scaling relationships with the most-to-least black hole masses.

\subsection{From galaxy mass}
\begin{figure*}
    \centering
    \includegraphics[width=0.32\linewidth]{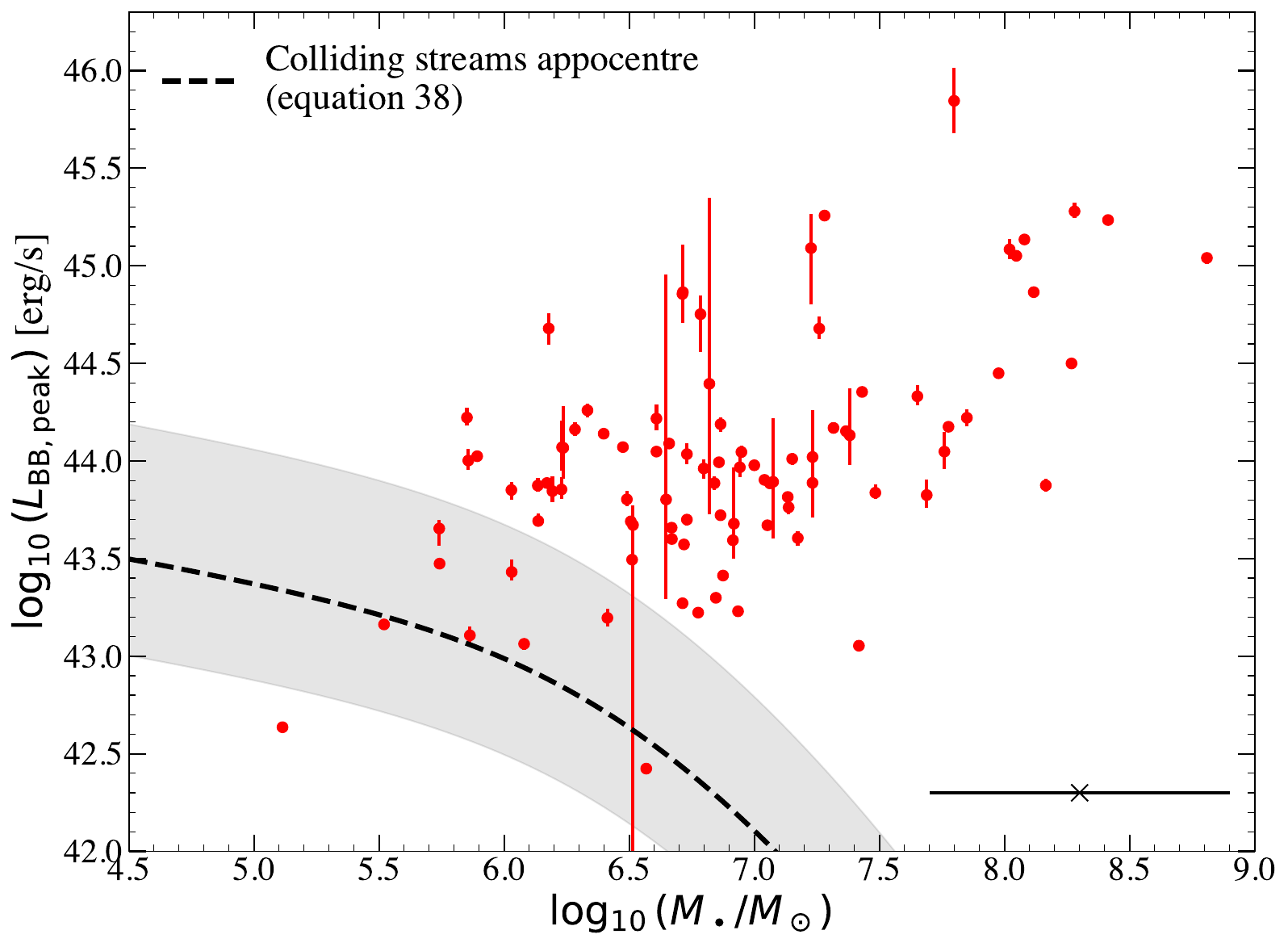}
    \includegraphics[width=0.32\linewidth]{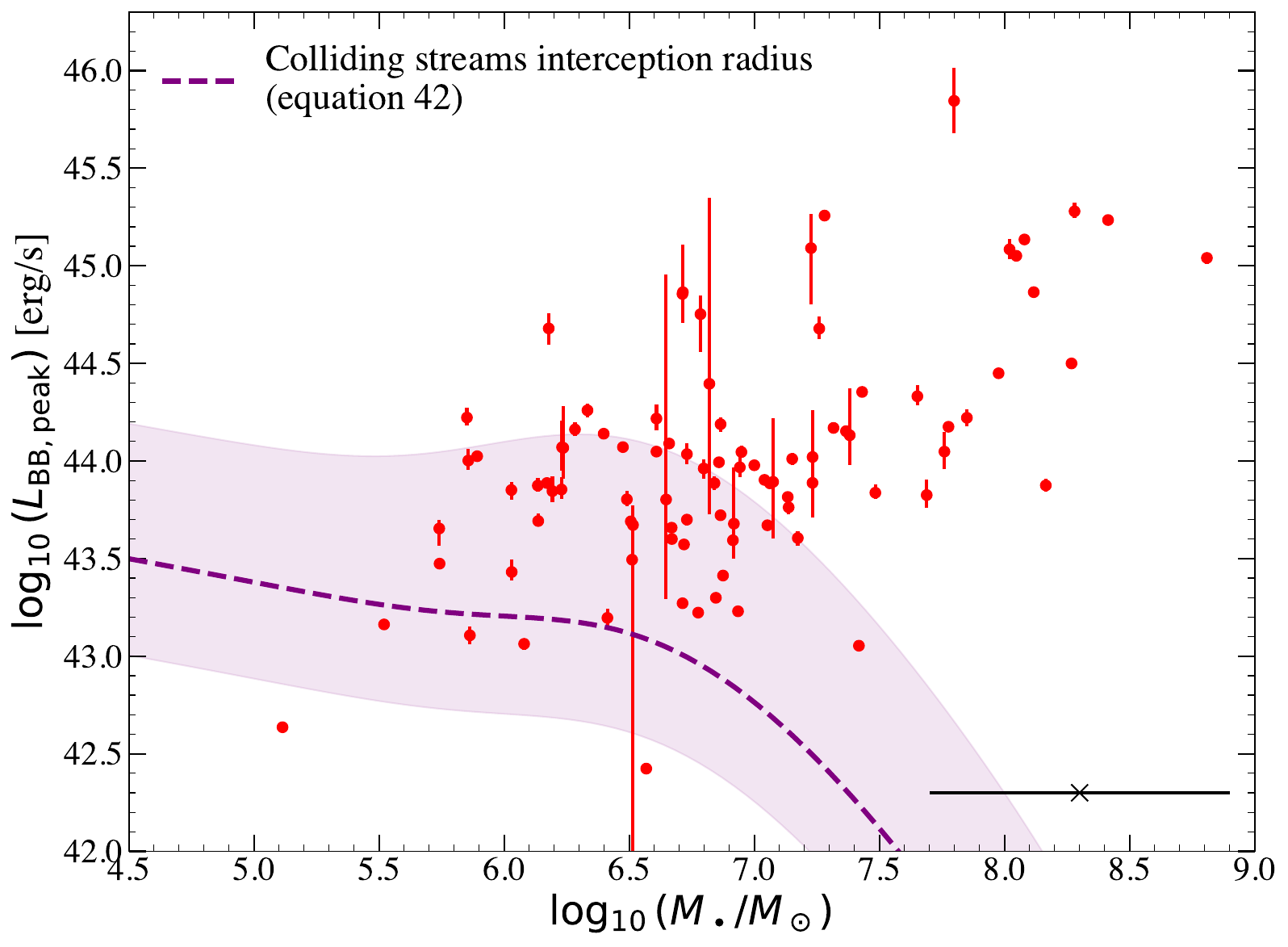}
    \includegraphics[width=0.32\linewidth]{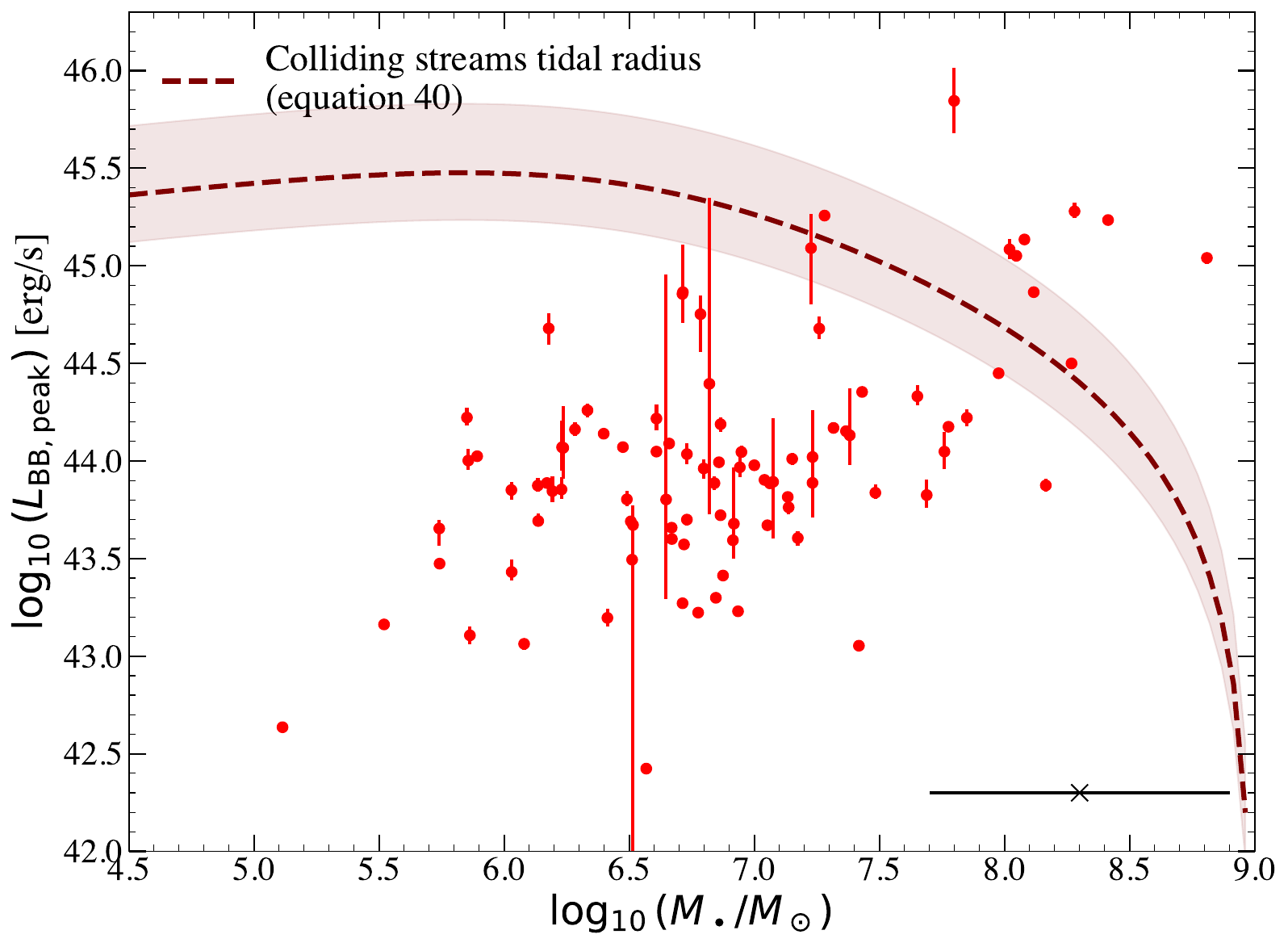}
    \includegraphics[width=0.32\linewidth]{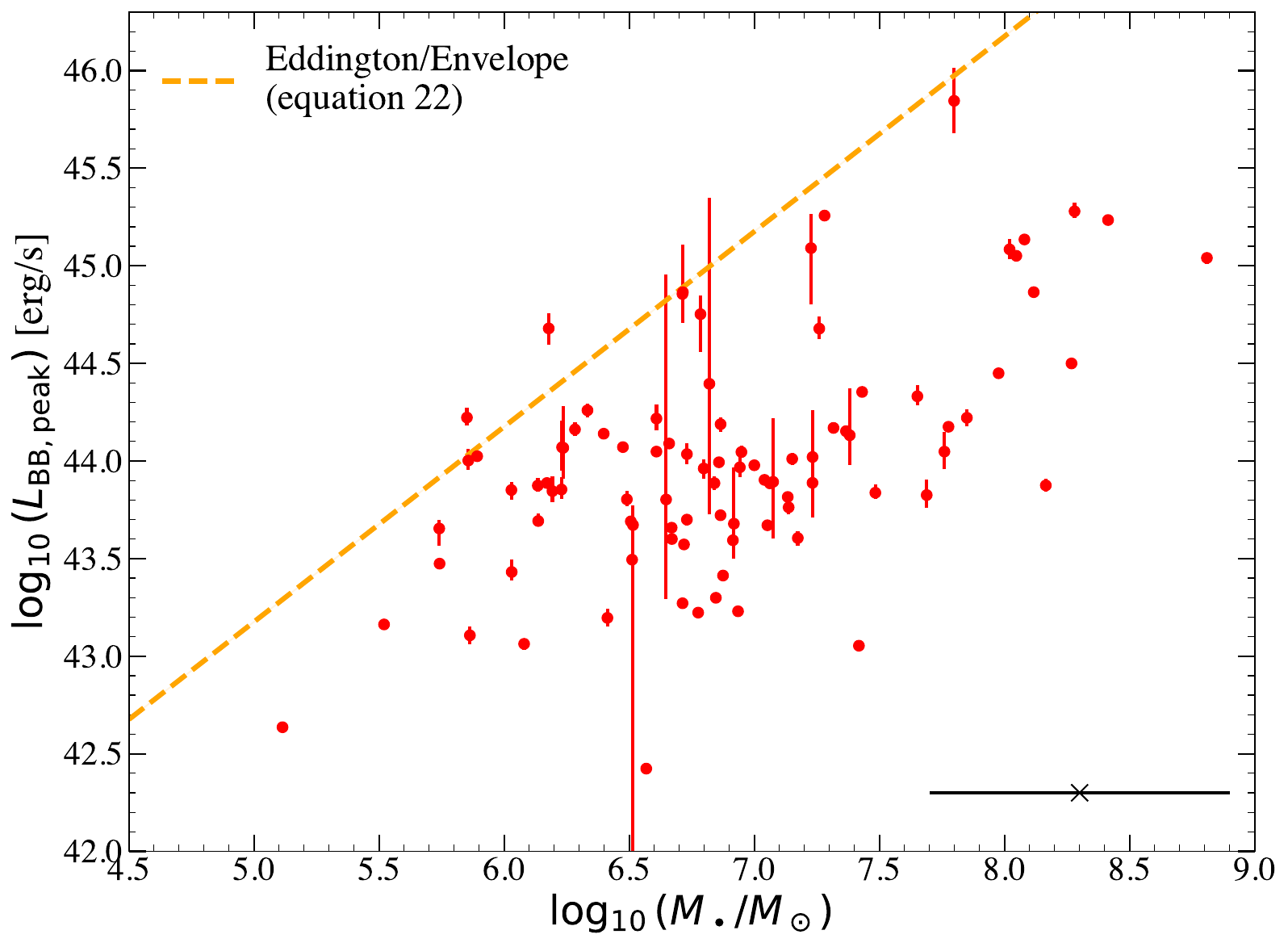}
    \includegraphics[width=0.32\linewidth]{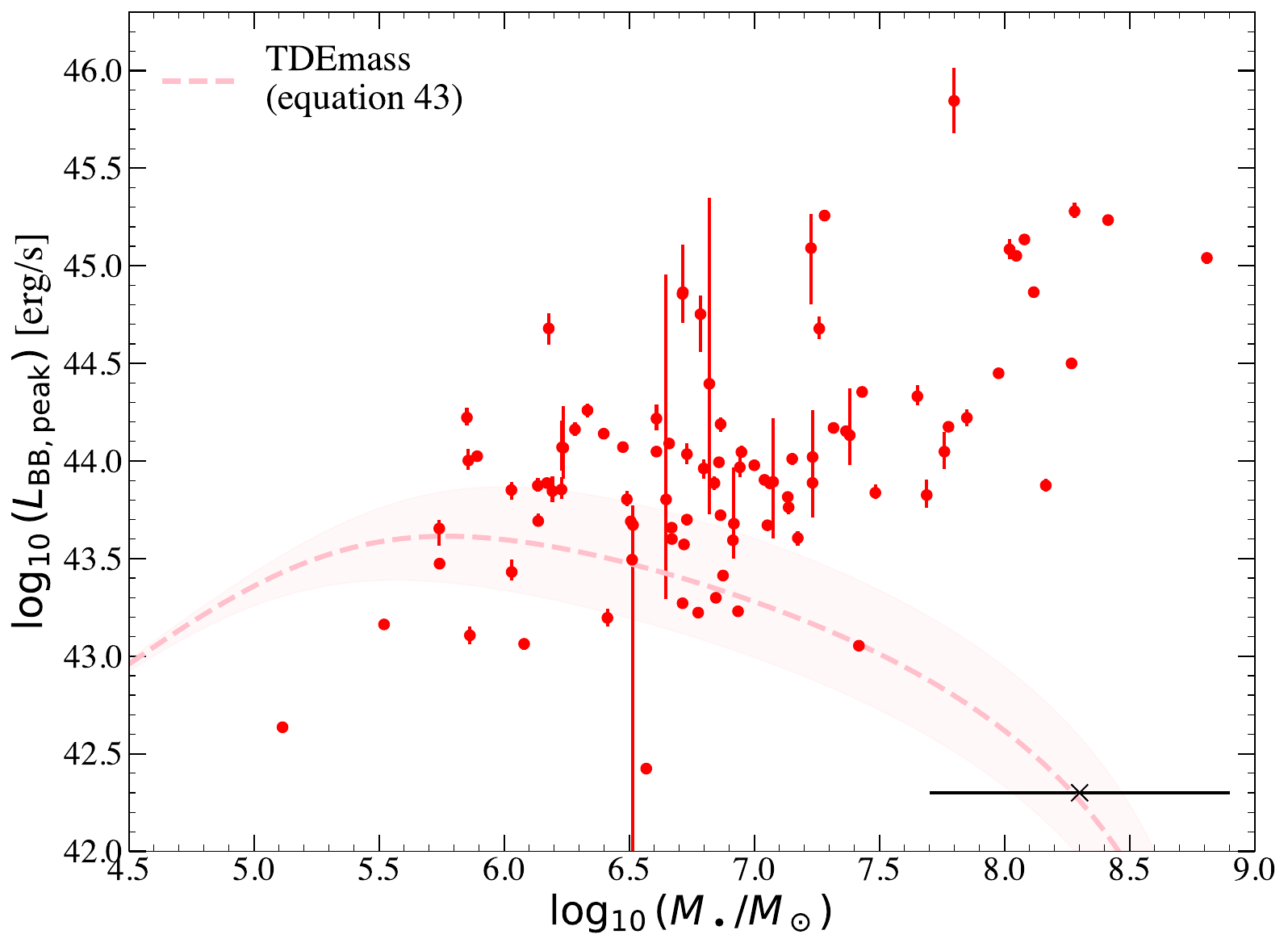}
    \includegraphics[width=0.32\linewidth]{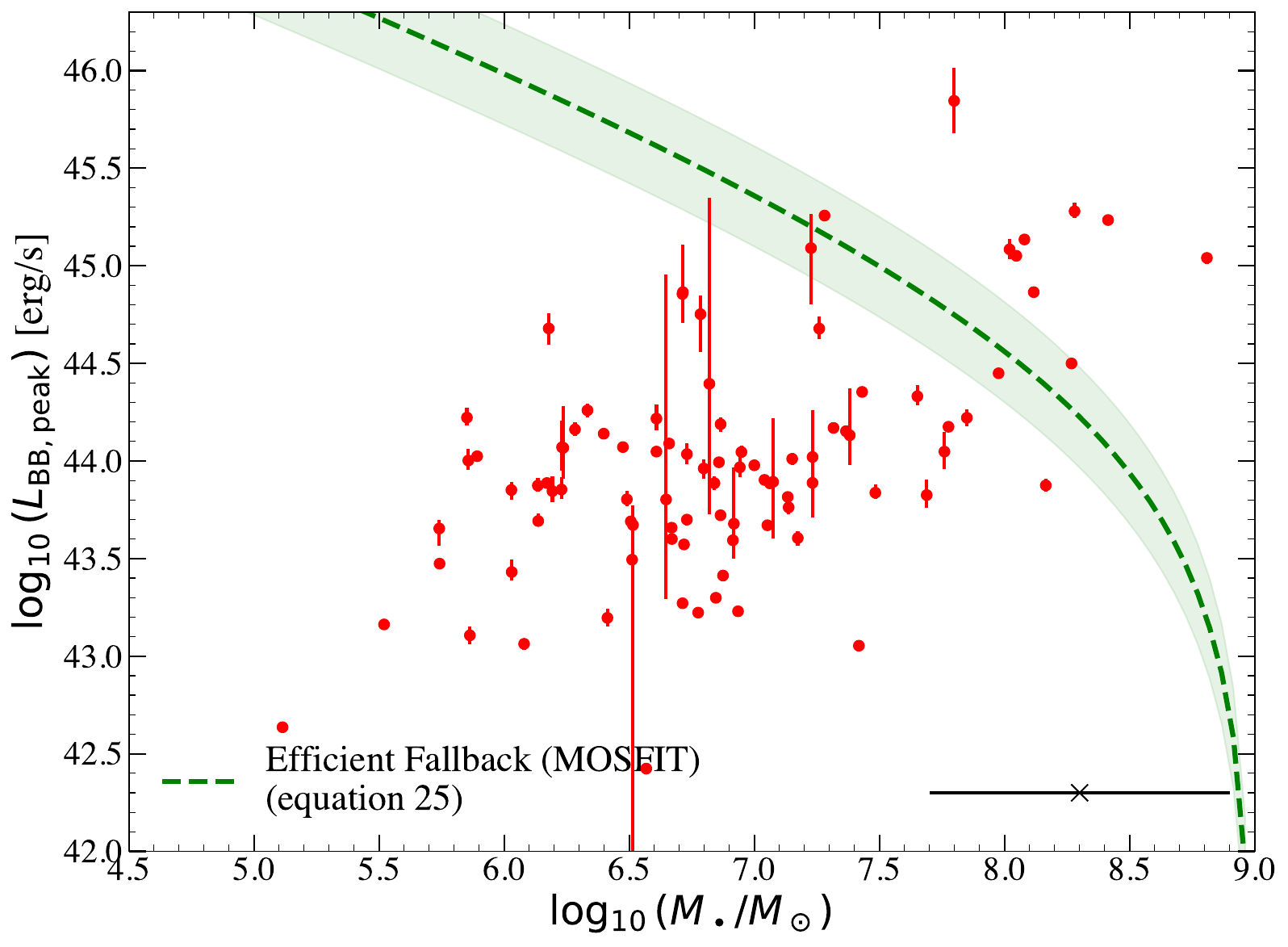}
    \includegraphics[width=0.32\linewidth]{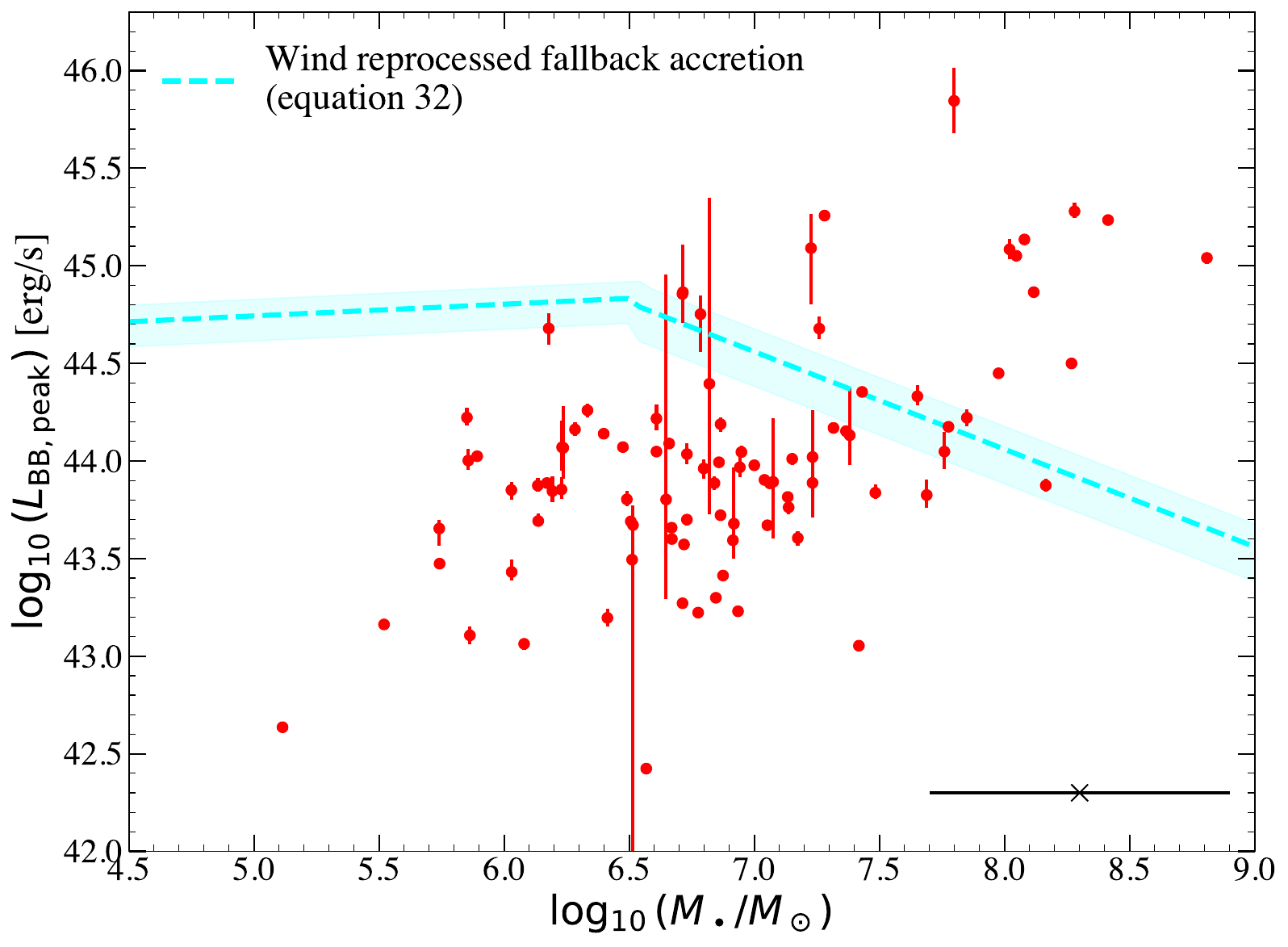}
    \caption{The same as Figure \ref{fig:lum_comp}, except now with mass inferred from the total galaxy mass.  The choice of mass inference technique makes no difference to any conclusion in this paper. }
    \label{fig:lum_comp_gal}
\end{figure*}

\begin{figure*}
    \centering
    \includegraphics[width=0.32\linewidth]{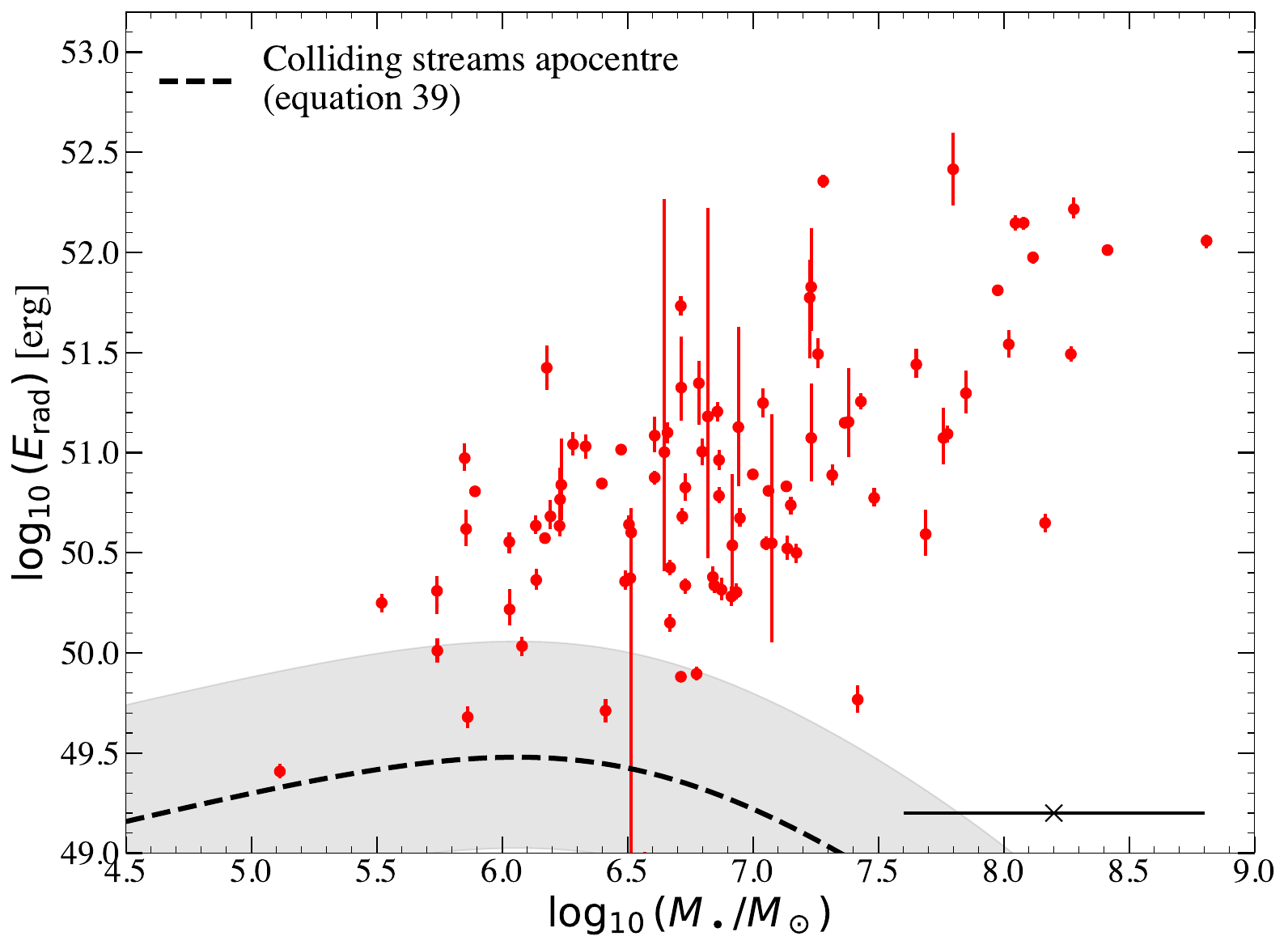}
    \includegraphics[width=0.32\linewidth]{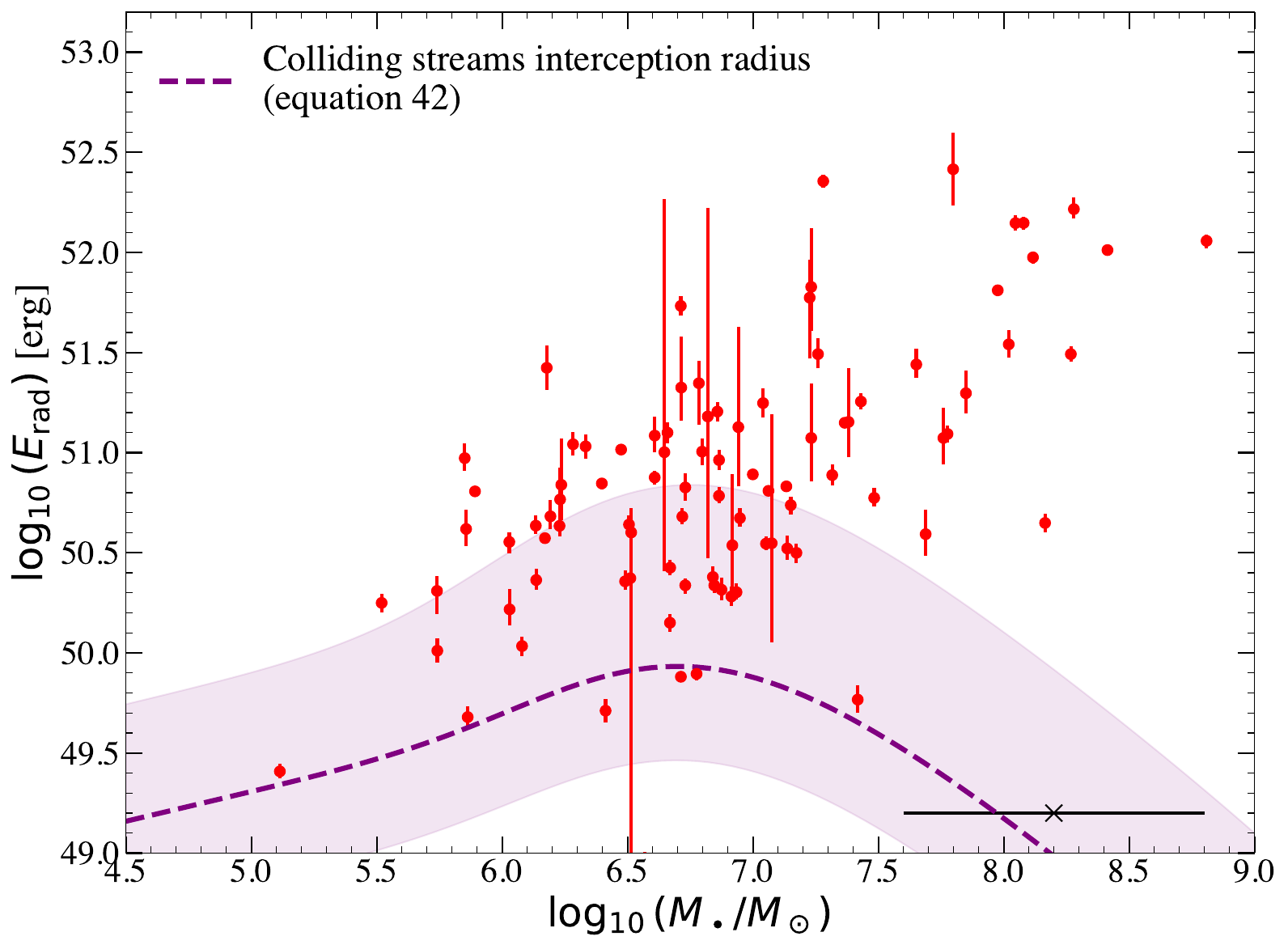}
    \includegraphics[width=0.32\linewidth]{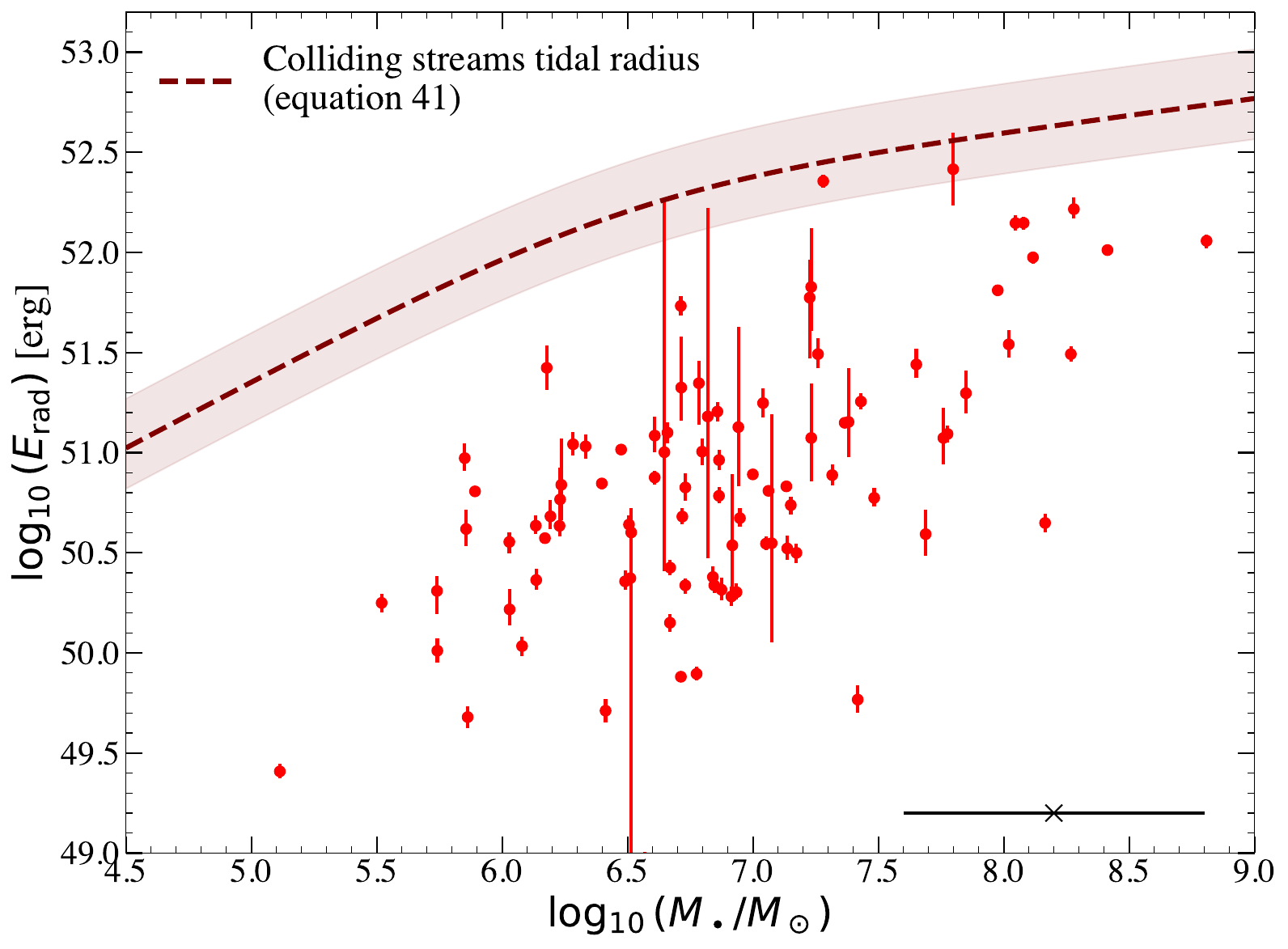}
    \includegraphics[width=0.32\linewidth]{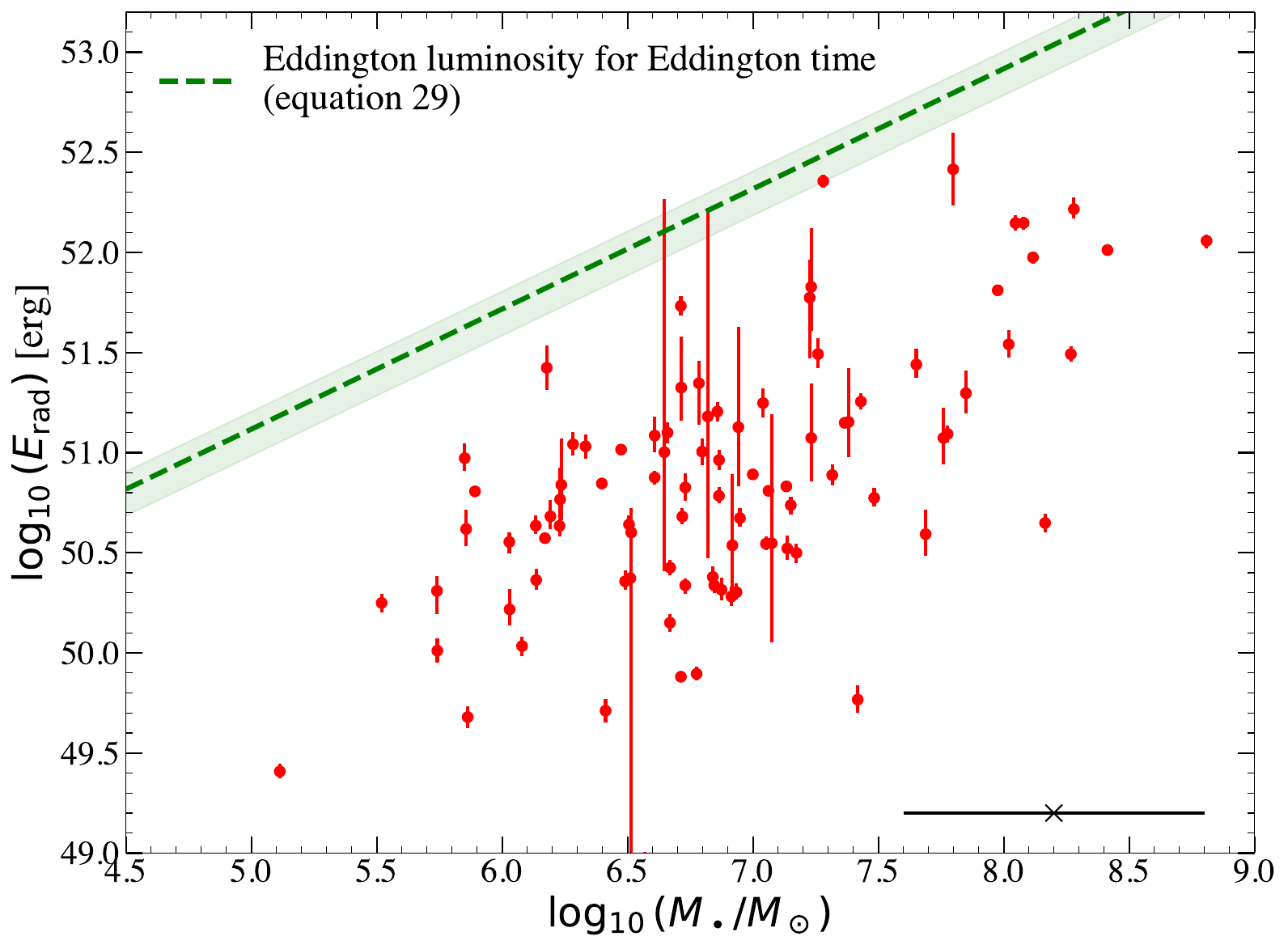}
    \includegraphics[width=0.32\linewidth]{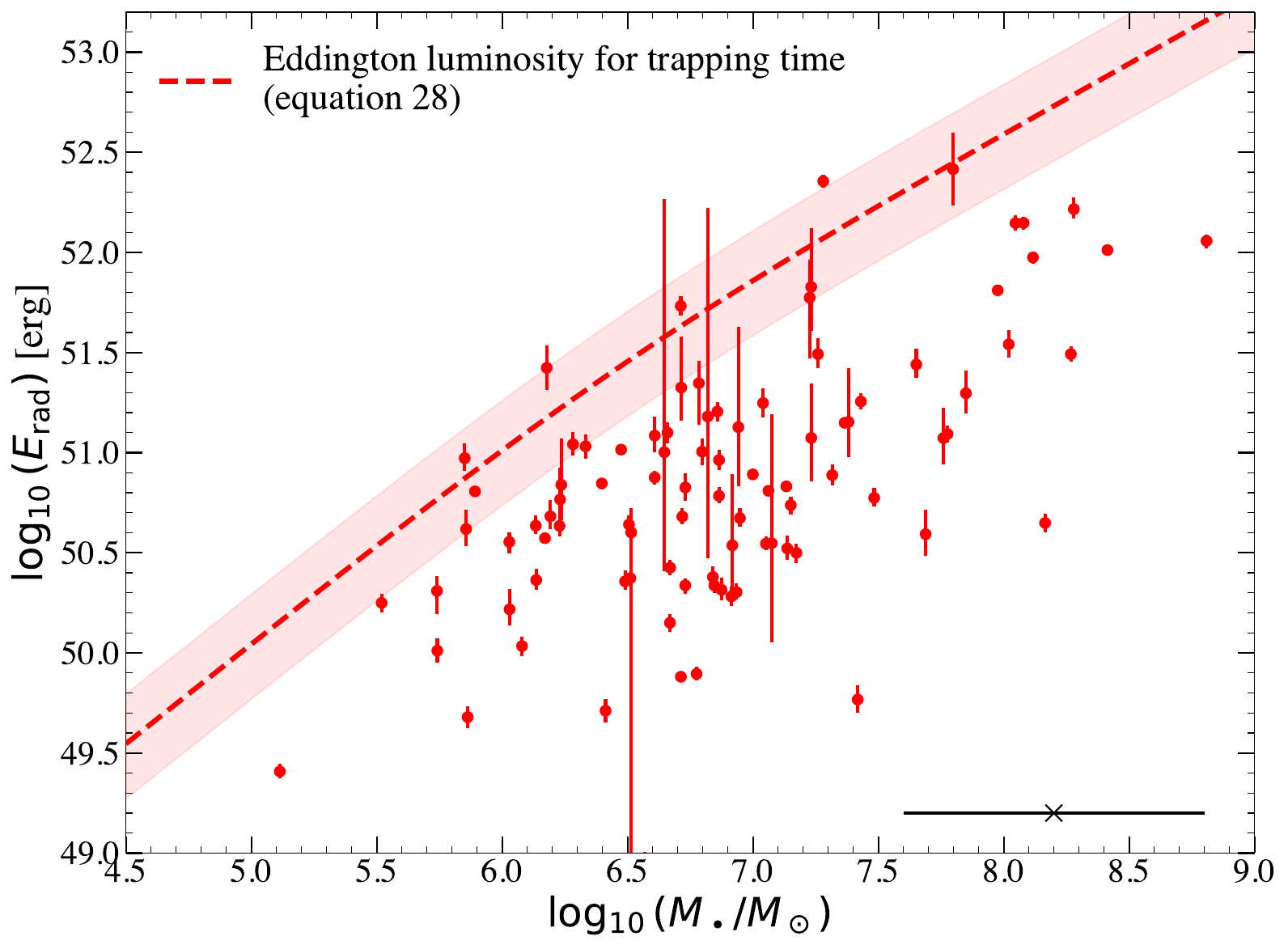}
    \includegraphics[width=0.32\linewidth]{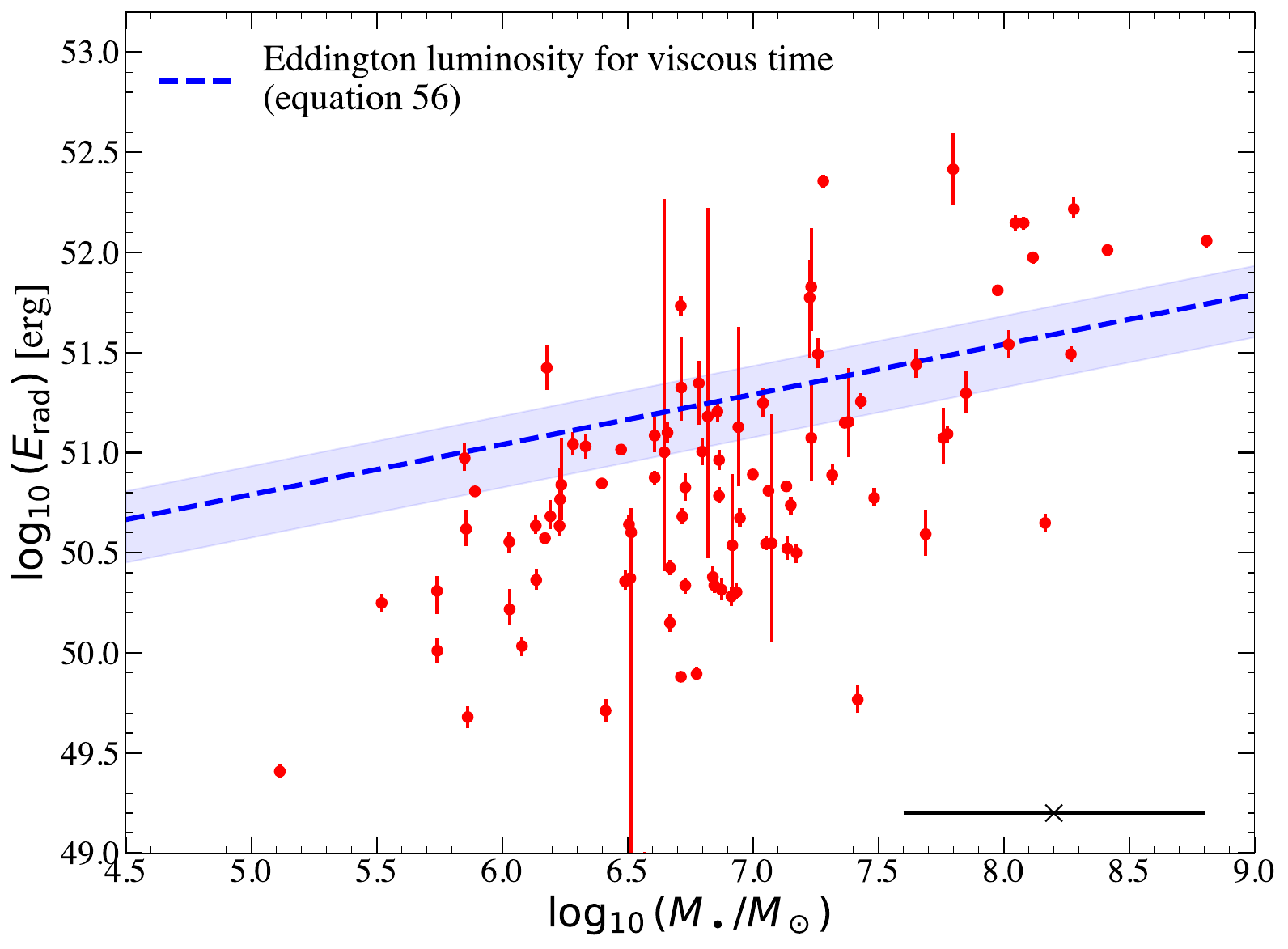}
    \includegraphics[width=0.32\linewidth]{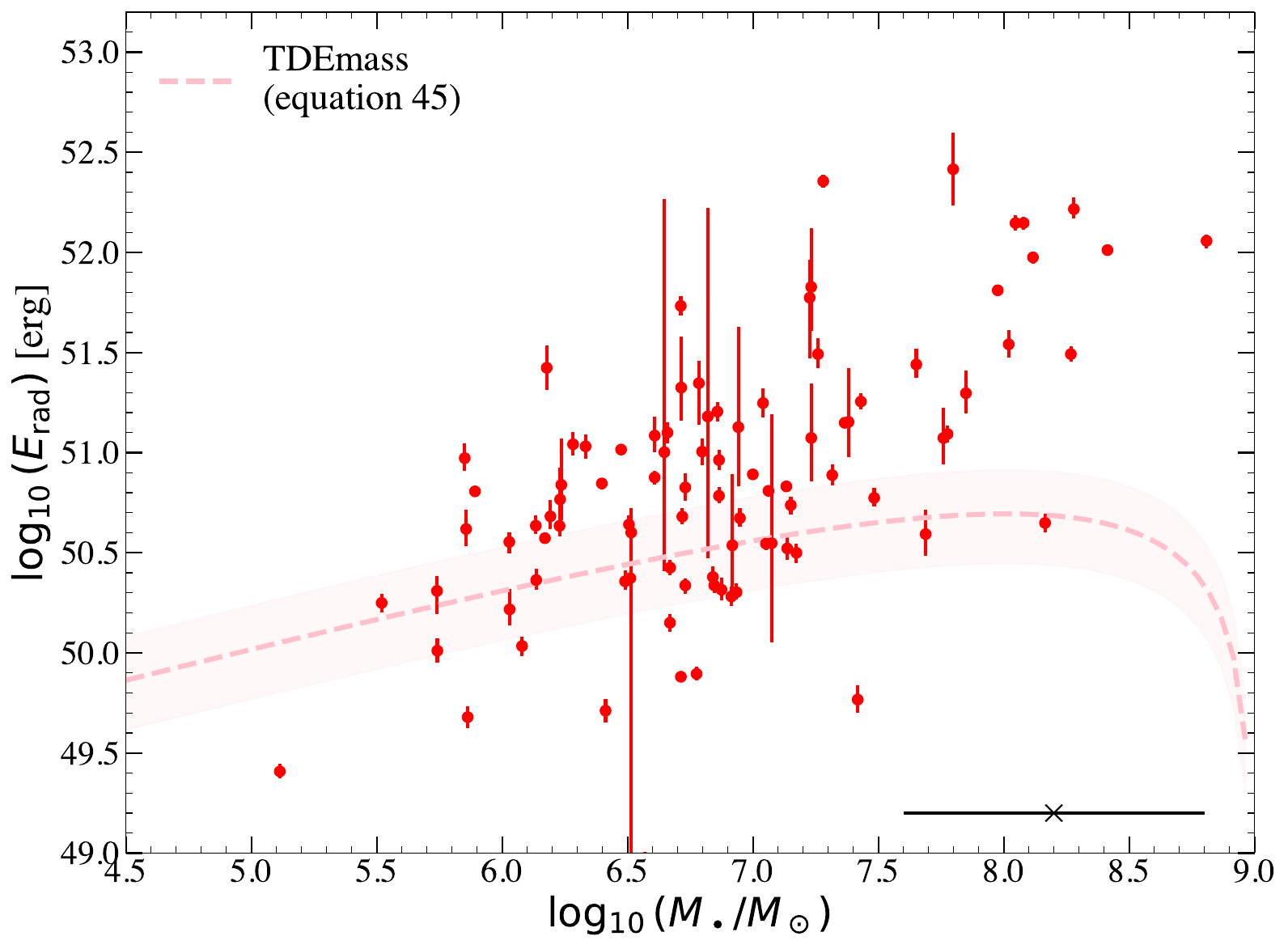}
    \includegraphics[width=0.32\linewidth]{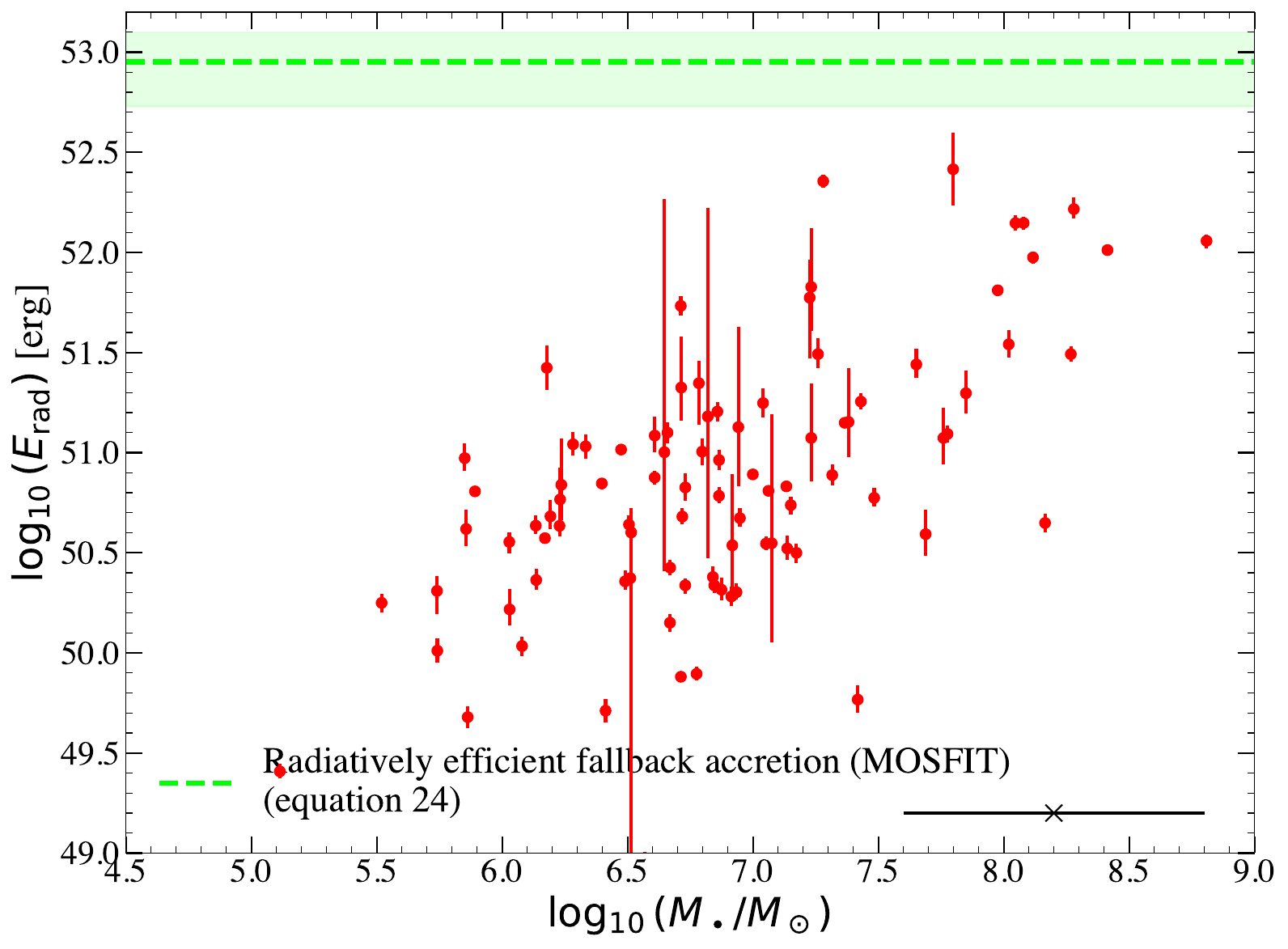}
    \includegraphics[width=0.32\linewidth]{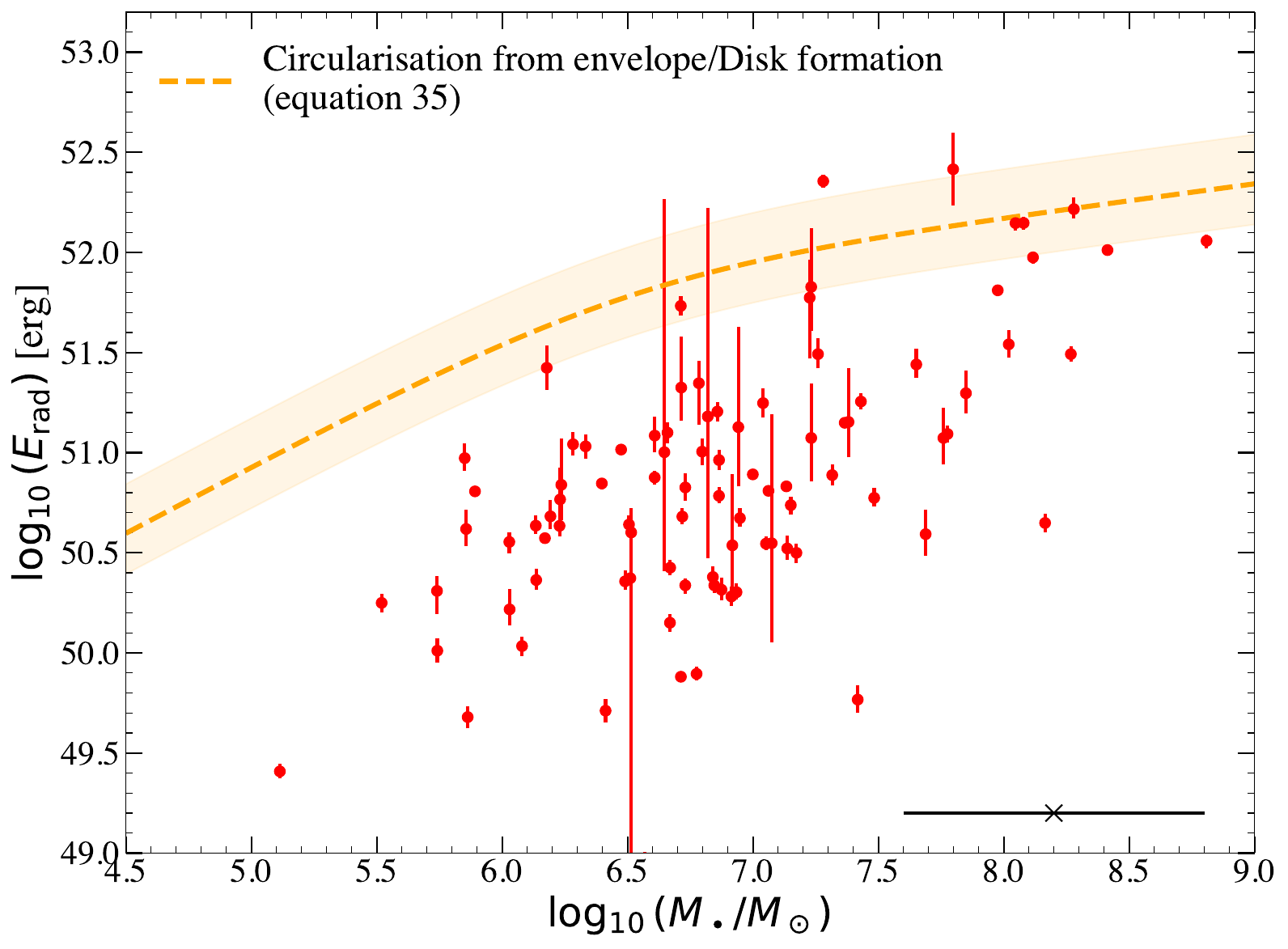}
    \includegraphics[width=0.32\linewidth]{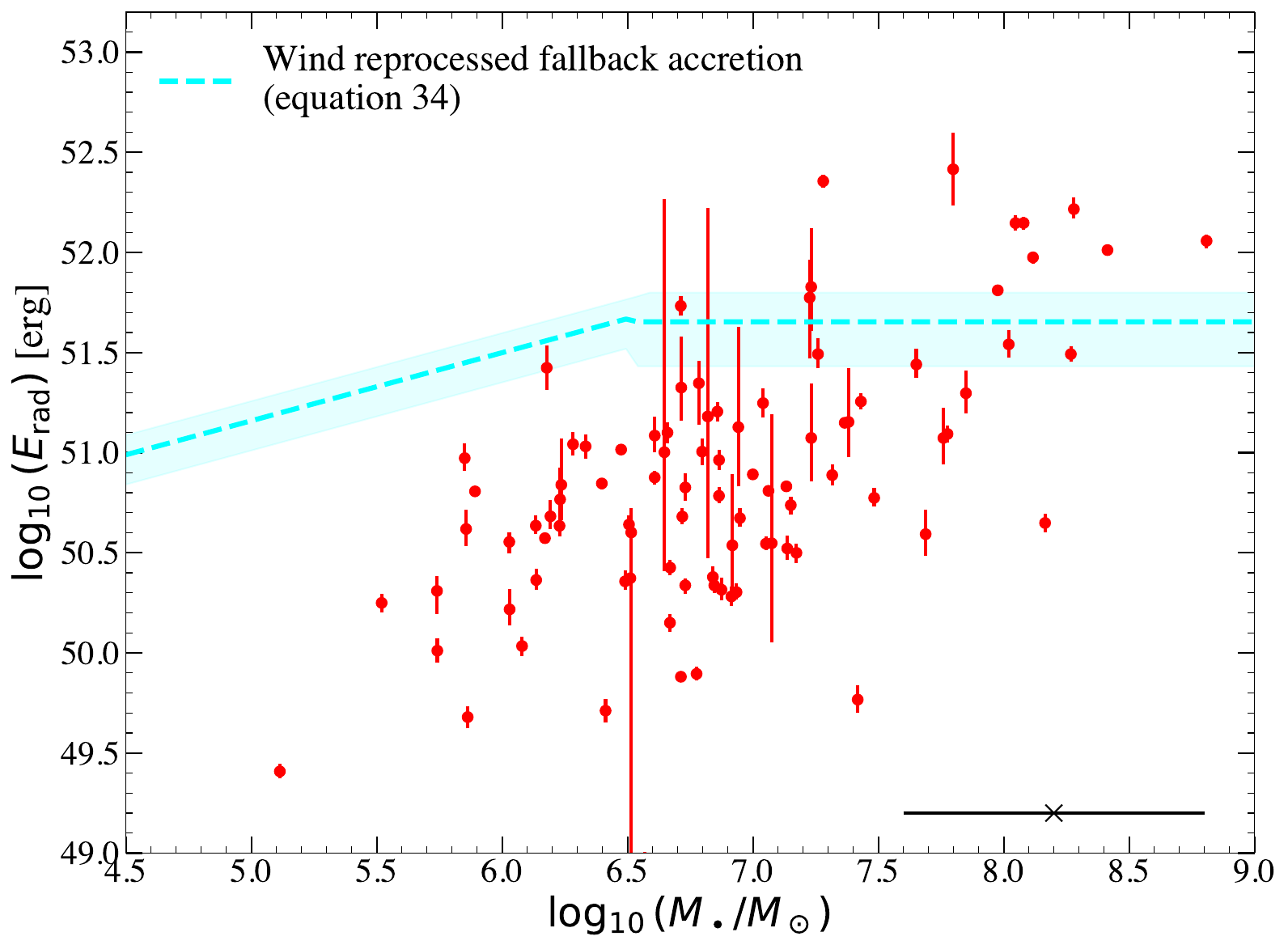}
    \caption{The same as Figure \ref{fig:en_comp}, except now with mass inferred from the total galaxy mass.  The choice of mass inference technique makes no difference to any conclusion in this paper. }
    \label{fig:en_comp_gal}
\end{figure*}

In Figures \ref{fig:lum_comp_gal} and \ref{fig:en_comp_gal} we show the correlation between peak blackbody luminosity (Fig. \ref{fig:lum_comp_gal}) and radiated energy (Fig. \ref{fig:en_comp_gal}) with black hole mass measured from the galaxy mass, a measurement available for every source in our sample (presented in \citealt{Mummery_et_al_2024, MummeryVV25}). Our results and interpretation are completely unchanged.

\subsection{From bulge mass}
\begin{figure*}
    \centering
    \includegraphics[width=0.32\linewidth]{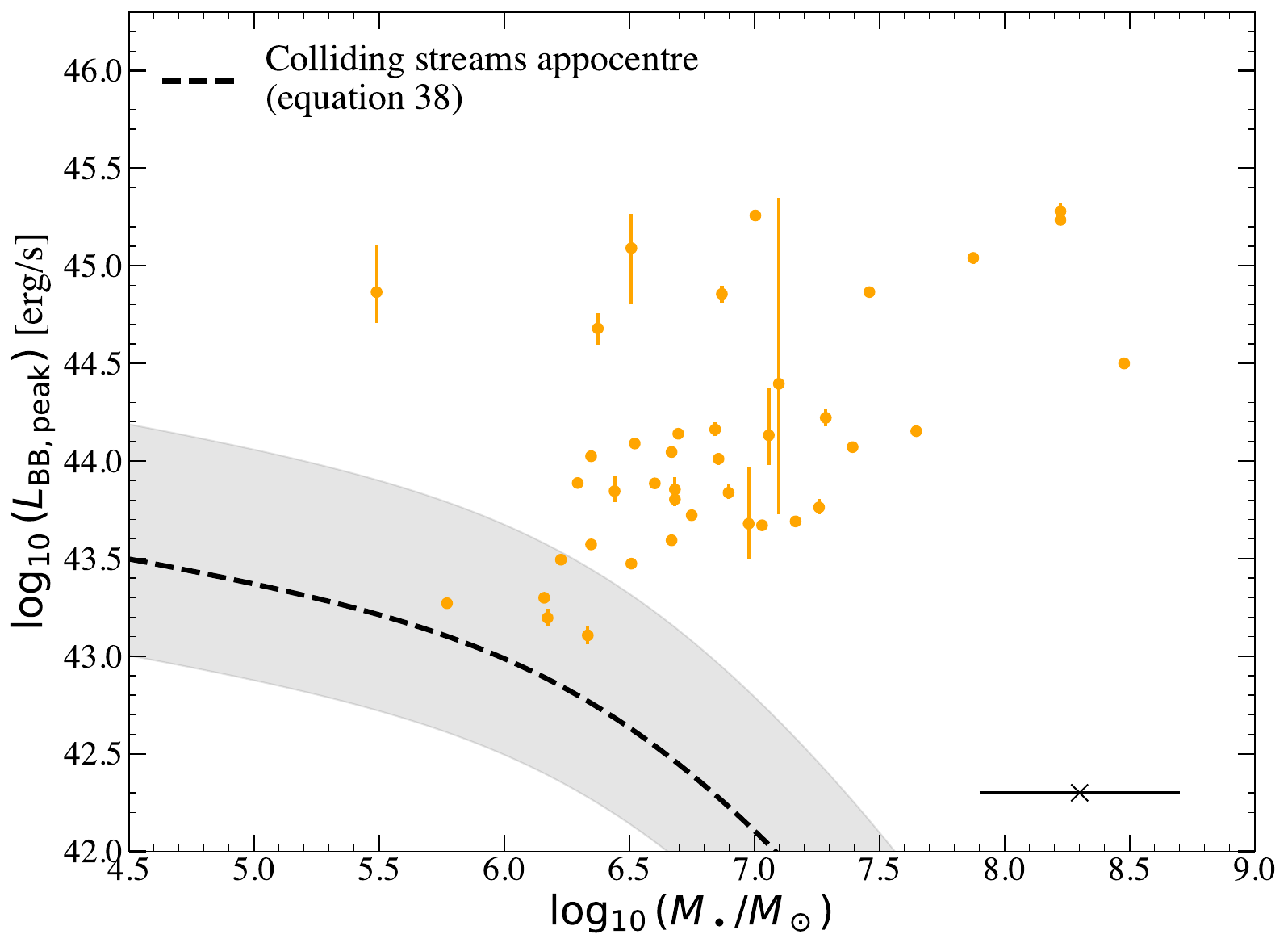}
    \includegraphics[width=0.32\linewidth]{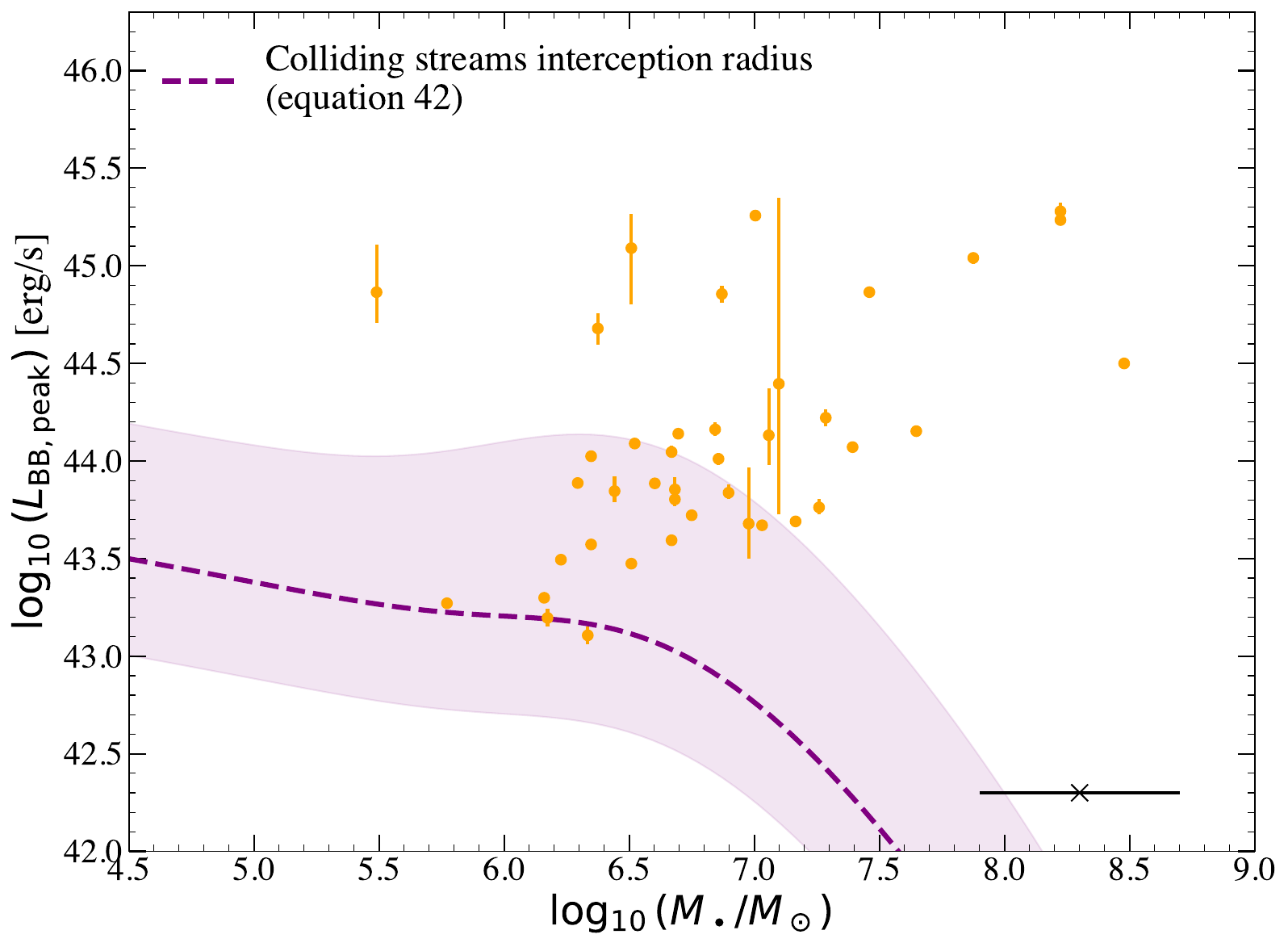}
    \includegraphics[width=0.32\linewidth]{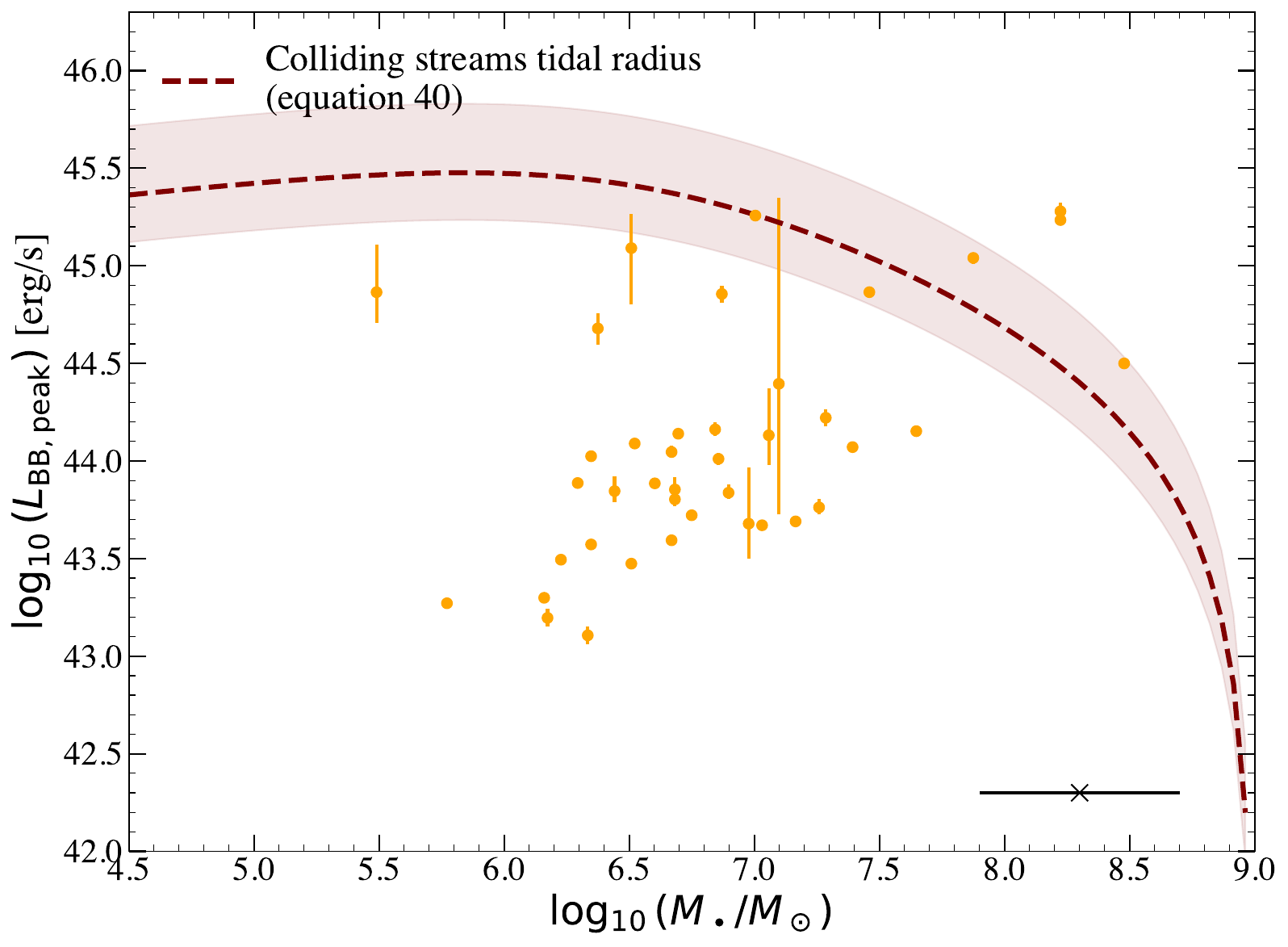}
    \includegraphics[width=0.32\linewidth]{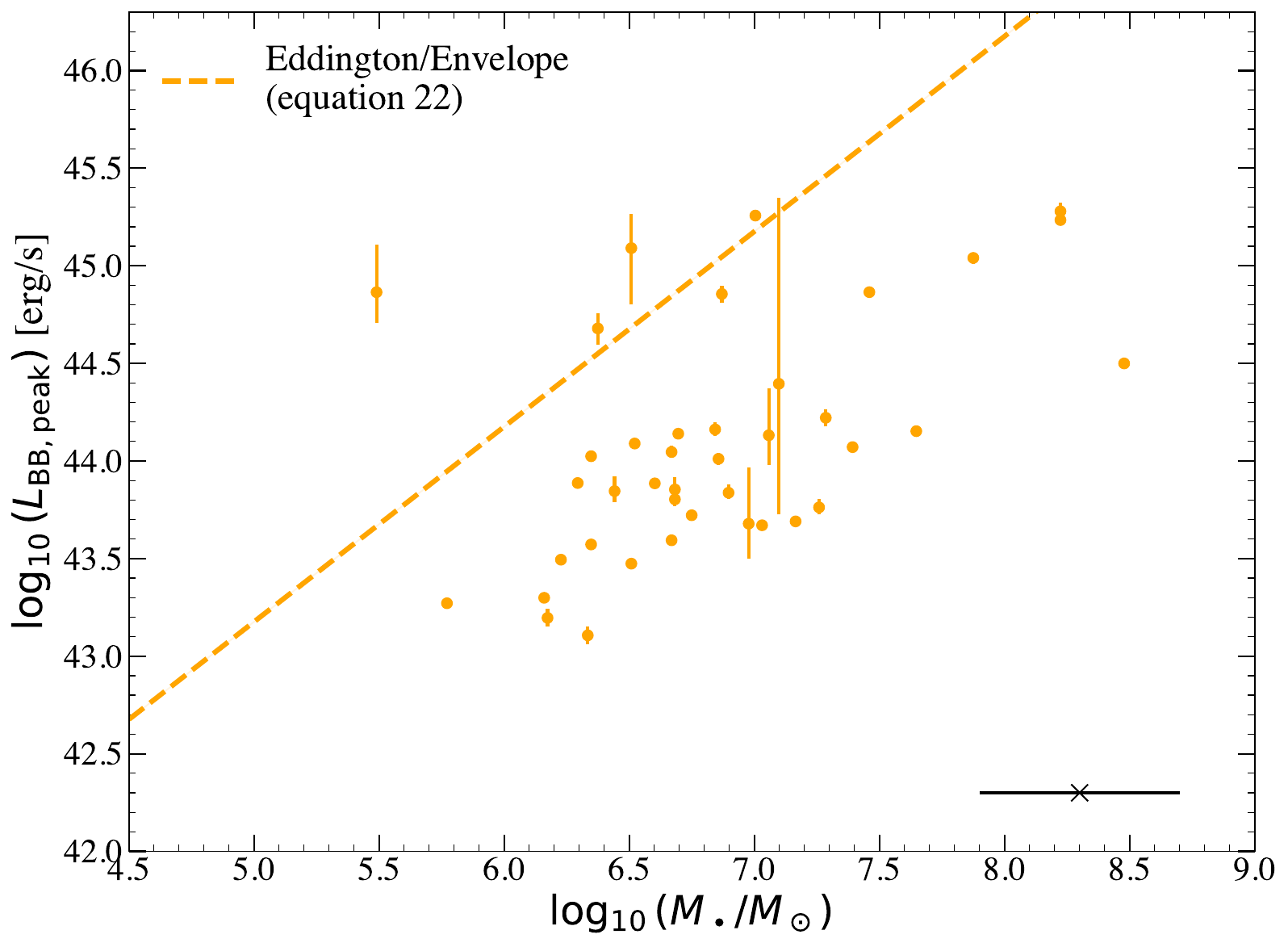}
    \includegraphics[width=0.32\linewidth]{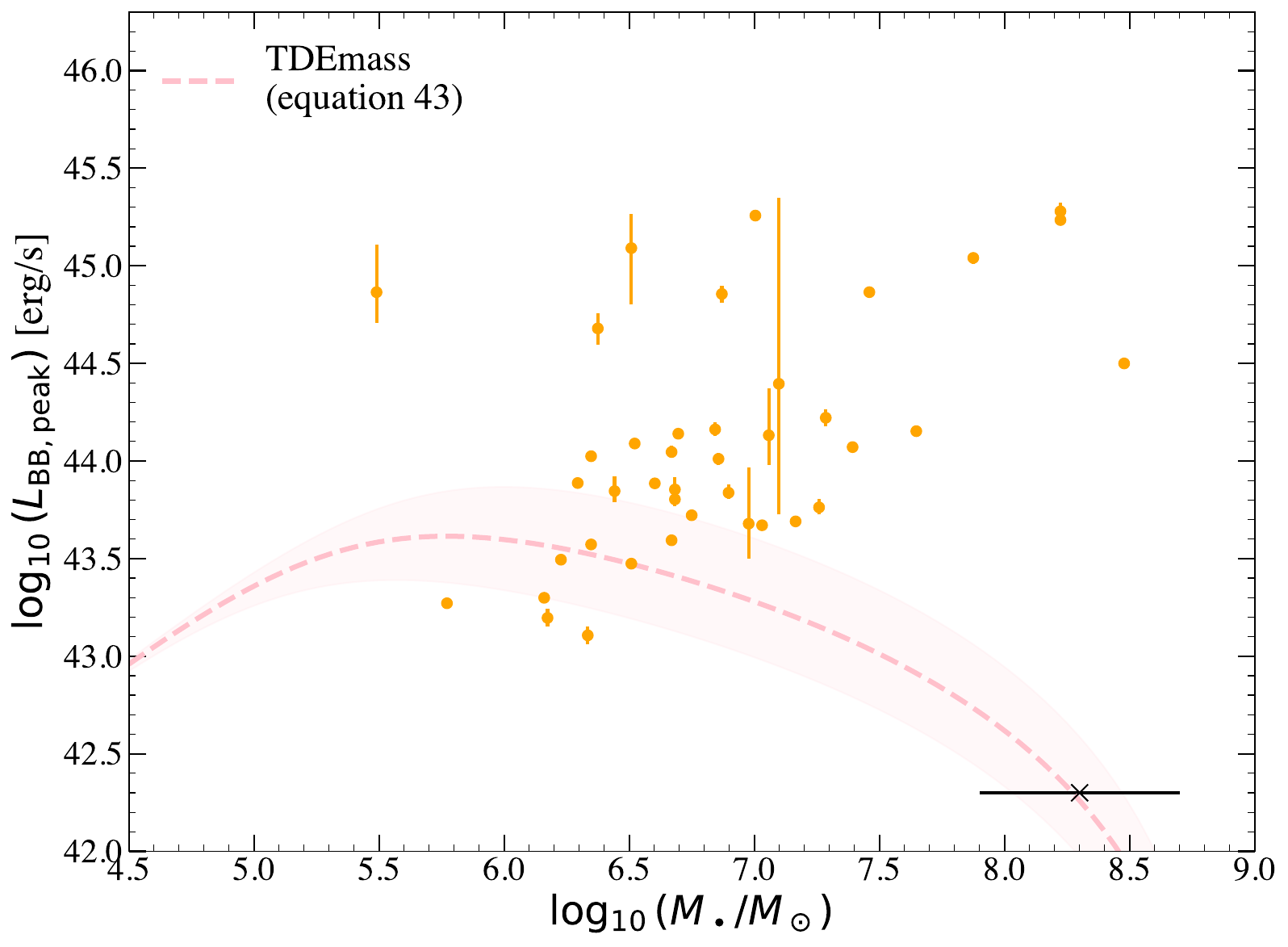}
    \includegraphics[width=0.32\linewidth]{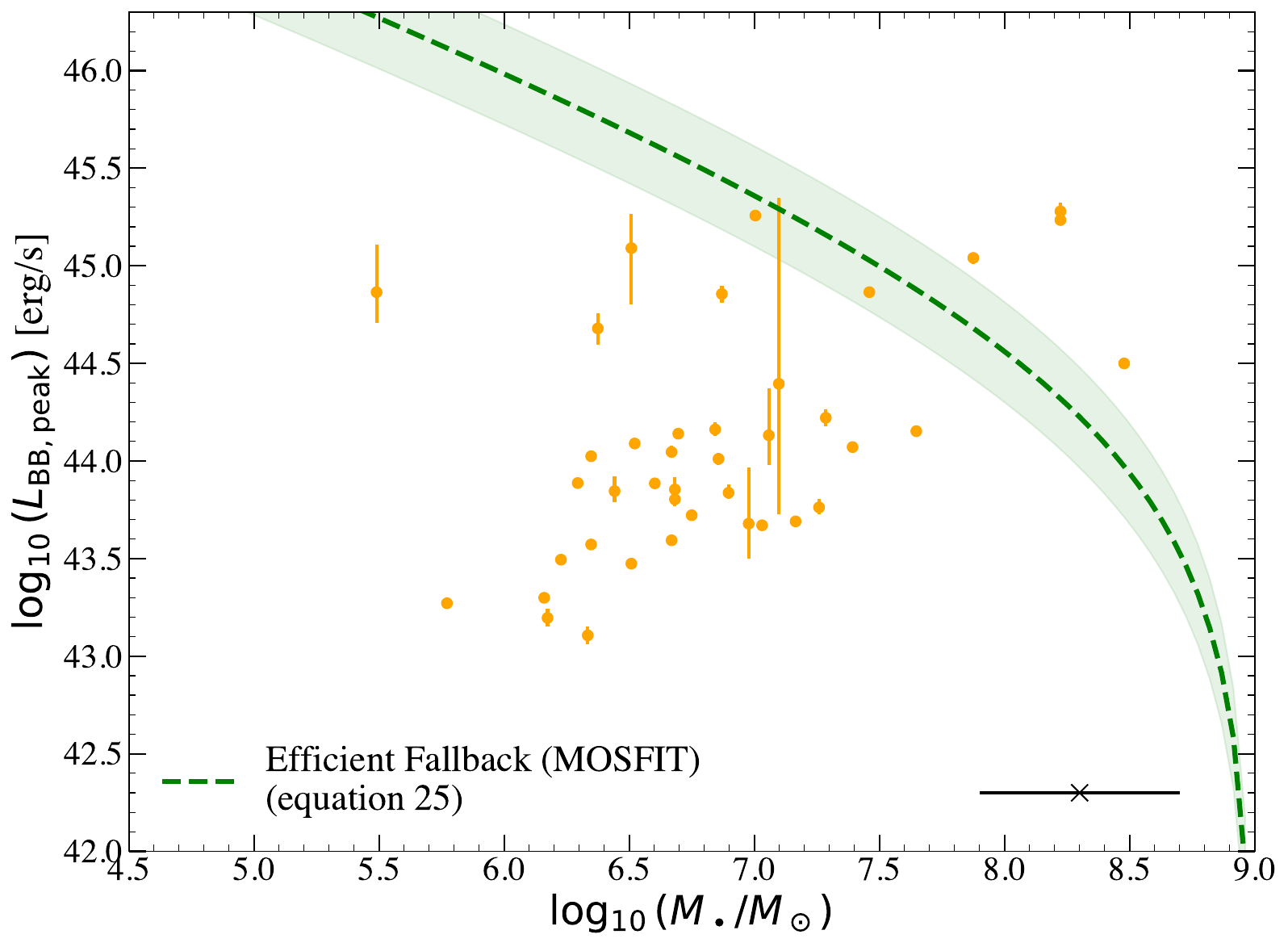}
    \includegraphics[width=0.32\linewidth]{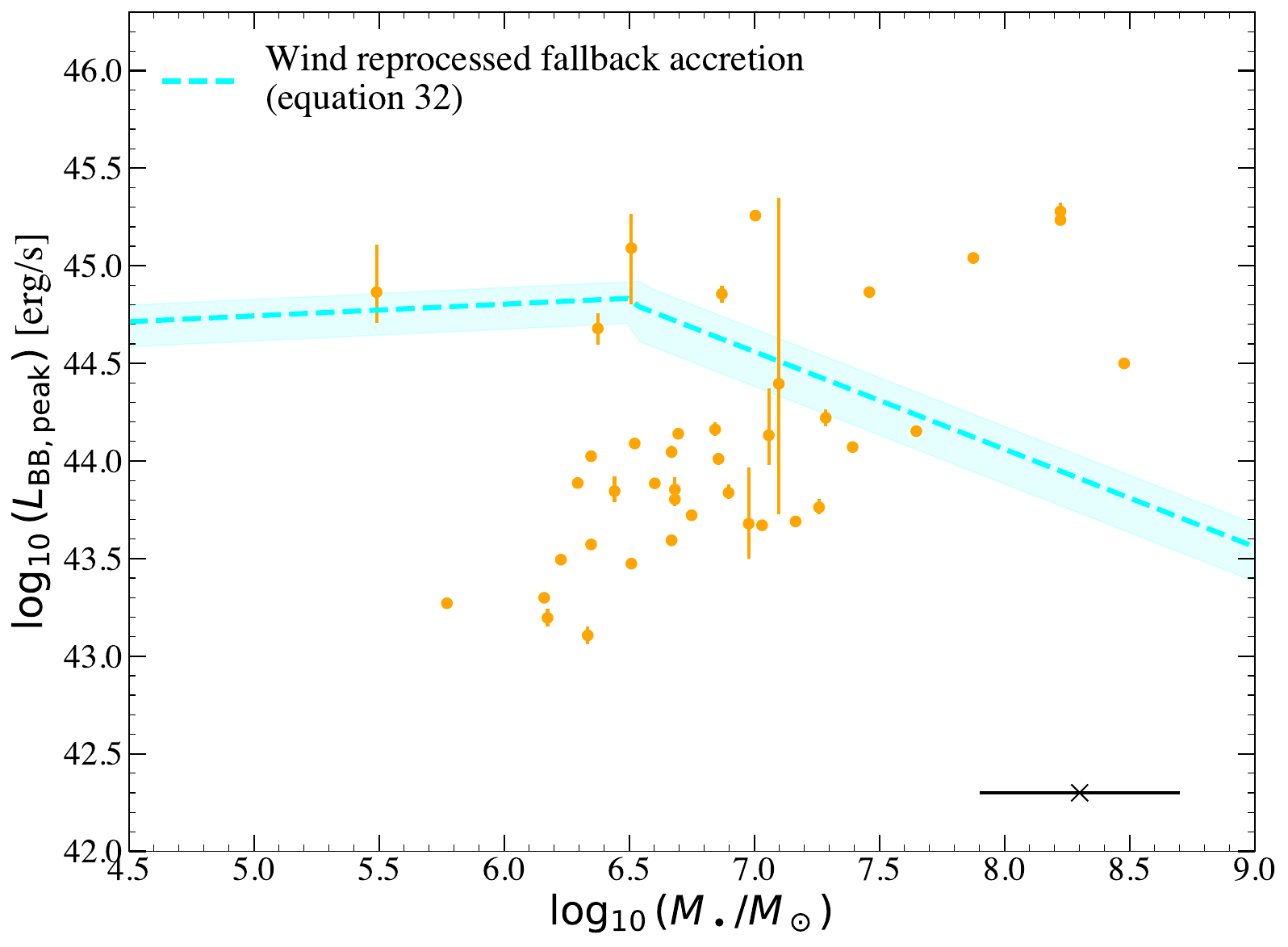}
    \caption{The same as Figure \ref{fig:lum_comp}, except now with mass inferred from the galaxy bulge mass.  The choice of mass inference technique makes no difference to any conclusion in this paper. }
    \label{fig:lum_comp_bulge}
\end{figure*}

\begin{figure*}
    \centering
    \includegraphics[width=0.32\linewidth]{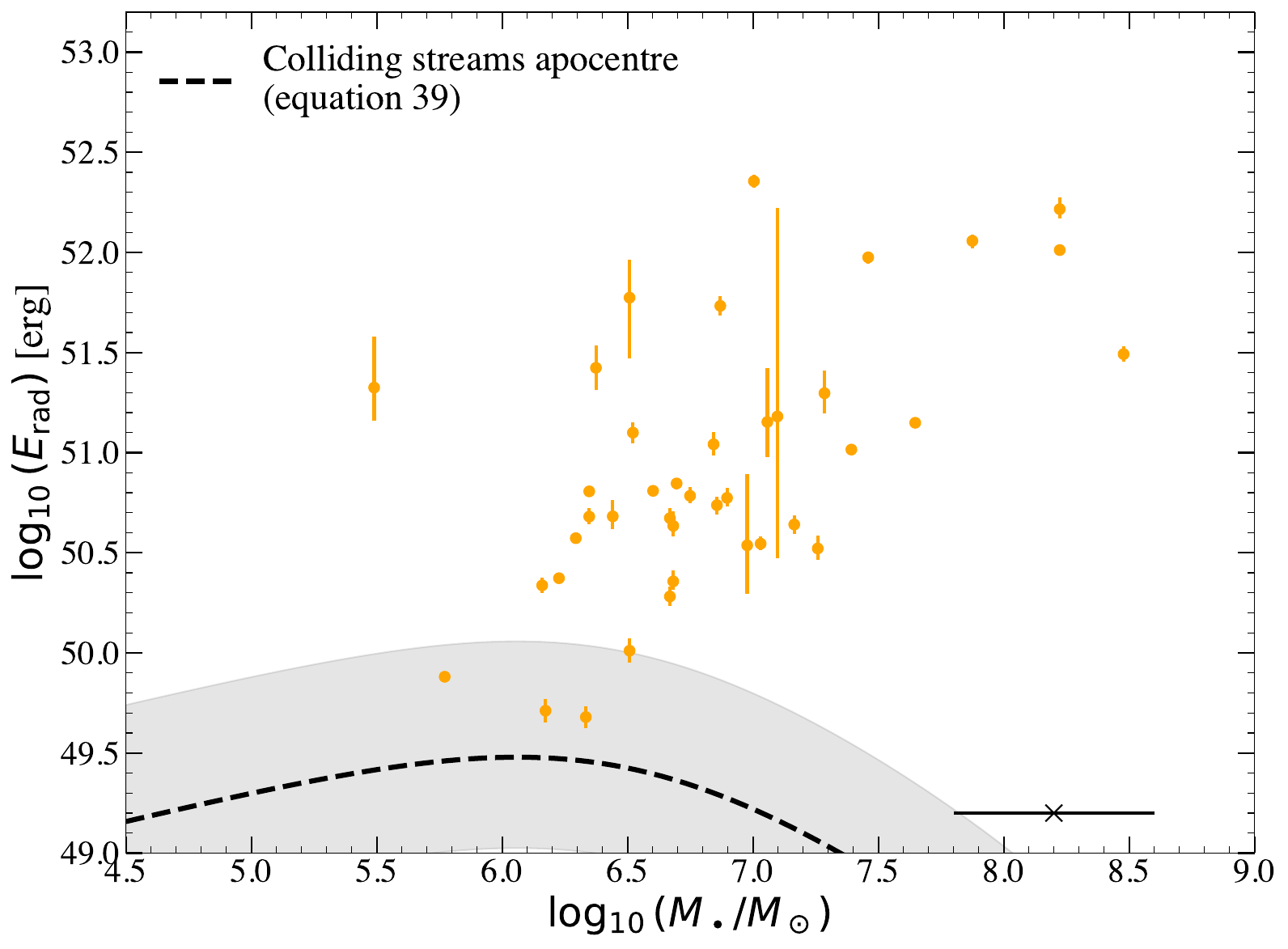}
    \includegraphics[width=0.32\linewidth]{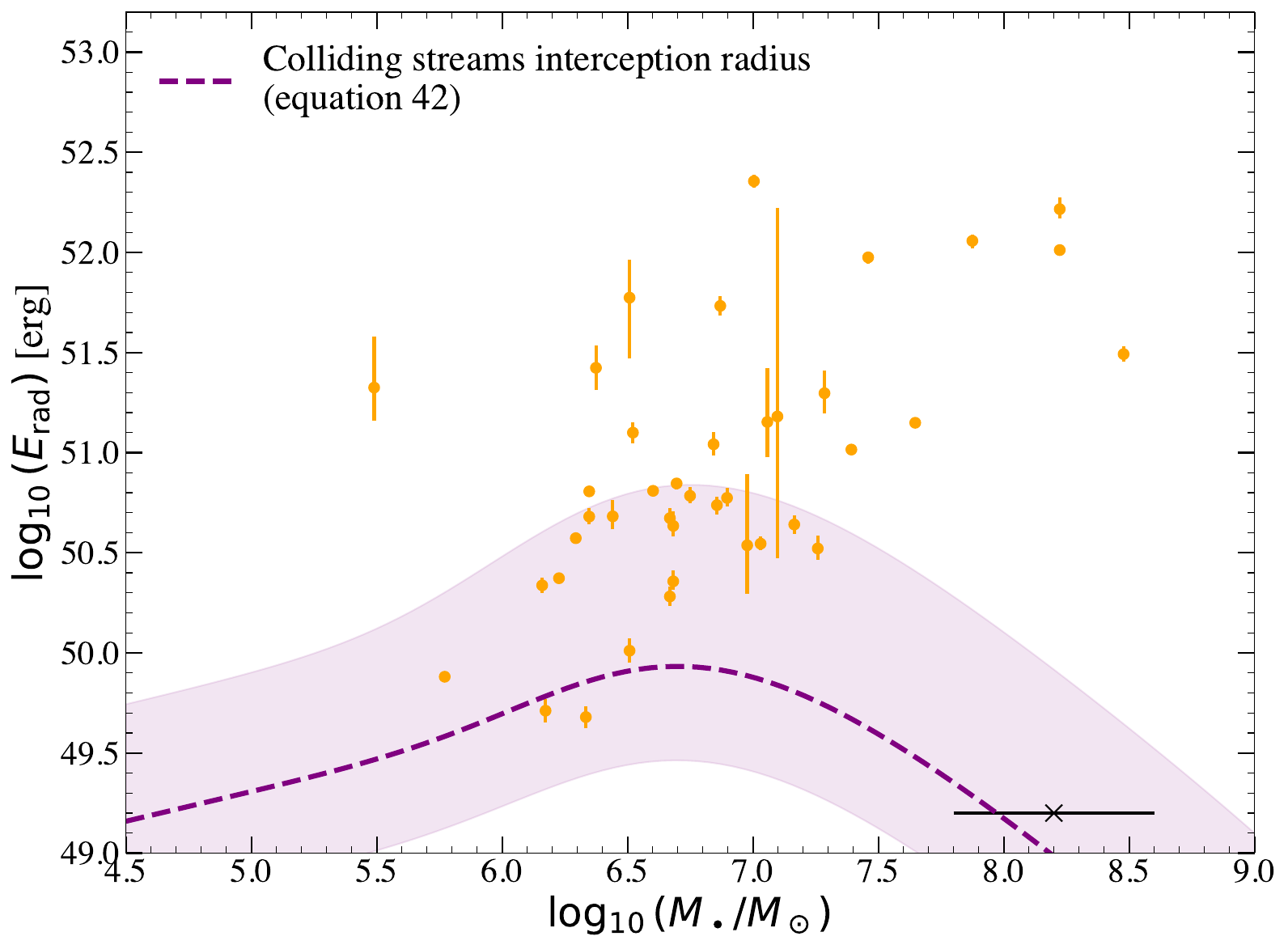}
    \includegraphics[width=0.32\linewidth]{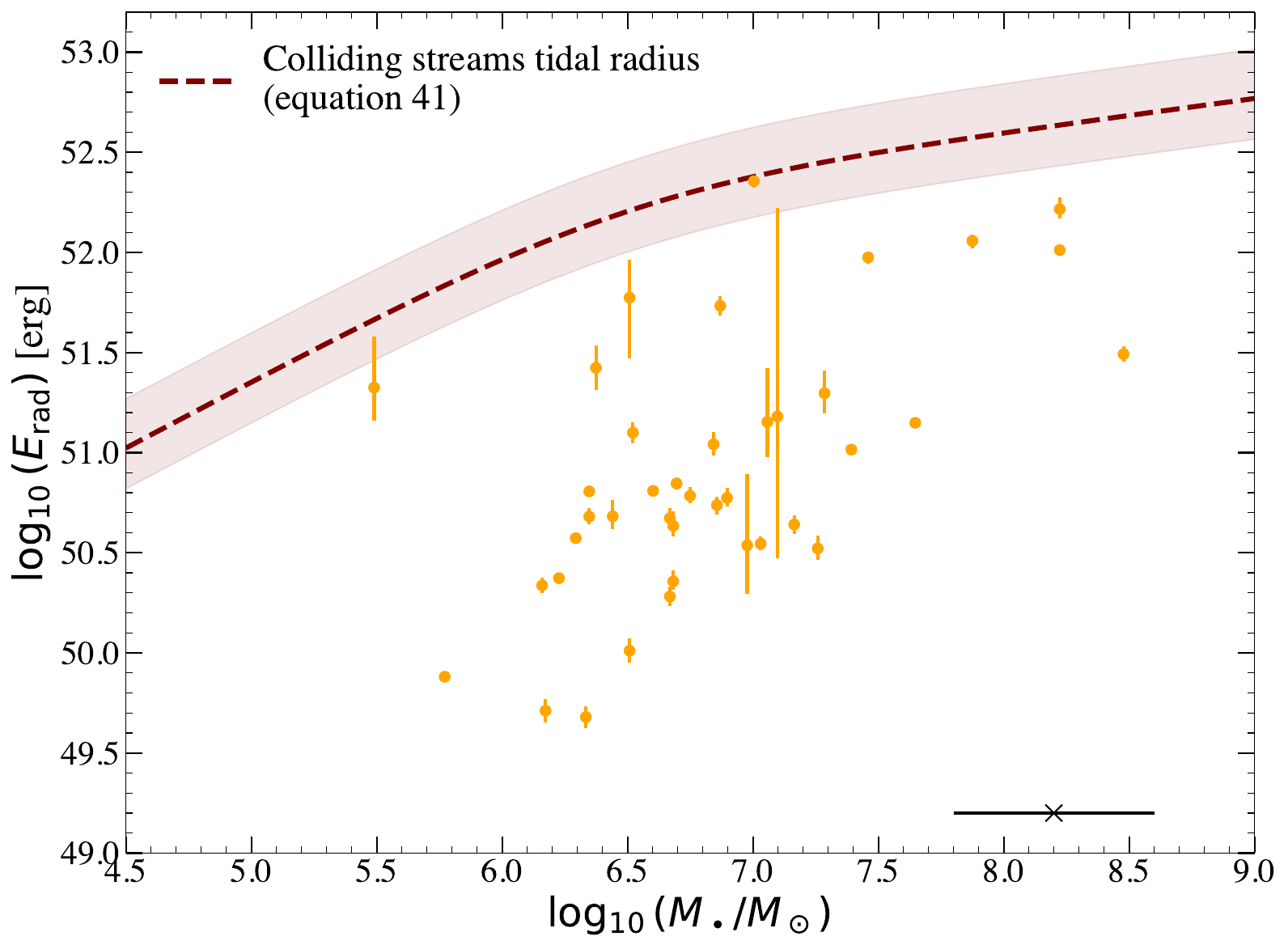}
    \includegraphics[width=0.32\linewidth]{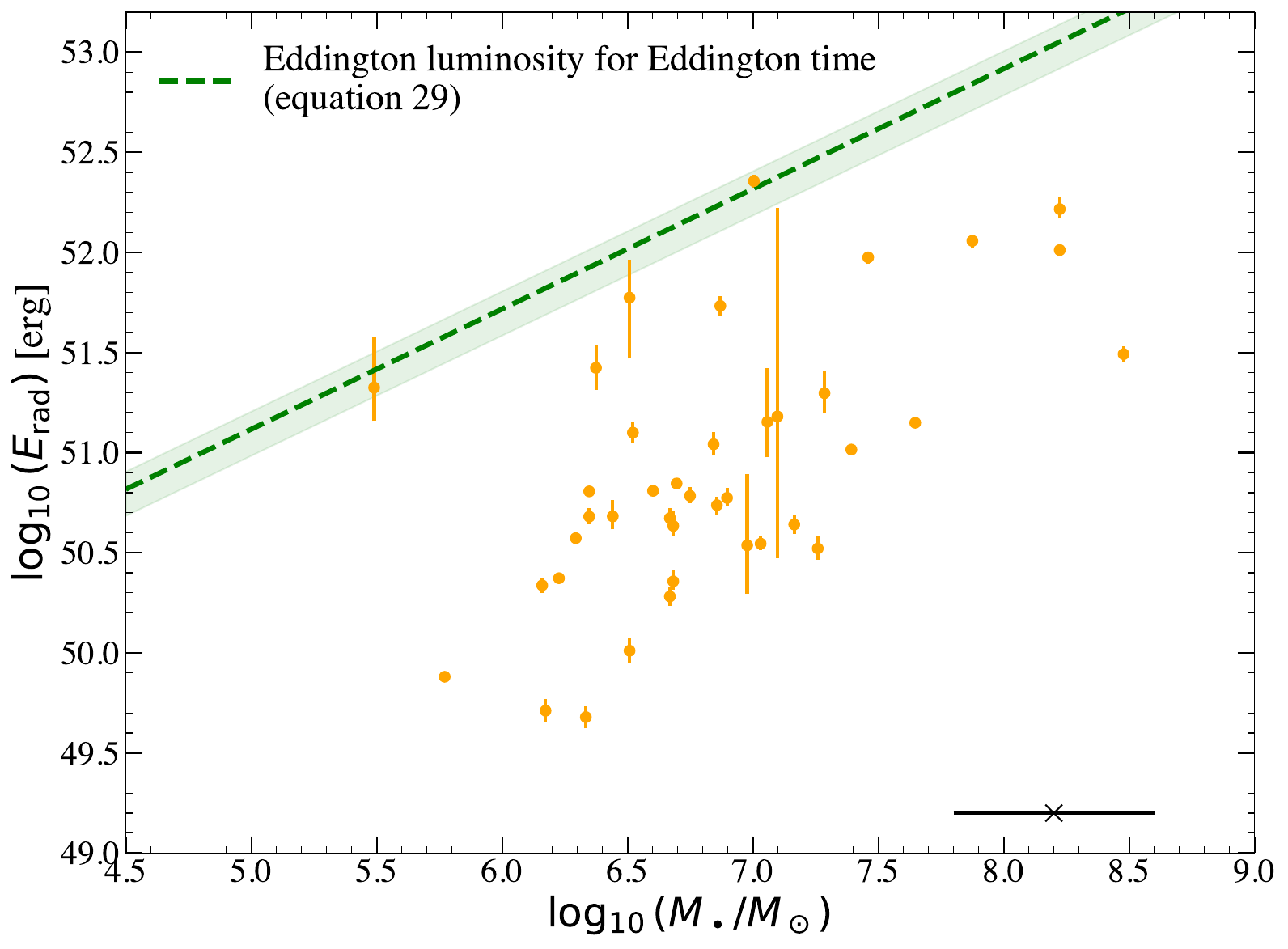}
    \includegraphics[width=0.32\linewidth]{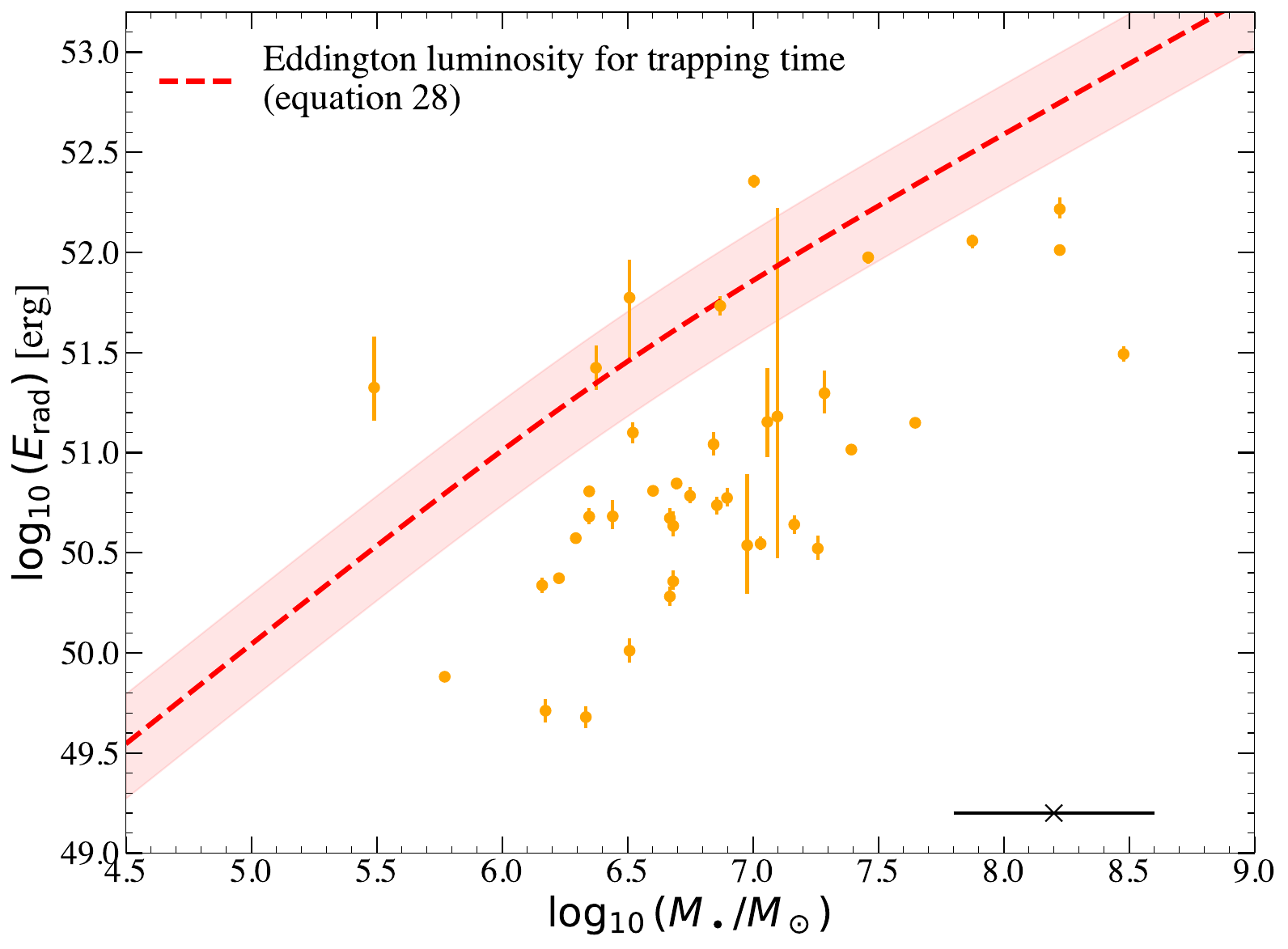}
    \includegraphics[width=0.32\linewidth]{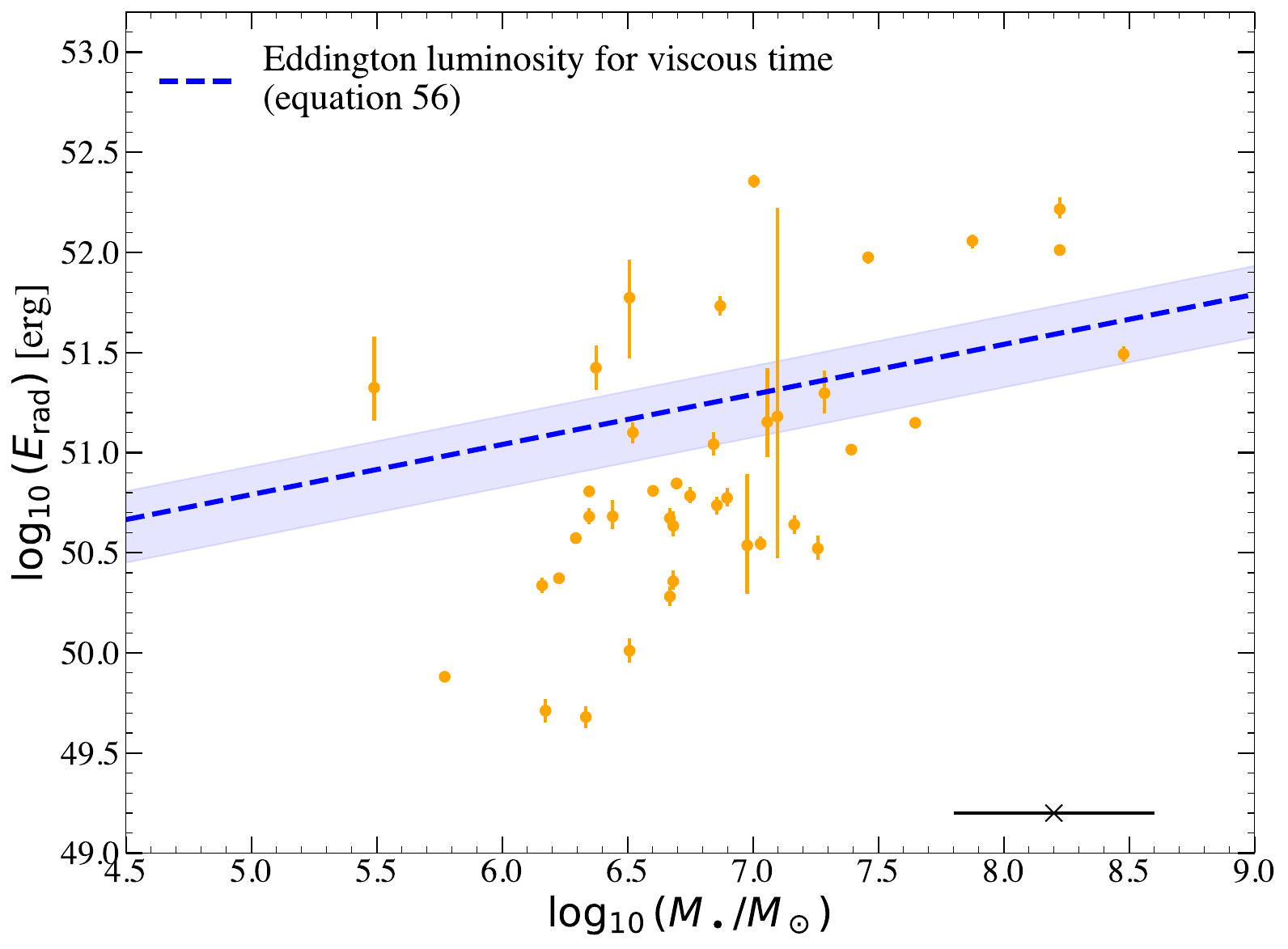}
    \includegraphics[width=0.32\linewidth]{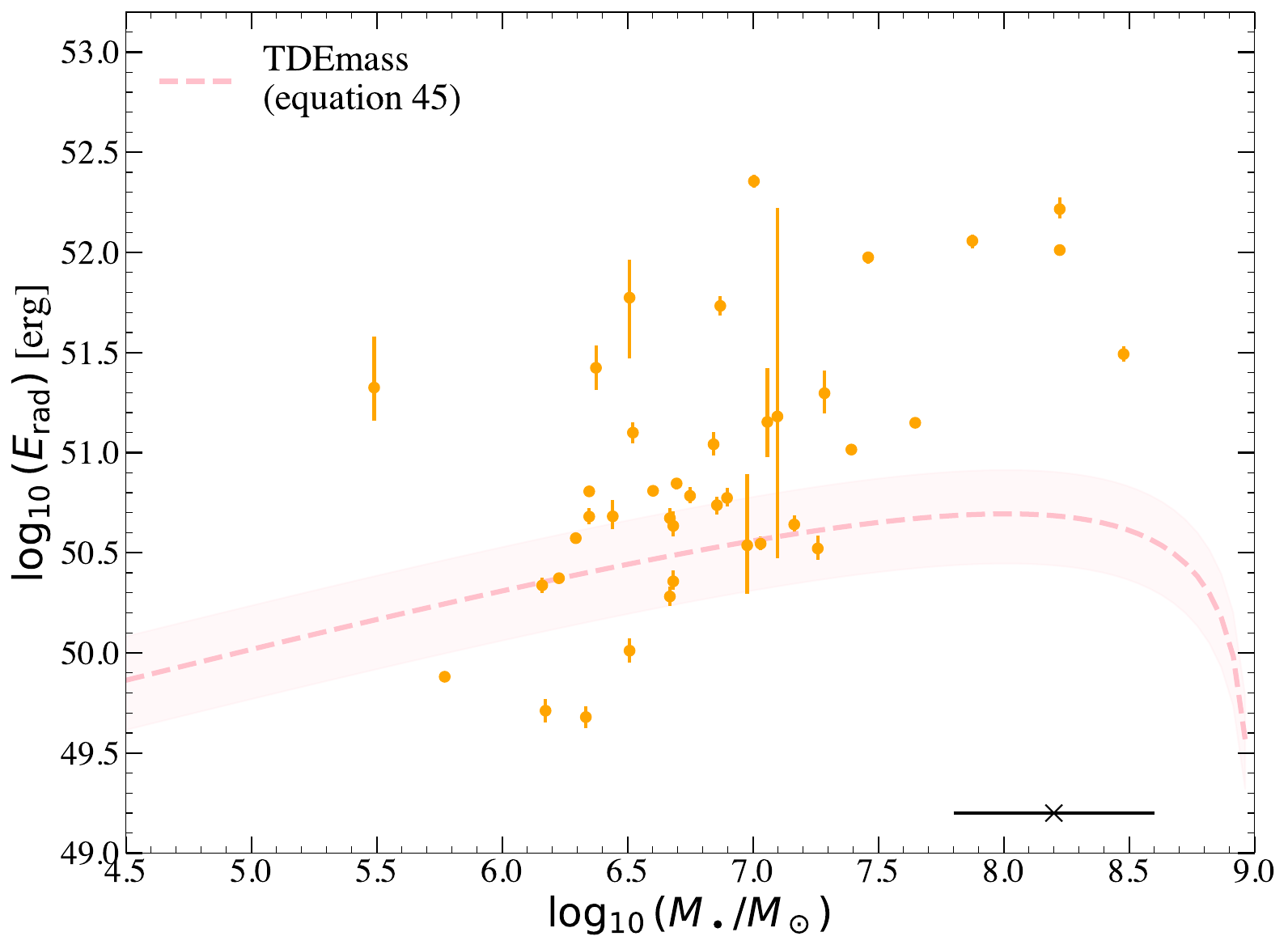}
    \includegraphics[width=0.32\linewidth]{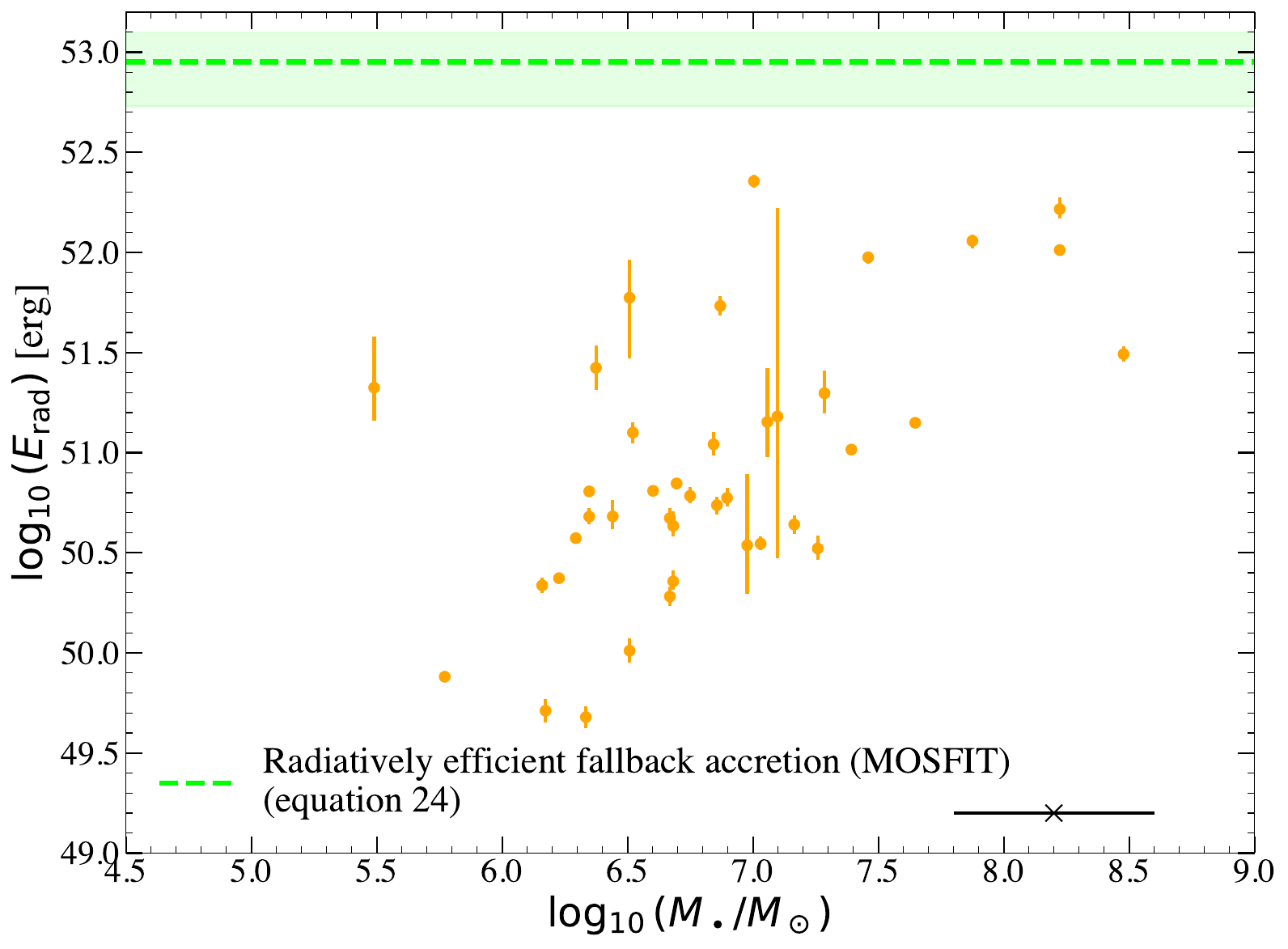}
    \includegraphics[width=0.32\linewidth]{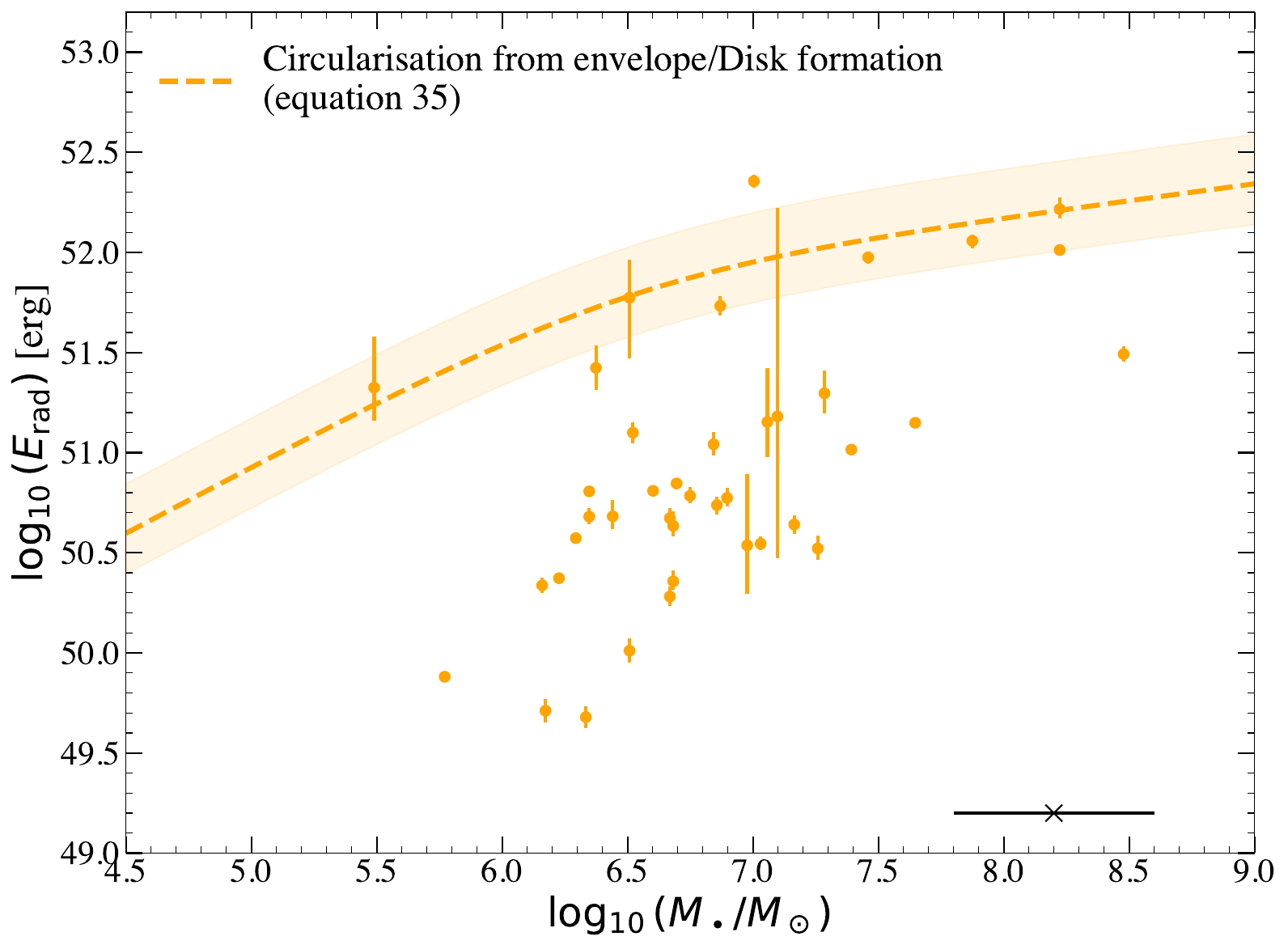}
    \includegraphics[width=0.32\linewidth]{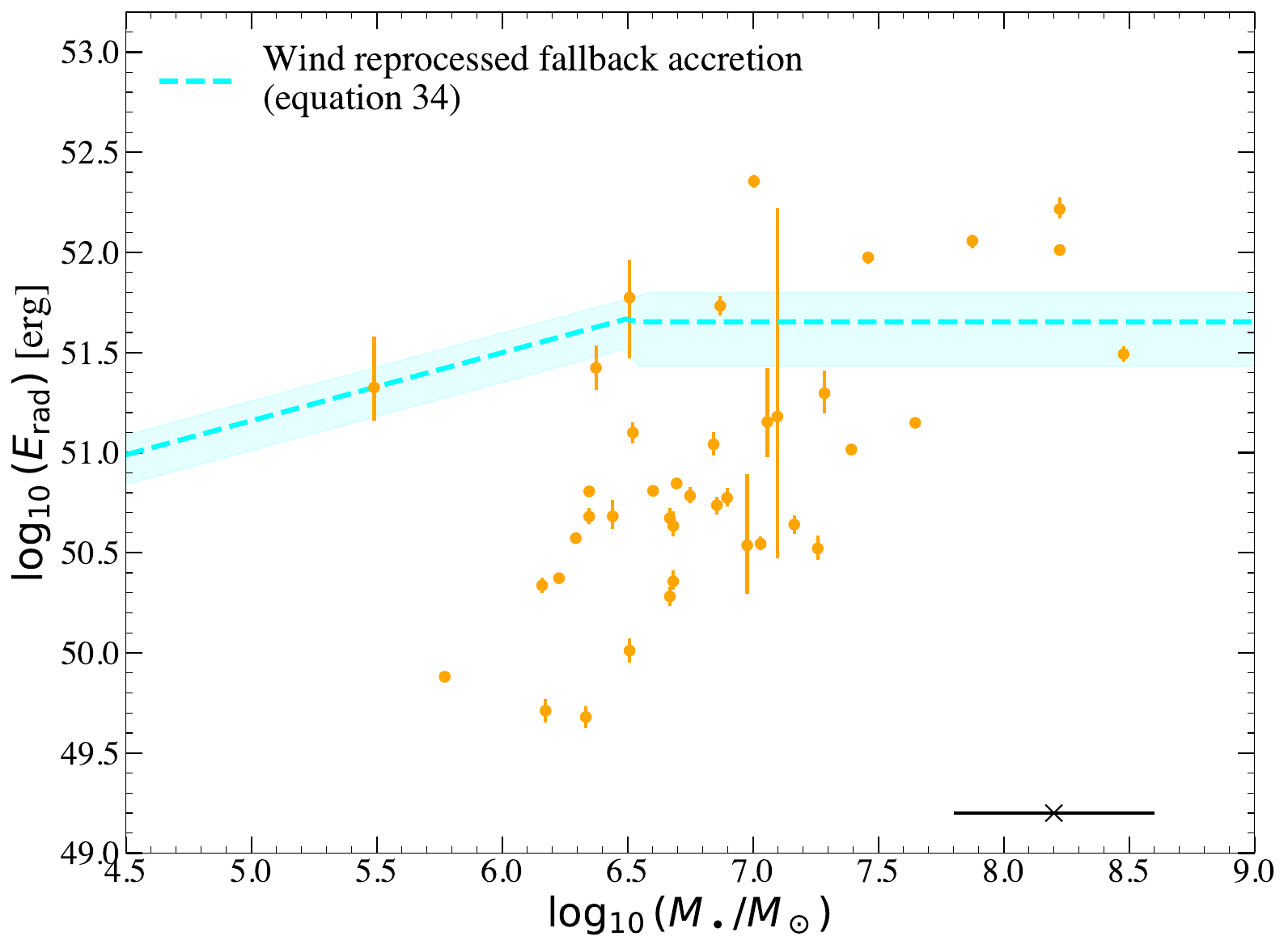}
    \caption{The same as Figure \ref{fig:en_comp}, except now with mass inferred from the galaxy bulge mass.  The choice of mass inference technique makes no difference to any conclusion in this paper. }
    \label{fig:en_comp_bulge}
\end{figure*}

In Figures \ref{fig:lum_comp_bulge} and \ref{fig:en_comp_bulge} we show the correlation between peak blackbody luminosity (Fig. \ref{fig:lum_comp_bulge}) and radiated energy (Fig. \ref{fig:en_comp_bulge}) with black hole mass measured from the galaxies bulge  mass, a measurement available for most sources in our sample (presented in \citealt{Ramsden25}). Our results and interpretation are completely unchanged.

\subsection{From velocity dispersion}
\begin{figure*}
    \centering
    \includegraphics[width=0.32\linewidth]{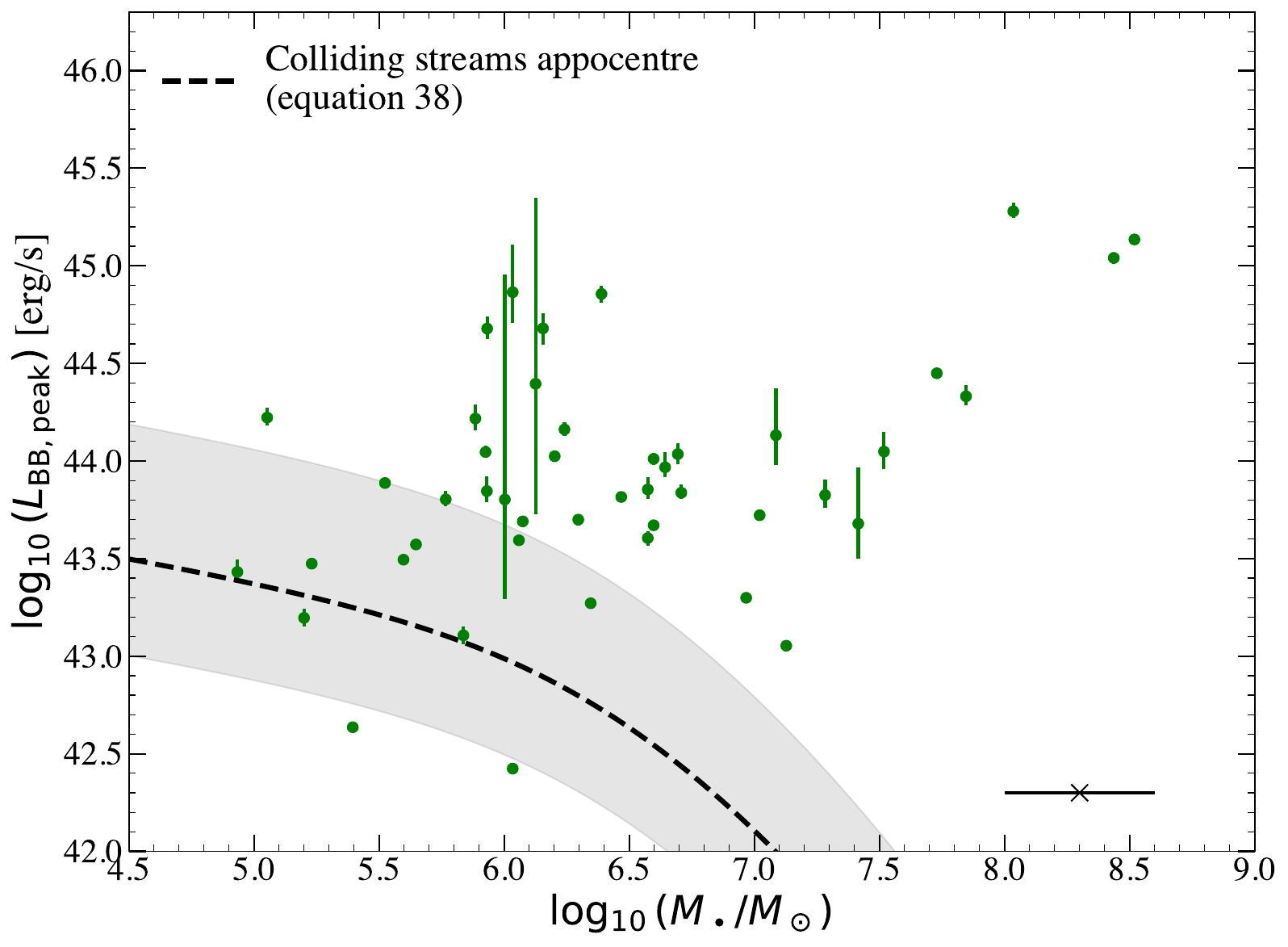}
    \includegraphics[width=0.32\linewidth]{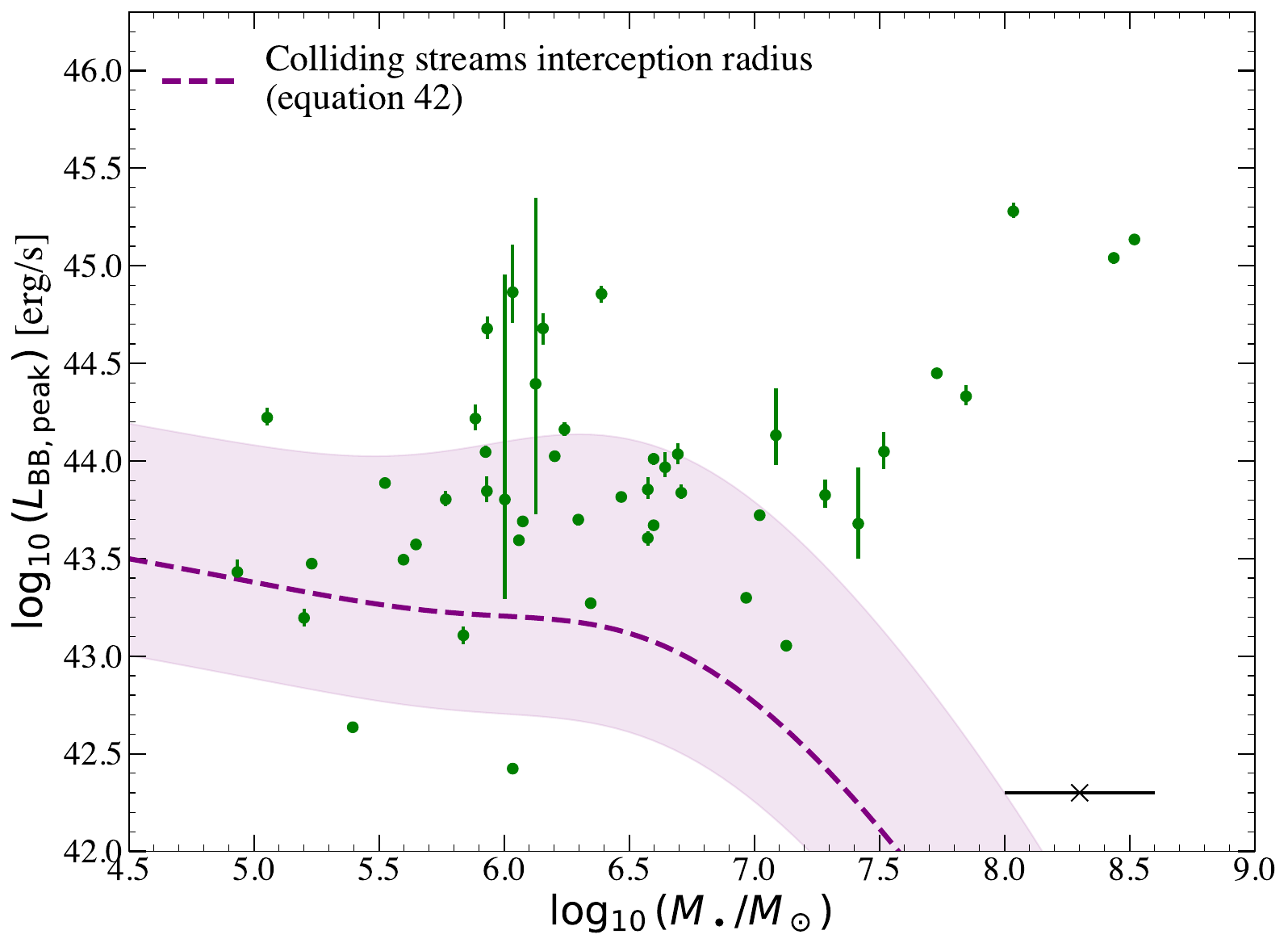}
    \includegraphics[width=0.32\linewidth]{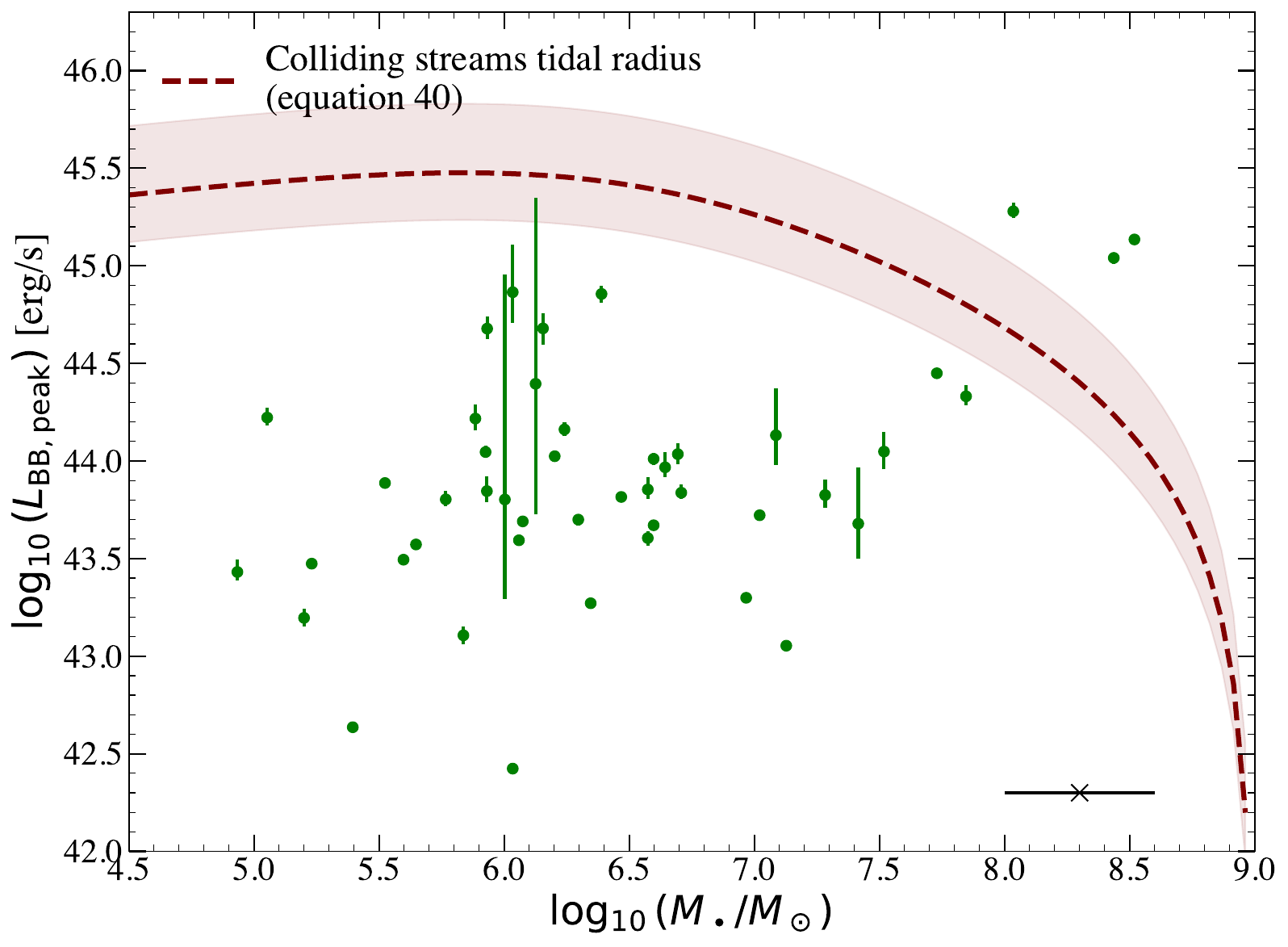}
    \includegraphics[width=0.32\linewidth]{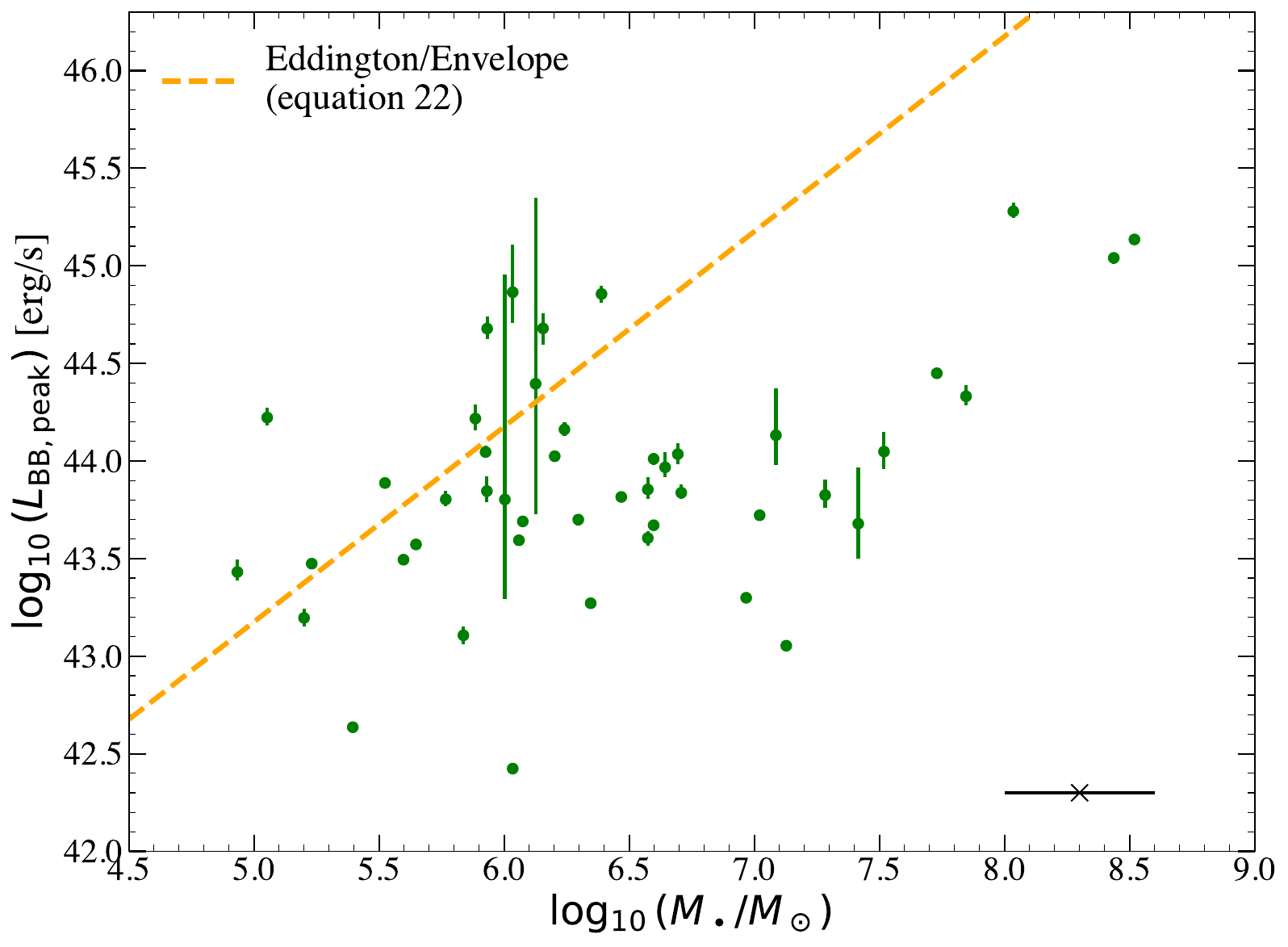}
    \includegraphics[width=0.32\linewidth]{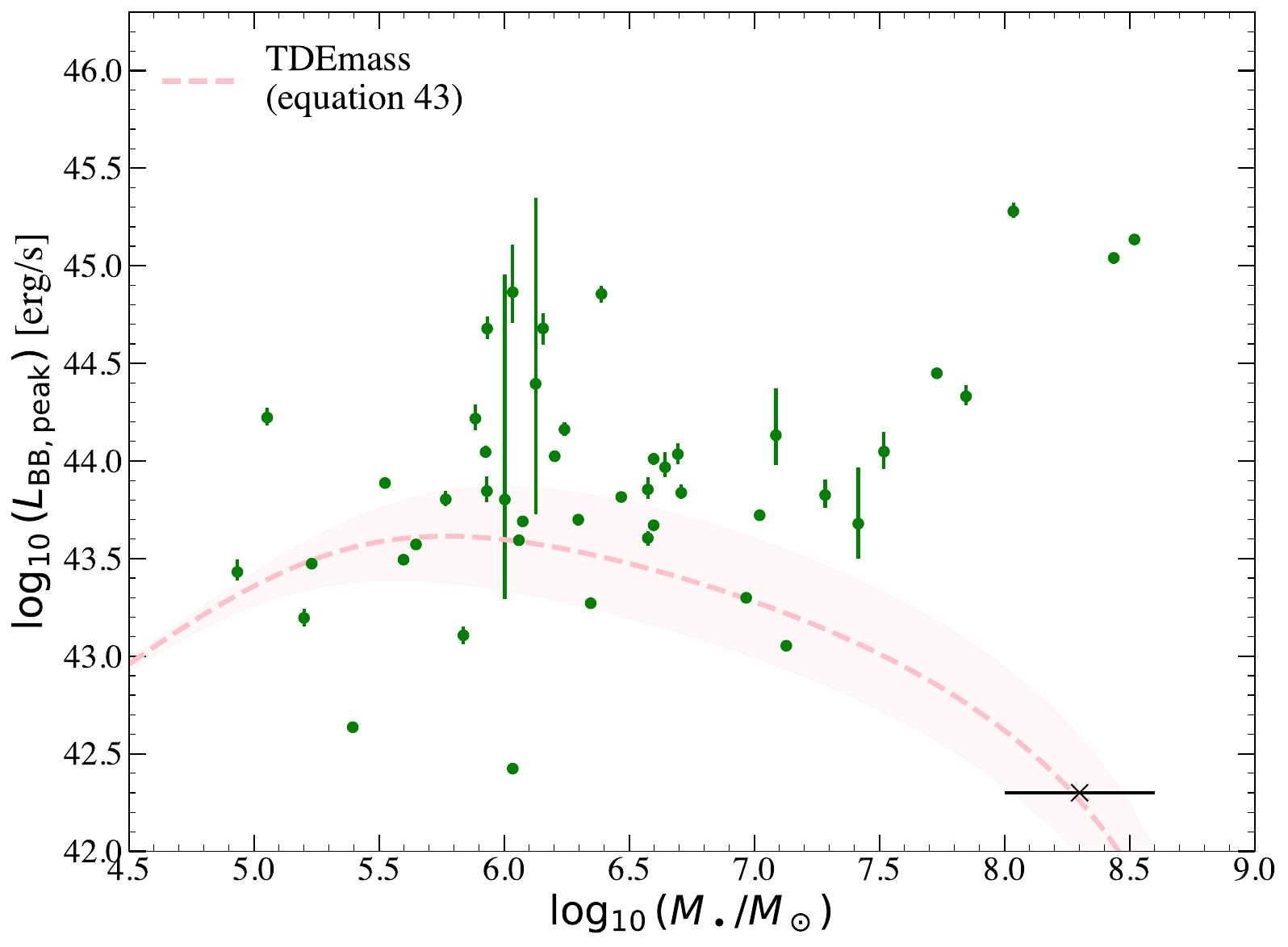}
    \includegraphics[width=0.32\linewidth]{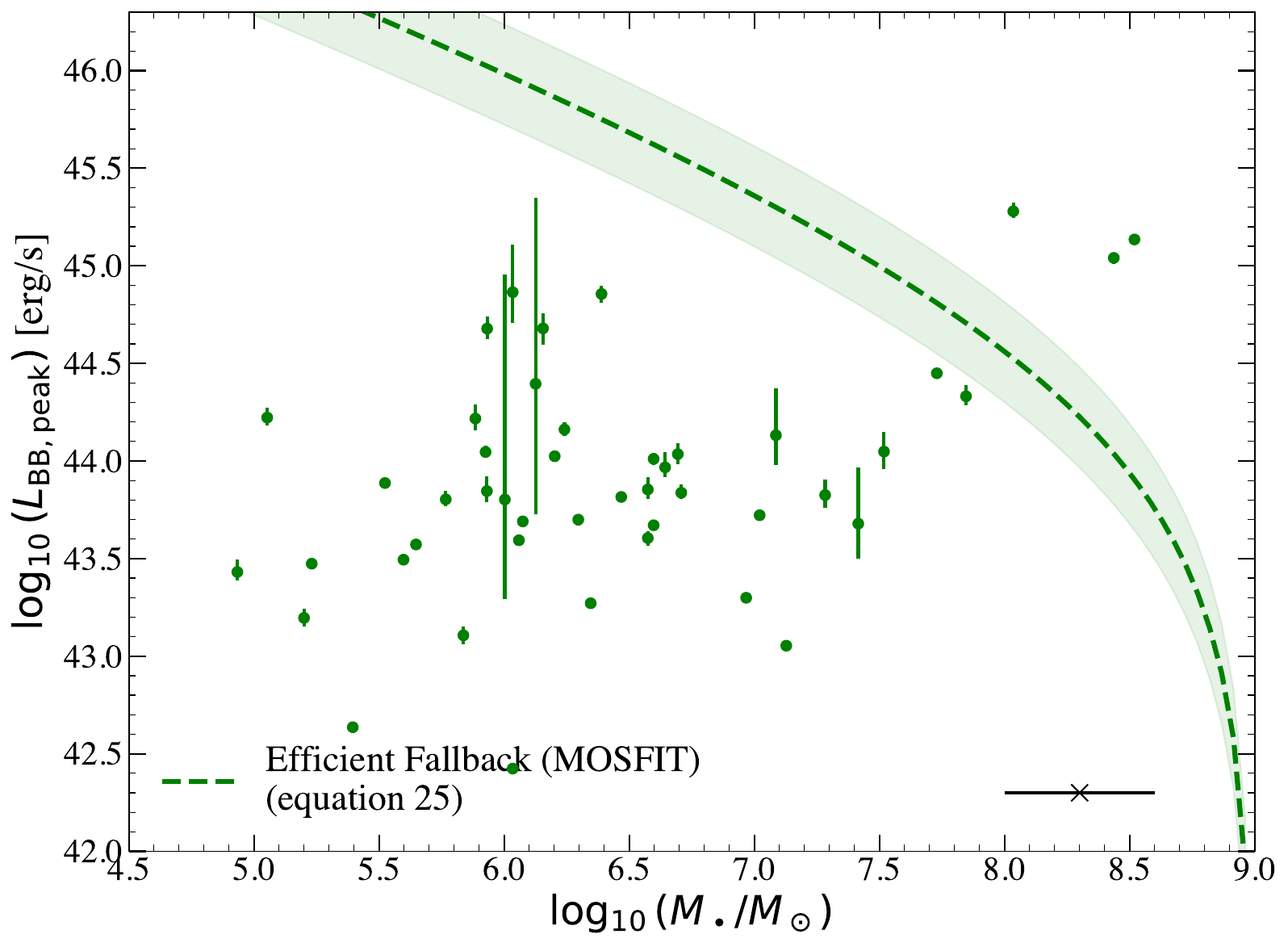}
    \includegraphics[width=0.32\linewidth]{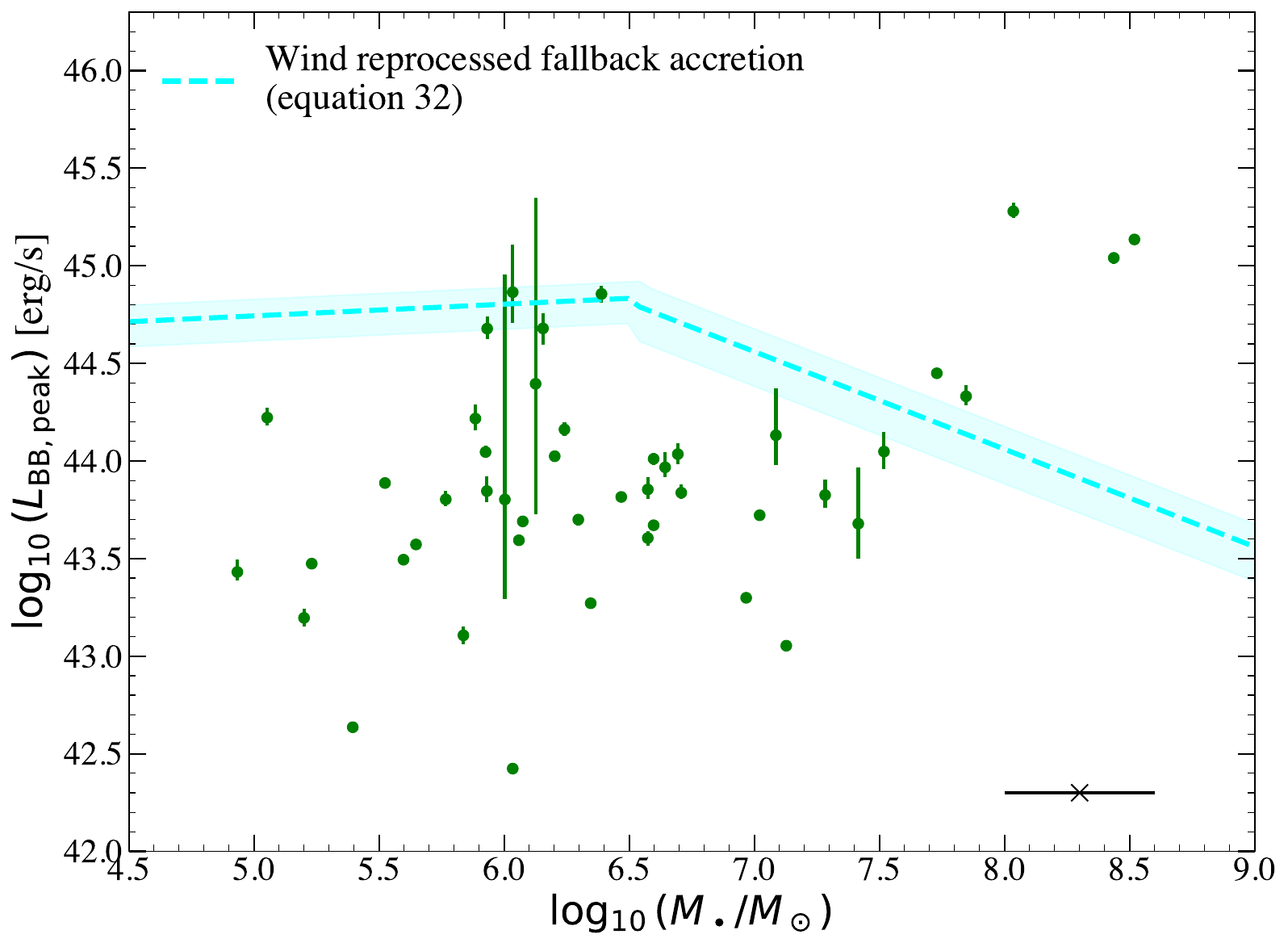}
    \caption{The same as Figure \ref{fig:lum_comp}, except now with mass inferred from the galaxy velocity dispersion.  The choice of mass inference technique makes no difference to any conclusion in this paper. }
    \label{fig:lum_comp_sigma}
\end{figure*}

\begin{figure*}
    \centering
    \includegraphics[width=0.32\linewidth]{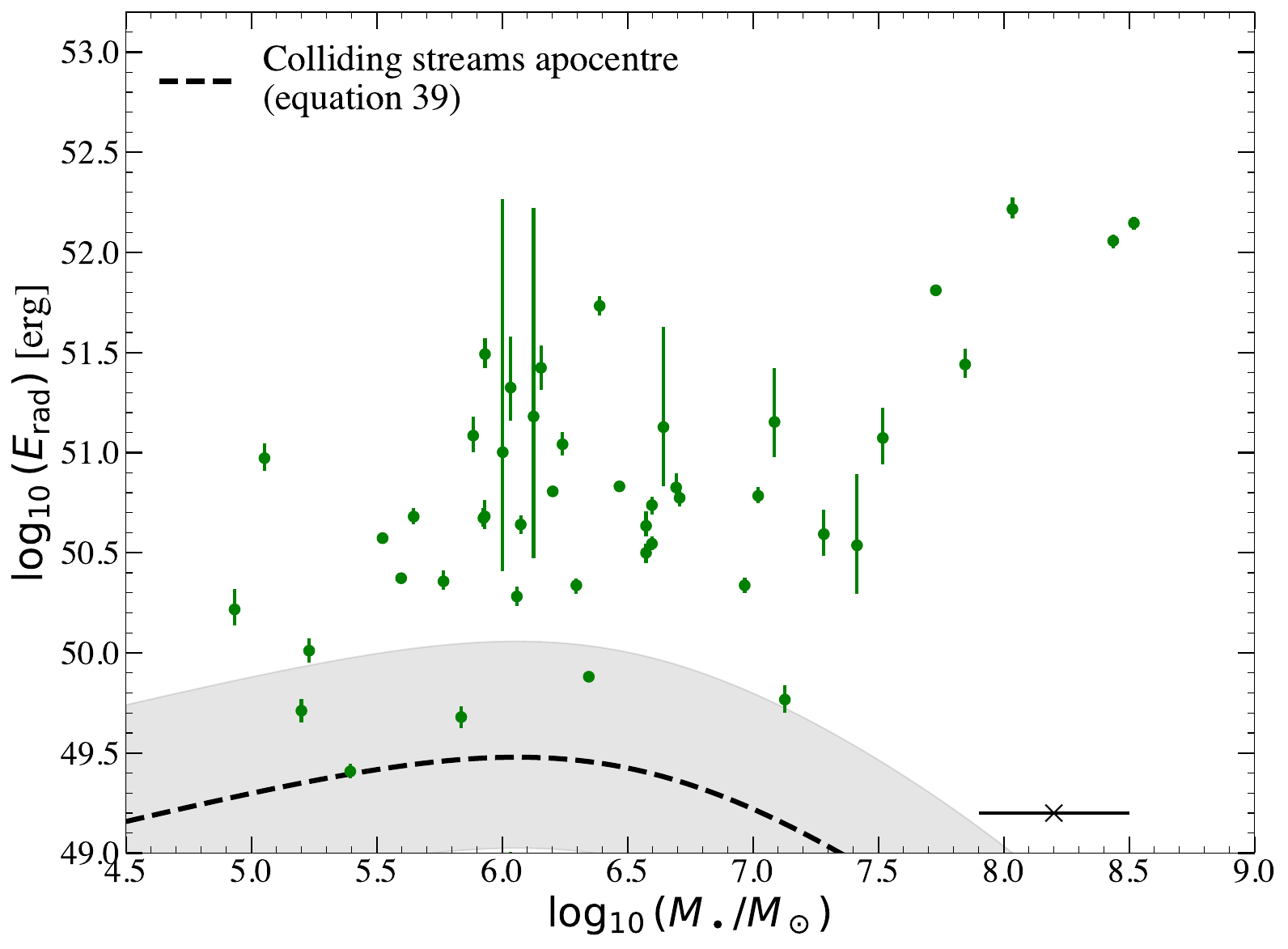}
    \includegraphics[width=0.32\linewidth]{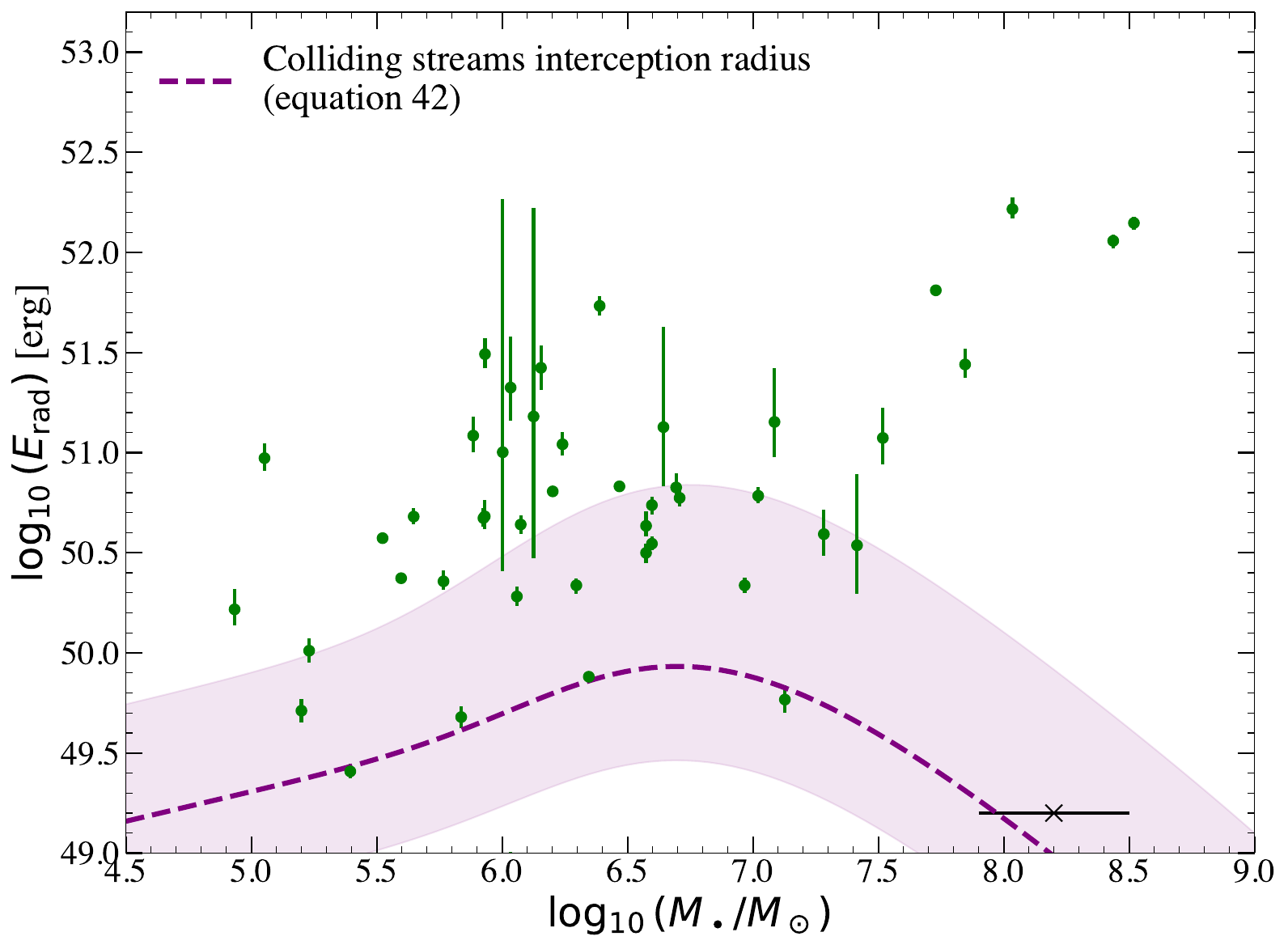}
    \includegraphics[width=0.32\linewidth]{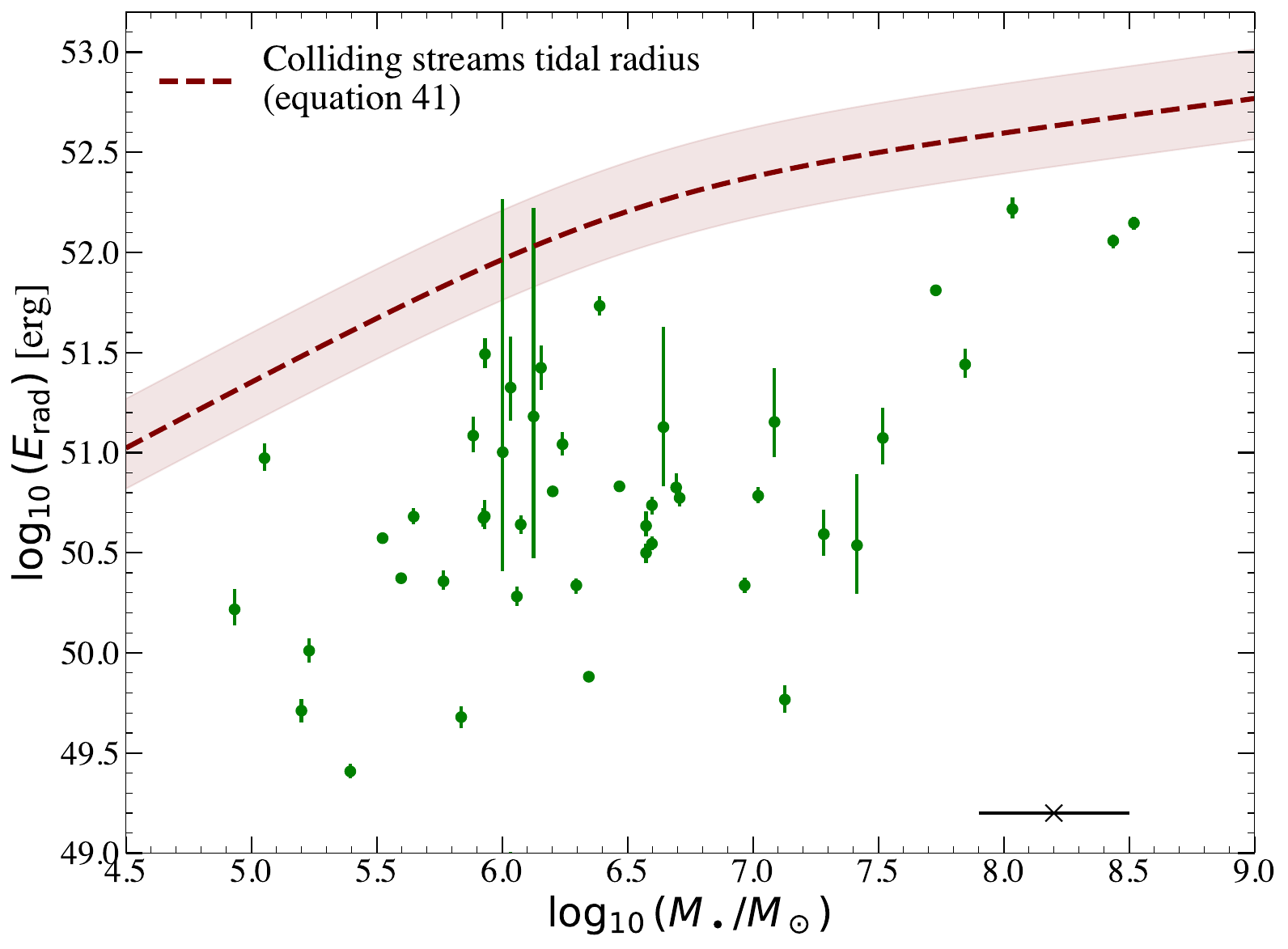}
    \includegraphics[width=0.32\linewidth]{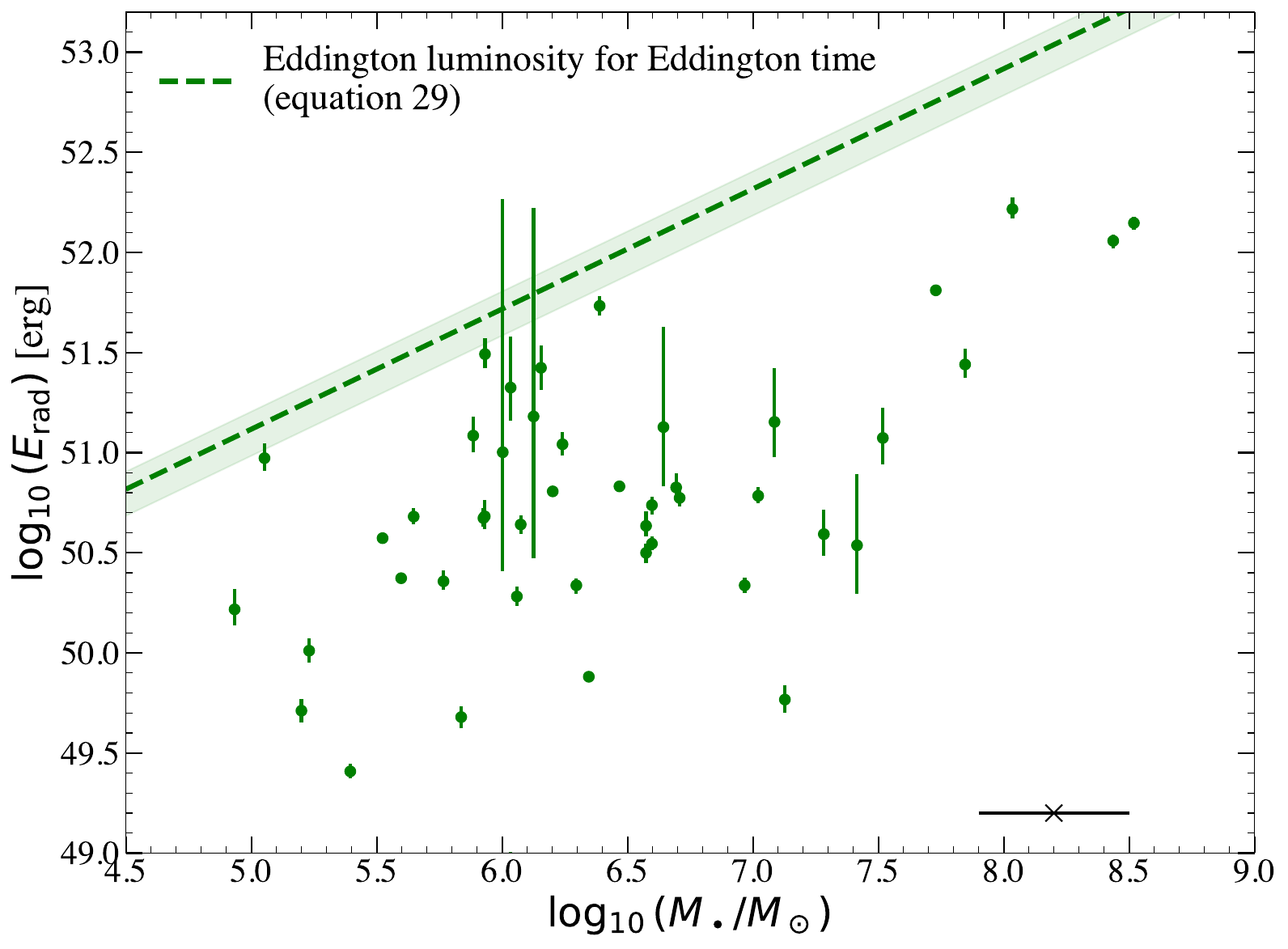}
    \includegraphics[width=0.32\linewidth]{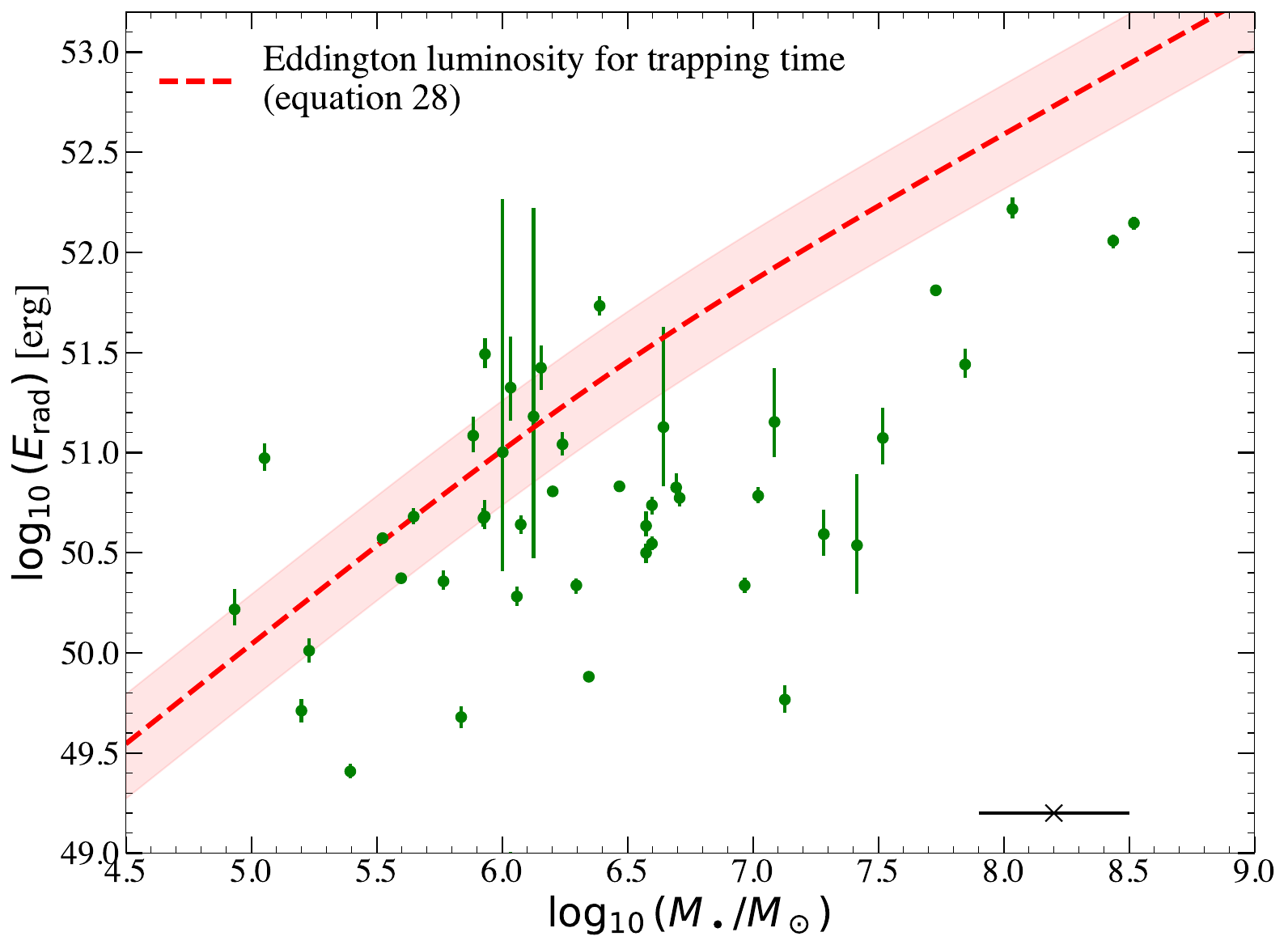}
    \includegraphics[width=0.32\linewidth]{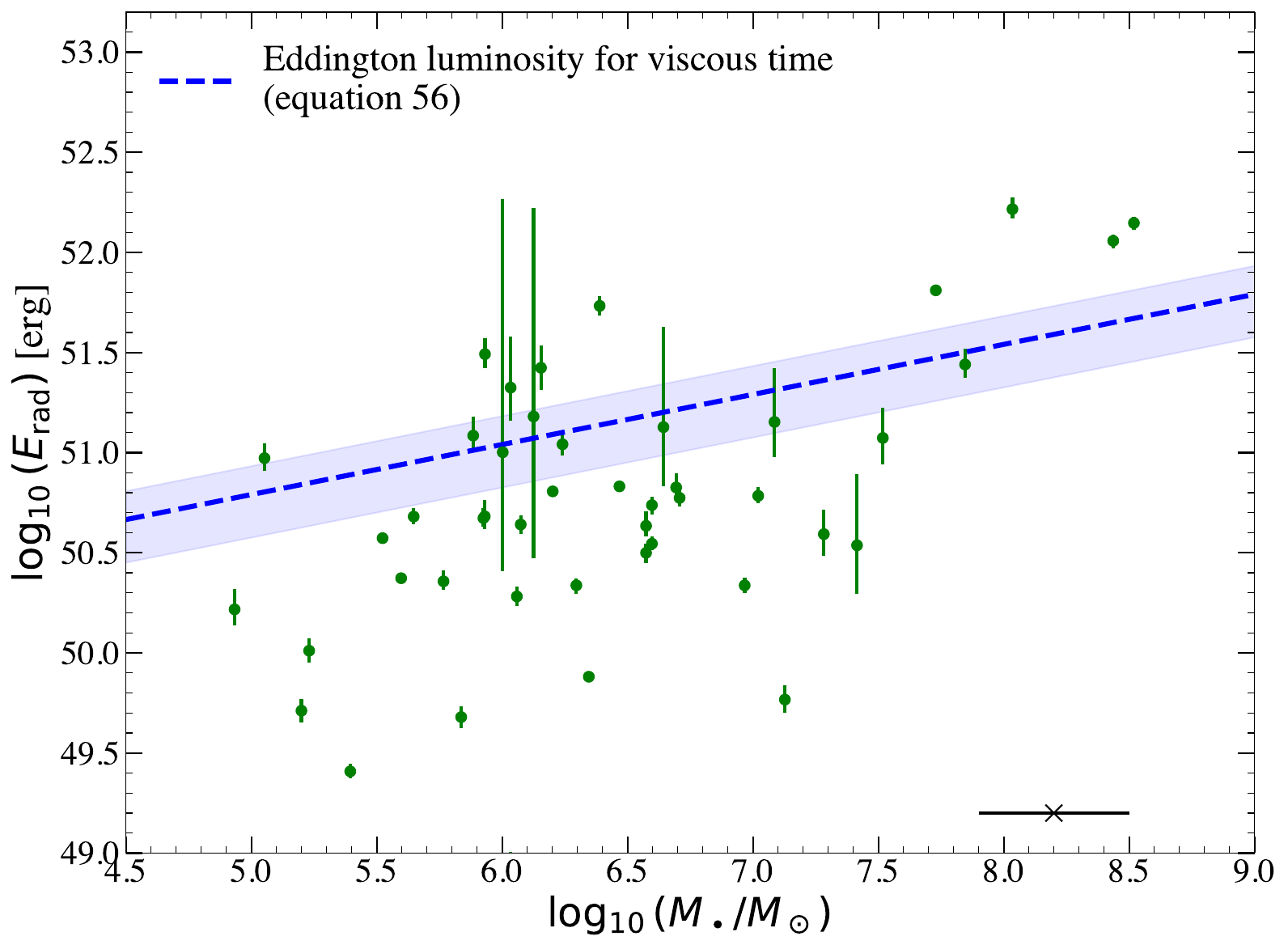}
    \includegraphics[width=0.32\linewidth]{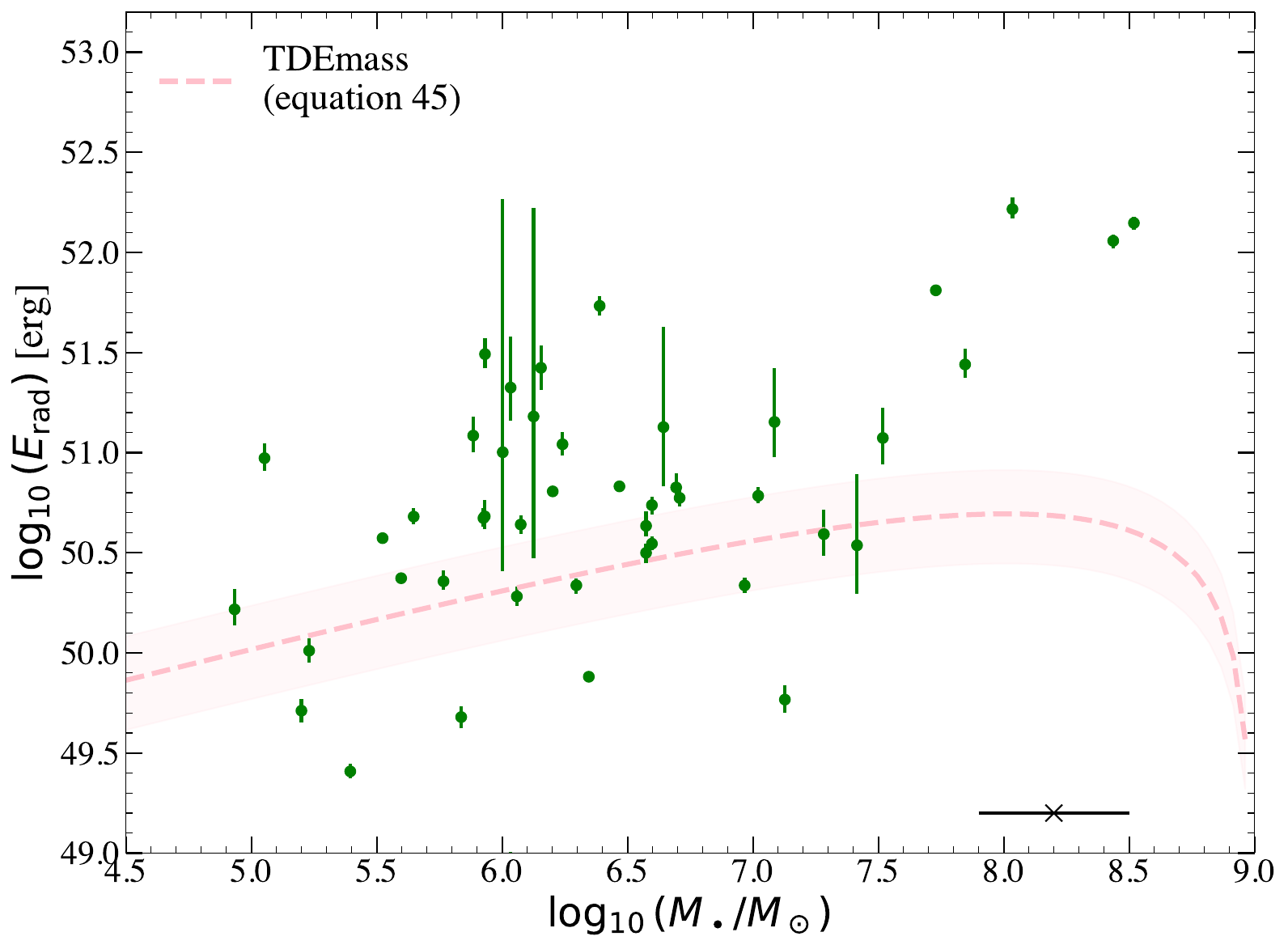}
    \includegraphics[width=0.32\linewidth]{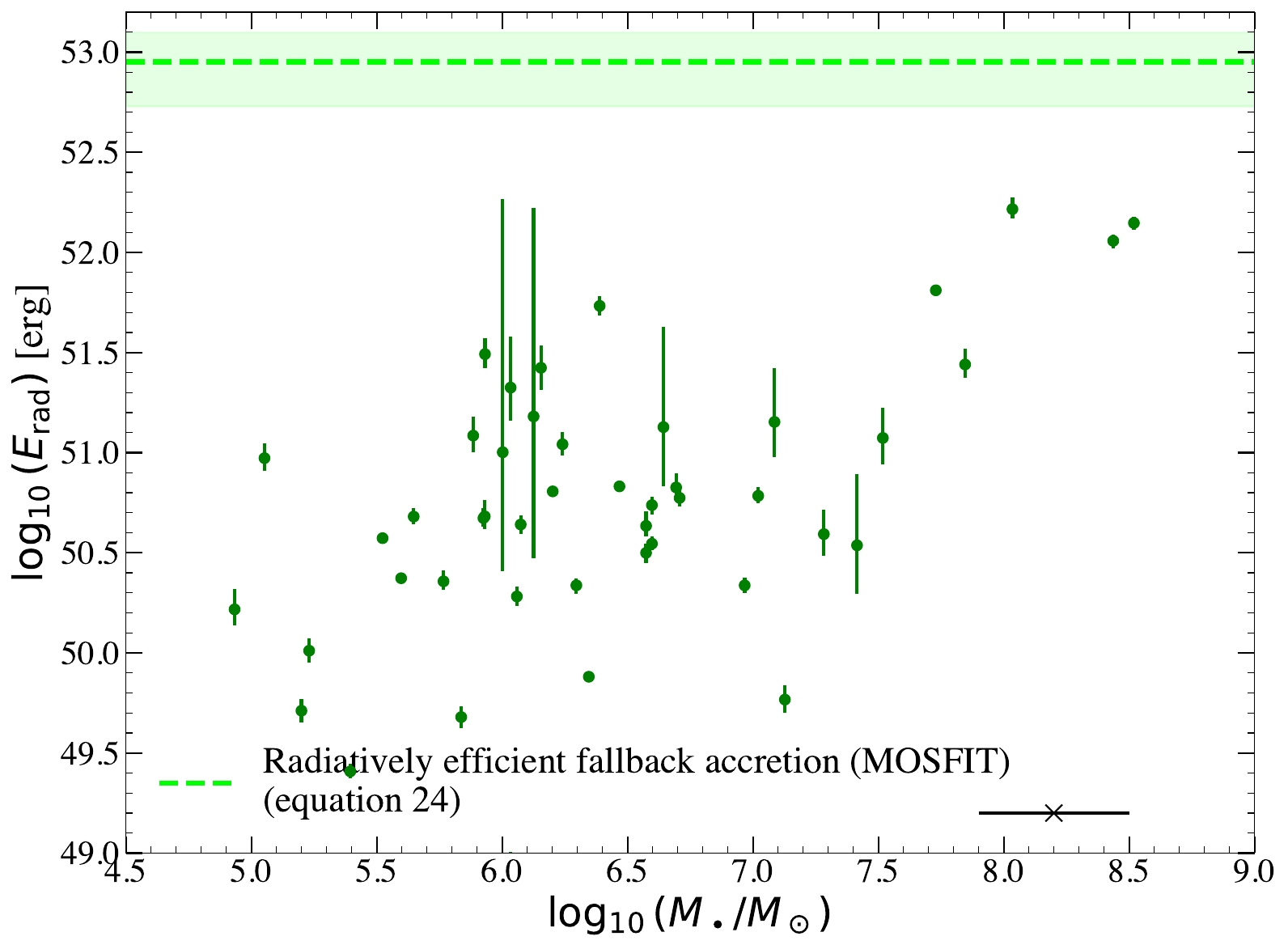}
    \includegraphics[width=0.32\linewidth]{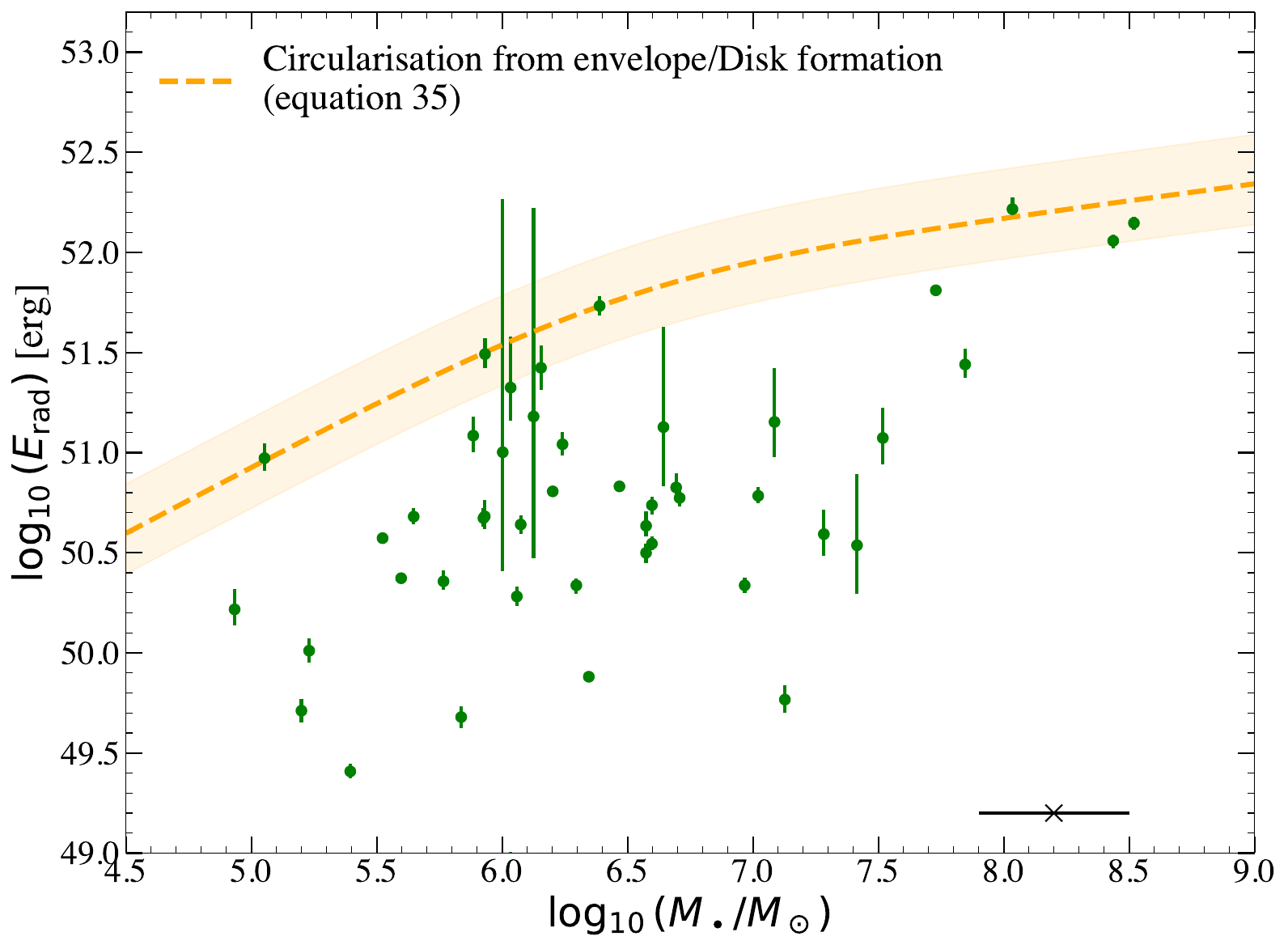}
    \includegraphics[width=0.32\linewidth]{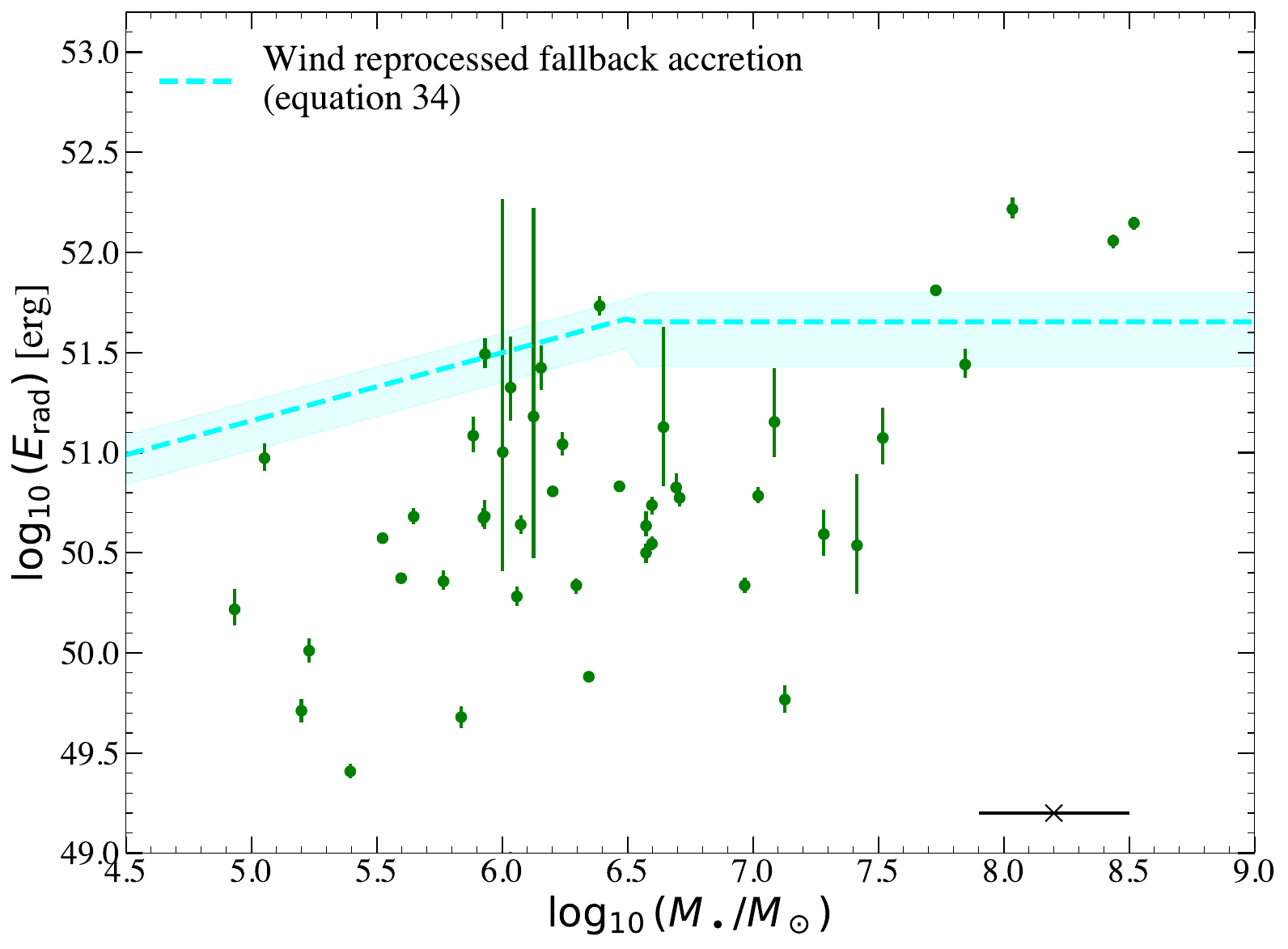}
    \caption{The same as Figure \ref{fig:en_comp}, except now with mass inferred from the galaxy velocity dispersion.  The choice of mass inference technique makes no difference to any conclusion in this paper. }
    \label{fig:en_comp_sigma}
\end{figure*}

In Figures \ref{fig:lum_comp_sigma} and \ref{fig:en_comp_sigma} we show the correlation between peak blackbody luminosity (Fig. \ref{fig:lum_comp_sigma}) and radiated energy (Fig. \ref{fig:en_comp_sigma}) with black hole mass measured from the galaxies velocity dispersion, a measurement available for a sub-set of our sources in our sample (presented in \citealt{Mummery_et_al_2024, MummeryVV25}). Our results and interpretation are completely unchanged.

\end{document}